\documentclass{elsarticle}

\usepackage[reviewcopy]{adndt}
\usepackage{longtable}

\usepackage{amsmath}

\usepackage{amssymb}

\biboptions{square,sort&compress}
\bibpunct[]{[}{]}{,}{n}{}{;}
\begin{document}

\begin{frontmatter}

\journal{Atomic Data and Nuclear Data Tables}


\title{Discovery of Isotopes of Elements with Z $\ge$ 100}

\author{M. Thoennessen\corref{cor1}}\ead{thoennessen@nscl.msu.edu}

 \cortext[cor1]{Corresponding author.}

 \address{National Superconducting Cyclotron Laboratory and \\ Department of Physics and Astronomy, Michigan State University, \\ East Lansing, MI 48824, USA}

\begin{abstract}
Currently, 163 isotopes of elements with Z $\ge$ 100 have been observed and the discovery of these isotopes is discussed here. For each isotope a brief synopsis of the first refereed publication, including the production and identification method, is presented.
\end{abstract}

\end{frontmatter}





\newpage
\tableofcontents
\listofDtables

\vskip5pc

\section{Introduction}\label{s:intro}

The discovery of isotopes of elements with Z $\ge$ 100 is discussed as part of the series summarizing the discovery of isotopes, beginning with the cerium isotopes in 2009 \cite{2009Gin01}. Guidelines for assigning credit for discovery are (1) clear identification, either through decay-curves and relationships to other known isotopes, particle or $\gamma$-ray spectra, or unique mass and Z-identification, and (2) publication of the discovery in a refereed journal. The authors and year of the first publication, the laboratory where the isotopes were produced as well as the production and identification methods are discussed. When appropriate, references to conference proceedings, internal reports, and theses are included. When a discovery includes a half-life measurement the measured value is compared to the currently adopted value taken from the NUBASE evaluation \cite{2003Aud01} which is based on the ENSDF database \cite{2008ENS01}. In cases where the reported half-life differed significantly from the adopted half-life (up to approximately a factor of two), the subsequent literature was searched for indications that the measurement was erroneous. If that was not the case the authors with the discovery in spite of the inaccurate half-life was credited.

The first criterium is not clear cut and in many instances debatable. Within the scope of the present project it is not possible to scrutinize each paper for the accuracy of the experimental data as is done for the discovery of elements \cite{1991IUP01}. In some cases an initial tentative assignment is not specifically confirmed in later papers and the first assignment is tacitly accepted by the community. The readers are encouraged to contact the authors if they disagree with an assignment because they are aware of an earlier paper or if they found evidence that the data of the chosen paper were incorrect. Measurements of half-lives of a given element without mass identification are not accepted.



In contrast to the criteria for the discovery of an element \cite{1976Har02,1990Sea01,1991IUP01} the criteria for the discovery or even the existence of an isotope are not well defined (see for example the discussion in reference \cite{2004Tho01}). Therefore it is possible, for example in the case of fermium, that the discovery of an element does not necessarily coincide with the first discovery of a specific isotope.

The ENSDF \cite{2008ENS01} and NSR \cite{2008NSR01,2011Pri01} databases of the National Nuclear Data Center at Brookhaven National Laboratory were used as the initial starting point for the literature searches. Additional excellent resources were the book ``The elements beyond uranium'' by Seaborg and Loveland \cite{1990Sea01}, the 1987 review article ``A history and analysis of the discovery of elements 104 and 105'' by Hyde, Hoffman, and Keller \cite{1987Hyd01}, and for the transfermium isotopes the technical reports of the International Union of Pure and Applied Chemistry (IUPAC) \cite{1993TWG01,2001Kar01,2003Kar01,2009Bar01,2011Bar01}. The original articles were researched and referenced for all nuclides. The cutoff date for publications included in the present overview was March 31$^{st}$, 2011.

In controversial cases, the judgement and recommendations of the IUPAC reports were followed. In cases where the initial mass assignment was incorrect, credit was given to the first subsequent publication which made the correct assignment.

The following sections describe details of the discovery of each nuclide in short paragraphs. These paragraphs contain quotes from the original papers where the original notation is maintained including the original element names even if they are not in use anymore. A summary of the discovery of each isotope  is presented in Table 1.


\section{Discovery of $^{241-259}$Fm}

The new element with Z = 100 was first identified by Ghiorso et al.\ in December 1952 from uranium which had been irradiated by neutrons in the ``Mike'' thermonuclear explosion on November 1, 1952 \cite{1955Ghi01}. However, the work was classified and could not be published. The authors realized the possibility that others could produce this new element independently, publish the results first and take credit for the discovery: ``At this juncture we began to worry that other laboratories might discover lighter isotopes of the elements 99 and 100 by the use of reactions with cyclotron-produced heavy ions. They would be able to publish that work without any problem and would feel that they should be able to name these elements. This might well happen before we could declassify the Mike work and it would make it difficult for us to claim priority in discovery. (Traditionally, the right to name a new element goes to the first to find it, but it is not clear that the world would accept that premise if the work is done secretly.)'' \cite{2000Hof01}.

The first observation of the fermium isotope $^{254}$Fm was submitted on January 15, 1954 by Harvey et al.\ \cite{1954Har01}. They did not want this observation to be regarded as the discovery of einsteinium and added the note: ``Because of the existence of unpublished information on element 100 the question of its first preparation should not be prejudged on the basis of this paper.'' The first identifications of $^{255}$Fm (March 19, 1954) and $^{256}$Fm (April 18, 1955 \cite{1955Cho01}) were also submitted  prior to the official announcement of the discovery of the new element. This announcement was finally made in the summer of 1955 with the publication of the article ``New Elements Einsteinium and Fermium, Atomic Numbers 99 and 100'' \cite{1955Ghi01}.

The early accounts of the events were not specific about the details: ``Without going into the details, it may be pointed out that such experiments involving the groups at the three laboratories led to the positive identification of isotopes of elements 99 and 100'' \cite{1958Sea01,1963Sea01,1990Sea01}. Only later were the difficult discussions regarding the publication strategy between the research groups involved described in detail \cite{2000Hof01}. It is interesting to note that the loss of life during the collection of samples from the thermonuclear explosion was only mentioned in the more recent accounts \cite{1978Sea01,1990Sea01} of the discovery of these transuranium elements: ``These samples cost the life of First Lieutenant Jimmy Robinson, who waited too long before he went home, tried to land on Eniwetok, and ditched about a mile short of the runway'' \cite{1978Sea01}.

In the discovery paper the authors suggested to name the new element with Z = 100 Fermium with the symbol ``Fm''. IUPAC adopted the name and the symbol at the 19$^{th}$ IUPAC Conference in Paris 1957 \cite{2005Kop01,1957IUP01}.

\subsection*{$^{241}$Fm}\vspace{0.0cm}
J. Khuyagbaatar et al.\ discovered $^{241}$Fm in ``Spontaneous fission of neutron-deficient fermium isotopes and the new nucleus $^{241}$Fm'' in 2008 \cite{2008Khu01}. An enriched $^{204}$Pb target was bombarded with 187$-$206~MeV $^{40}$Ar beams from the GSI UNILAC accelerator forming $^{241}$Fm in the (3n) fusion-evaporation reaction. Recoil products were separated with the velocity filter SHIP and implanted in a position-sensitive 16-strip Si detector. Subsequent emission of $\alpha$-particles and spontaneous fission were detected in the implantation detector as well as in a box detector mounted in the backward hemisphere. ``We observed a total number of 145 ER-SF events. The time distribution of these events is shown in [the figure]. The resulting lifetime of (1.05$\pm$0.09)~ms, T$_{1/2}$ = (0.7$\pm$±0.06)~ms, is definitely shorter than that of the other fermium isotopes shown in [the figure].''

\subsection*{$^{242}$Fm}\vspace{0.0cm}
In the 1975 paper ``Synthesis of the new neutron-deficient isotopes $^{250}$102, $^{242}$Fm, and $^{254}$Ku'' Ter-Akopyan et al.\ reported the first observation of $^{242}$Fm \cite{1975Ter01}. $^{40}$Ar beams with energies up to 225~MeV from the Dubna U-300 cyclotron bombarded $^{204}$Pb and $^{206}$Pb targets and $^{242}$Fm was populated in the (2n) and (4n) fusion-evaporation reactions, respectively. Spontaneous fission fragments were measured with mica detectors. ``A comparison of the $^{204}$Pb and $^{206}$Pb target yields permits the assignment of the 0.8~msec activity to the isotope $^{242}$Fm.'' This half-life has been recommended in an IUPAC technical report \cite{2000Hol01}, however, more recently, the data could not be reproduced \cite{2008Khu01}.

\subsection*{$^{243}$Fm}\vspace{0.0cm}
M\"unzenberg et al.\ reported the discovery of $^{243}$Fm in 1981 in ``The new isotopes $^{247}$Md, $^{243}$Fm,$^{239}$Cf, and investigation of the evaporation residues from fusion of $^{206}$Pb, $^{208}$Pb, and $^{209}$Bi with $^{40}$Ar'' \cite{1981Mun01}. A $^{206}$Pb target was bombarded with a 4.8~MeV/u $^{40}$Ar beam from the GSI UNILAC accelerator to form $^{243}$Fm in the (3n) fusion-evaporation reaction. Recoil products were separated with the velocity filter SHIP and implanted in an array of position sensitive surface-barrier detector which also recorded subsequent $\alpha$ decay and spontaneous fission. ``Correlated to
this decay we observed a daughter decay of (7,630$\pm$25)keV and a half1ife of (39$^{+37}_{-12}$)s. We assign these two decays to $^{243}$Fm and its daughter $^{239}$Cf.''

\subsection*{$^{244,245}$Fm}\vspace{0.0cm}
Nurmia et al.\ reported the discovery of $^{244}$Fm and $^{245}$Fm in the 1967 article ``Spontaneous fission of light fermium isotopes; New nuclides $^{244}$Fm and $^{245}$Fm'' \cite{1967Nur01}. A $^{16}$O beam from the Berkeley heavy-ion accelerator Hilac bombarded a $^{233}$U target forming $^{244}$Fm and $^{245}$Fm in (5n) and (4n) fusion-evaporation reactions, respectively. Spontaneous fission events from $^{244}$Fm were recorded using mica to scan a rotating drum, recoil-collection device and a half-life of 3.3(5)~ms was measured: ``The activity was not produced in bombardments of the same target with $^{14}$N ions. This was taken as indicating that the activity is unlikely due to an isotope or isomeric state of an element other than fermium. The above evidence suggests the assignment of $^{244}$Fm to this activity.'' For $^{245}$Fm recoil products were moved in front of semiconductor detectors with a conveyor-gas system to measure $\alpha$-particles and spontaneous fission and an $\alpha$-decay half-life of 4.2(13)~s was measured: ``The activity was assigned to $^{245}$Fm on the basis of its production in the cross-bombardments with the expected excitation functions and from $\alpha$ decay systematics.''

\subsection*{$^{246}$Fm}\vspace{0.0cm}
The observation of $^{246}$Fm was reported in ``Synthesis of several isotopes of fermium and determination of their radioactive properties'' by Akapev et al.\ in 1966 \cite{1966Aka01}. $^{16}$O beams with energies of 80$-$105~MeV from the Dubna 310-cm cyclotron bombarded an enriched $^{235}$U target and $^{246}$Fm was produced in the (5n) fusion-evaporation reaction. Recoil products were transported in front of a semiconductor detector with a helium gas stream to measure subsequent $\alpha$ decay. ``For O$^{16}$ ion energies corresponding to the estimated values of the maxima of the excitation function of the U$^{235}$(O$^{16}$,5n) reaction, we obtained an activity with T$_{1/2}$=1.4$\pm$0.6~sec and an energy of E$_\alpha$=8.23$\pm$0.02~MeV... As the energy of the bombarding ions is increased, the yield of this activity decreases, in accordance with the behavior of the excitation function of a complete-fusion reaction with the evaporation of five neutrons. It must be assumed that this $\alpha$ activity belongs to Fm$^{246}$.''

\subsection*{$^{247}$Fm}\vspace{0.0cm}
In 1967, Flerov et al.\ identified $^{247}$Fm in the paper ``Synthesis of isotopes of fermium with mass numbers 247 and 246'' \cite{1967Fle01}. A $^{239}$Pu target was bombarded with 72$-$74~MeV $^{12}$C from the Dubna 310-cm heavy-ion cyclotron and $^{247}$Fm was formed in (4n) fusion-evaporation reactions. Recoil products were collected with an oriented gas jet and subsequent $\alpha$ decay was measured with Si(Au) detectors. ``Based on the coincidence of the half lives and excitation functions, it can be concluded that the activities with E$_0$ equal to 7.87$\pm$0.05 and 7.93$\pm$0.05~MeV are associated with the decay of Fm$^{247}$ from a single state. For a more accurate determination of the half life of Fm$^{247}$ in this state, measurements were carried out in the cycle with $\tau$=200~sec. The results of the measurements are shown in [the figure] from which a value of T$_{1/2}$ = 35$\pm$4~sec is obtained.''

\subsection*{$^{248}$Fm}\vspace{0.0cm}
Ghiorso et al.\ reported the observation of $^{248}$Fm in the 1958 paper ``Element No.\ 102'' \cite{1958Ghi01}. A $^{240}$Pu target was bombarded with a $^{12}$C beam from the Berkeley heavy ion linear accelerator HILAC forming $^{248}$Fm in the (4n) fusion-evaporation reaction. Recoil products were transported with a helium gas stream and a conveyor belt onto catcher foils which were analysed in a multiplex assembly consisting of five Frisch grid chambers. ``The method was first successfully used in bombardments of Pu$^{240}$ with C$^{12}$ ions to identify a new isotope of element 100, Fm$^{248}$. It was shown to have a half-life of 0.6 minutes by analysis of the amounts of the 20-minute Cf$^{244}$ caught on the catcher foils.''

\subsection*{$^{249}$Fm}\vspace{0.0cm}
Perelygin et al.\ described the observation of $^{249}$Fm in 1960 in ``Experiments in the production of a new fermium isotope'' \cite{1960Per01}. A $^{238}$U target was bombarded with 84$-$98~MeV $^{16}$O beams from the Moscow 1.5~m cyclotron and $^{249}$Fm was produced in the (5n) fusion-evaporation reaction. Recoil products were stopped in an aluminum foil which was quickly moved to NIKFIT-1 photoplates which served as $\alpha$ detectors. ``Evidence has been obtained of the formation of a new fermium isotope Fm$^{249}$, which has a half-life of about 150~sec and an $\alpha$-particle energy of 7.9$\pm$0.3~Mev.''

\subsection*{$^{250}$Fm}\vspace{0.0cm}
$^{250}$Fm was observed by Atterling et al.\ as described in ``Element 100 produced by means of cyclotron-accelerated oxygen ions'' in 1954 \cite{1954Att01}. The Stockholm 225-cm cyclotron was used to bombard uranium targets with a $^{16}$O beam of energies up to 180~MeV. Subsequent $\alpha$ decay was measured with an ionization chamber following chemical separation. ``In the element-100 eluate fraction, up to 20 alpha disintegrations of energy 7.7~Mev were usually found, decaying with a half-life of about half an hour. According to alpha systematics a probable mass number corresponding to these data is 250.''

\subsection*{$^{251}$Fm}\vspace{0.0cm}
In the 1957 paper ``Production and properties of the nuclides fermium-250, 251, and 252'' Amiel et al.\ described the observation of $^{251}$Fm \cite{1957Ami01}. A $^{249}$Cf target was bombarded with 20$-$40~MeV $\alpha$ particles from the Berkeley 60-in. cyclotron forming $^{251}$Fm in ($\alpha$,2n) reactions. Subsequent $\alpha$ decay was measured following chemical separation. ``The element identification was established by means of a cation exchange column separation using alphahydroxy isobutyric acid as eluant. Mass assignments were based on the excitation functions. The properties of these nuclides are summarized in [the table]. The half-lives given are good to about $\pm$10\% and the alpha particle energies to $\pm$0.05~Mev.'' The measured half-life for $^{251}$Fm was 7~h.

\subsection*{$^{252}$Fm}\vspace{0.0cm}
Friedman et al.\ identified $^{252}$Fm in 1956 in ``Properties of Fm$^{252}$'' \cite{1956Fri01}. A californium target was bombarded with 34 and 42~MeV $\alpha$ particles from the Argonne 60-in.\ cyclotron. Subsequent $\alpha$ decay was measured following chemical separation. ``The 7.04-Mev alpha activity was observed to decay with a half-life of 22.7$\pm$0.7~hours... The observed increase in the yield of the 7.04-Mev alpha emitter relative to Fm$^{254}$ with increased cyclotron energy further agrees with the assignment of the 7.04-Mev alpha energy to Fm$^{252}$.'' This half-life agrees with the presently adopted value of 25.39(4)~h.

\subsection*{$^{253}$Fm}\vspace{0.0cm}
Amiel described the identification of $^{253}$Fm in ``Properties of fermium-253'' in 1957 \cite{1957Ami02}. A 40~MeV $\alpha$ beam from the Berkeley 60-in.\ cyclotron bombarded a $^{252}$Cf target forming $^{253}$Fm in the reaction ($\alpha$,3n). Alpha particle spectra were measured with an ionization-grid-chamber following chemical separation. ``The decay of the 6.94$\pm$0.04~Mev peak of Fm$^{253}$ was followed by a corresponding growth of a 6.64$\pm$0.03~Mev peak of E$^{253}$. The alpha-particle energy emitted by the Fm$^{253}$ was found in the range of 6.90$-$6.98~Mev... The growth of E$^{253}$ was found to result from the decay of Fm$^{253}$ with a 4.5$\pm$1.0~day half-life.'' An earlier tentative assignment of a half-life of $>$10~d \cite{1956Fri01} was evidently incorrect.

\subsection*{$^{254}$Fm}\vspace{0.0cm}
``Further production of transcurium nuclides by neutron irradiation'' was the first unclassified publication reporting the observation of $^{254}$Fm in 1954 by Harvey et al.\ \cite{1954Har01}. Targets of heavy californium isotopes were irradiated with neutrons in the Materials Testing Reactor. Alpha-particle spectra were measured following chemical separation. ``The isotope of element 100 emitting the approximately 7.2-Mev alpha particles is tentatively assigned as 100$^{254}$, and a possible reaction sequence leading to its production might be the following: Cf$^{252}$(n,$\gamma$)Cf$^{253}\stackrel{\beta^-}{\rightarrow}$99$^{253}$(n,$\gamma$)99$^{254}\stackrel{\beta^-}{\rightarrow}$100$^{264}$. Because of the existence of unpublished information on element 100 the question of its first preparation should not be prejudged
on the basis of this paper.''

\subsection*{$^{255}$Fm}\vspace{0.0cm}
In the 1954 paper ``Nuclear properties of some isotopes of californium, elements 99 and 100'' Choppin et al.\ identified $^{255}$Fm
\cite{1954Cho01}. Plutonium was irradiated with neutrons in the Materials Testing Reactor and $\alpha$-decay spectra were measured following chemical separation.
``Two alpha activities assigned to element 100 have been observed. The more abundant, probably due to 100$^{254}$, decays with a half-life of 3.2 hours by emission of 7.22$\pm$0.03-Mev alpha-particles, as previously reported. The other, probably due to 100$^{255}$, decays with a half-life of about 15 hours by emission of 7.1-Mev alpha particles.''

\subsection*{$^{256}$Fm}\vspace{0.0cm}
Choppin et al. described the identification of $^{256}$Fm in the 1955 paper ``Nuclear properties of 100$^{256}$ \cite{1955Cho01}. $^{255}$Es was irradiated with neutrons in the Materials Testing Reactor. Alpha-particles and spontaneous fission were measured following chemical separation. ``However, a total of 33 spontaneous fission events occurred in the 100 fraction which was well
outside the probability of the number of such events (10.8$\pm$3) expected from 100$^{254}$ based on the measured alpha-to-spontaneous-fission ratio of 1550 for this nuclide. The additional events are attributed to the nuclide 100$^{256}$. The spontaneous fission half-life was found to be approximately 3 to 4 hours.''

\subsection*{$^{257}$Fm}\vspace{0.0cm}
$^{257}$Fm was observed by Hulet et al.\ in ``79-day fermium isotope of mass 257'' in 1964 \cite{1964Hul01}. A target consisting of $^{242}$Pu, $^{243}$Am, and $^{244}$Cm was irradiated for four years in the Idaho Materials Testing Reactor and $^{257}$Fm was formed primarily by neutron capture on $^{256}$Fm. Alpha-decay spectra and spontaneous fission were measured following chemical separation. ``The decay of the complex 6.6-MeV alpha peak is shown in [the figure]. A best fit was made to the data obtained thus far by use of the least-squares method. The best value for the Fm$^{257}$ T$_{1/2}$ is 79 $\pm$ 8 days, provided the half-lives of Cf$^{253}$ and Es$^{253}$ are 17.6 days and 20.0 days, respectively.'' Previously a 11$^{+10}_{-6}$~d half-life had been assigned to either $^{257}$Fm or $^{258}$Fm \cite{1963Gat01}.

\subsection*{$^{258}$Fm}\vspace{0.0cm}
In 1971, $^{258}$Fm was identified by Hulet et al. in ``Spontaneous-fission half-life of $^{258}$Fm and nuclear instability'' \cite{1971Hul01}. The Berkeley heavy-ion linear accelerator HILAC was used to bombard a $^{257}$Fm target with 12.5~MeV deuterons to produce $^{258}$Fm in the (d,p) reaction. Recoil products were collected on the rim of a fast rotating drum which was surrounded by stationary strips of muscovite mica which recorded the tracks from spontaneous fission events. ``In summary, a 380$\pm$60-$\mu$sec (3$\sigma$) SF activity belonging to the ground-state decay of $^{258}$Fm has been identified.''

\subsection*{$^{259}$Fm}\vspace{0.0cm}
Hulet et al.\ reported the observation of $^{259}$Fm in the 1980 article ``Spontaneous fission of $^{259}$Fm'' \cite{1980Hul01}. A 16~MeV triton beam from the Los Alamos Tandem Van de Graaf accelerator bombarded a $^{257}$Fm target forming $^{259}$Fm in the reaction ($^3$H,p). Recoil products were caught on a rotating wheel which moved the activities in front of two Si(Au) surface barrier detectors which measured spontaneous fission events. ``A 1.5-s spontaneous fission activity has been produced by irradiating $^{257}$Fm with 16-MeV tritons. On the basis of formation cross sections, fission half-life systematics, and the identification of other possible products, this 1.5-s activity has been attributed to $^{259}$Fm formed by the reaction $^{257}$Fm(t,p)$^{259}$Fm.''


\section{Discovery of $^{245-260}$Md}
The first observation of a mendelevium isotope was reported in 1955 by Ghiorso et al.\ describing the formation and decay of $^{256}$Md \cite{1955Ghi02}. They suggested the name mendelevium: ``We would like to suggest the name mendelevium, symbol Mv, for the new element in recognition of the pioneering role of the great Russian chemist, Dmitri Mendeleev, who was the first to use the periodic system of the elements to predict the chemical properties of undiscovered elements, a principle which has been the key to the discovery of the last seven transuranium (actinide) elements'' \cite{1955Ghi02}. The discovery of mendelevium was officially accepted by the IUPAC-IUPAP Transfermium Working Group in 1993: ``Element 101 was discovered by the Berkeley group - with certainty in 1958 \cite{1958Phi01} following strong indications in 1955 \cite{1955Ghi02}'' \cite{1992Bar01,1993TWG01}. IUPAC adopted the name mendelevium but changed the symbol from ``Mv'' to ``Md'' at the 19$^{th}$ IUPAC Conference in Paris 1957 \cite{2005Kop01,1957IUP01}. Sixteen mendelevium isotopes have been reported so far.

\subsection*{$^{245,246}$Md}\vspace{0.0cm}
In 1996, Ninov et al.\ discovered $^{245}$Md and $^{246}$Md as reported in the paper ``Identification of new mendelevium and einsteinium isotopes in bombardments of $^{209}$Bi with $^{40}$Ar'' \cite{1996Nin01}. $^{40}$Ar beams were accelerated to 4.78, 4.93, and 5.12 A$\cdot$MeV with the GSI UNILAC accelerator and bombarded $^{209}$Bi targets to form $^{245}$Md and $^{246}$Md in (4n) and (3n) fusion-evaporation reactions, respectively. Recoil products were separated with the velocity filter SHIP and implanted in a position sensitive PIPS detector which also recorded subsequent $\alpha$ decay and spontaneous fission. ``Since the chosen bombarding energy further coincided with the maximum of the 4n deexcitation channel, we assign this decay sequence to $^{245}$Md and its daughter $^{241}$Es. From these events in E$_{\alpha m1}$ and E$_{\alpha m2}$ we obtained a mean half-life of T$_{1/2}$ = (0.35$^{+0.23}_{-0.16}$)~s for $^{245}$Md... From these data we obtained for $^{246}$Md an $\alpha$ energy of E$_\alpha$ = (8740$\pm$20)~keV and a half-life of T$_{1/2}$ = (1.0$\pm$0.4)~s, and for $^{242}$Es E$_\alpha$ = (7920$\pm$20)~keV and T$_{1/2}$ = (16$^{+6}_{-4}$)~s.''

\subsection*{$^{247}$Md}\vspace{0.0cm}
M\"unzenberg et al.\ reported the discovery of $^{247}$Md in 1981 in ``The new isotopes $^{247}$Md, $^{243}$Fm,$^{239}$Cf, and investigation of the evaporation residues from fusion of $^{206}$Pb, $^{208}$Pb, and $^{209}$Bi with $^{40}$Ar'' \cite{1981Mun01}. A $^{209}$Bi target was bombarded with a 4.8~MeV/u $^{40}$Ar beam from the GSI UNILAC accelerator to form $^{247}$Md in the (2n) fusion-evaporation reaction. Recoil products were separated with the velocity filter SHIP and implanted in an array of position sensitive surface-barrier detectors which also recorded subsequent $\alpha$ decay and spontaneous fission. ``Three new $\alpha$-emitting isotopes were found: $^{247}$Md, $^{243}$Fm, and $^{239}$Cf. $^{247}$Md was identified by correlation to its known daughter decay $^{243}$Es.'' The measured half-life was 2.9$^{+1.7}_{-1.2}$~s.

\subsection*{$^{248-252}$Md}\vspace{0.0cm}
Eskola discovered $^{248}$Md, $^{249}$Md, $^{250}$Md, $^{251}$Md, and $^{252}$Md in the 1973 paper ``Studies of mendelevium isotopes with mass numbers 248 through 252'' \cite{1973Esk02}. $^{12}$C and $^{13}$C beams with a maximum energy of 10.4~MeV/u from the Berkeley heavy-ion linear accelerator bombarded $^{241}$Am and $^{243}$Am targets. Recoil products were transported with a rapid flowing helium gas onto a wheel which periodically rotated in front of a series of Si-Au surface barrier detectors. ``In bombardments of the $^{241}$Am target with $^{12}$C ions two new $\alpha$ activities were observed: a 8.32-MeV, 7-sec activity assigned to $^{248}$Md, and a 8.03-MeV, 24-sec activity which was also observed in bombardments of $^{243}$Am with $^{12}$C ions and which was assigned to $^{249}$Md... In bombardments of the $^{243}$Am target with $^{13}$C and $^{12}$C ions, two new $\alpha$ activities were observed: a 7.75-MeV, 52-sec activity which was assigned to $^{250}$Md and a 7.55-MeV, 4.0-min activity assigned to $^{251}$Md... Because of the long half-life of $^{252}$Fm most of the counts in 7.04-MeV peak originate from $^{252}$Md produced in previous bombardments with $^{13}$C ions. The decay curve of the $^{252}$Fm in daughter spectra combined from four bombardments with 72$-$88~MeV $^{13}$C ions is plotted in [the figure]. A value of 140$\pm$50~sec is derived for the half-life of $^{252}$Md by a least-squares analysis. This is considerably shorter than the 8-min value reported by Donets, Schegolev, and Ermakov.'' This earlier reported half-life mentioned in the quote \cite{1966Don01} was not credited with the discovery because of the large discrepancy with the currently accepted value of 2.3(8)~min measured by Eskola.

\subsection*{$^{253}$Md}\vspace{0.0cm}
Kadkhodayan et al.\ discovered $^{253}$Md in the 1992 article ``Identification of $^{253}$Md'' \cite{1992Kad01}. $^{243}$Am targets were bombarded with 66$-$74~MeV $^{13}$C beams from the Berkeley 88-in.\ cyclotron and $^{253}$Md was produced in the (3n) fusion-evaporation reaction. Alpha-particles and spontaneous fission were recorded with a Si(Au) surface barrier detector following chemical separation. ``An increase in the length of irradiation will cause a corresponding increase in the amount of the new isotope $^{253}$Md and hence, in the amount of $^{253}$Es produced, provided the length of irradiations are not very long compared to the half-life of $^{253}$Md. In this way, the Md half-life was estimated to be about 6 minutes with a production cross section of the order of 50 nb.''

\subsection*{$^{254}$Md}\vspace{0.0cm}
In ``Nuclear properties of $^{254}$Md, $^{255}$Md, $^{256}$Md, $^{257}$Md and $^{258}$Md'', Fields et al.\ reported the observation of $^{254}$Md in 1970 \cite{1970Fie01}. An einsteinium target was bombarded with 46~MeV $\alpha$-particles from the Argonne 152~cm cyclotron forming $^{254}$Md in the reaction $^{253}$Es($\alpha$,3n). The half-life of $^{254}$Md was determined from the growth of $^{254}$Fm due to the EC decay of $^{254}$Md following chemical separation. ``The nuclide $^{254}$Md was observed for the first time and was found to decay by EC with a half-life of 10$\pm$3 min. Successive chemical separations of Fm indicate the existence of another $^{254}$Md isomer having a half-life of 28$\pm$8~min.''

\subsection*{$^{255}$Md}\vspace{0.0cm}
Phillips et al.\ discovered $^{255}$Md as described in ``Discovery of a new mendelevium isotope'' in 1958 \cite{1958Phi01}. A $^{253}$Es target was bombarded with 24$-$42~MeV $\alpha$ particles from the Berkeley 60-in.\ cyclotron forming $^{255}$Md in the reaction $^{253}$Es($\alpha$,2n). Alpha-particle spectra were recorded with a 50-channel alpha pulse-height analyzer following chemical separation. ``The 7.08-MeV alpha group was observed in both new fractions. It was concluded that in the time interval between cation columns, Fm$^{255}$ grew in as a result of the electron-capture decay of Mv$^{255}$. A half-life of the order of 1/2 hour was estimated for the electron capture decay of Mv$^{255}$.''

\subsection*{$^{256}$Md}\vspace{0.0cm}
$^{256}$Md was discovered by Ghiorso et al. in ``New element mendelevium, atomic number 101'' in 1955 \cite{1955Ghi02}. A $^{253}$Es target was bombarded with 48~MeV $\alpha$ particles from the Berkeley 60-in.\ cyclotron forming $^{256}$Md in the reaction $^{253}$Es($\alpha$,n). Alpha-particles and spontaneous fission were measured following chemical separation.``By an ($\alpha$,n) reaction on 99$^{253}$ we have produced the isotope 101$^{256}$ which decays by electron capture with a half-life of the order of a half hour to 100$^{256}$; this isotope then decays by spontaneous fission with a half-life of the order of 3 to 4 hours.'' The half-life was later revised to 1.5~h \cite{1958Phi01}.

\subsection*{$^{257}$Md}\vspace{0.0cm}
Sikkeland et al.\ reported the observation of $^{257}$Md in 1965 in ``Decay properties of the nuclides fermium-256 and -257 and mendelevium-255, -256, and -257'' \cite{1965Sik01}. A californium target was bombarded with $^{11}$B, $^{12}$C, and $^{13}$C beams with a maximum energy of 10.5 MeV/u from the Berkeley heavy-ion linear accelerator HILAC. Alpha-particles and spontaneous fission were measured with a 20-sample solid-state detection device and two Frisch-grid ionization chambers. ``The 3-h component of the 7.07- and 7.24-MeV alpha groups was assigned to the previously unobserved isotope Md$^{257}$.''

\subsection*{$^{258}$Md}\vspace{0.0cm}
In ``Nuclear properties of $^{254}$Md, $^{255}$Md, $^{256}$Md, $^{257}$Md and $^{258}$Md'', Fields et al.\ reported the observation of $^{258}$Md in 1970 \cite{1970Fie01}. An einsteinium target was bombarded with 46~MeV $\alpha$-particles from the Argonne 152~cm cyclotron forming $^{258}$Md in the reaction $^{255}$Es($\alpha$,n). Alpha-particle spectra were measured with a Au-Si surface-barrier detector following chemical separation. ``The $\alpha$-spectrum of $^{258}$Md as measured with a 25~mm$^2$ detector for 8~d is shown in [the figure]. Two $\alpha$ groups with energies 6.716$\pm$0.005 and 6.79$\pm$0.01~MeV were observed. The half-life of $^{258}$Md was determined from the decay of the 6.716~MeV $\alpha$ group and was found to be 56$\pm$7~d.''
Fields et al.\ did not consider the observation of $^{258}$Md a new discovery referring to a conference abstract \cite{1968Hul01}.

\subsection*{$^{259}$Md}\vspace{0.0cm}
Wild et al.\ identified $^{259}$Md as described in the 1982 paper ``Unusually low fragment energies in the symmetric fission of $^{259}$Md'' \cite{1982Wil01}. A $^{248}$Cm target was bombarded with a 97~MeV $^{18}$O beam from the Berkeley 88-in.\ cyclotron forming $^{259}$No in an ($\alpha$3n) fusion-evaporation reaction. $^{259}$Md was then populated be electron capture. Spontaneous fission events were recorded with a surface barrier detector following chemical separation. ``Thus, the $^{259}$No remained essentially at the top of the column while the daughter atoms $^{255}$Fm and $^{259}$Md, produced by the $\alpha$ and EC decay of $^{259}$No between elutions, were removed rapidly... We calculated a weighted-average half-life of 103$\pm$12~min for $^{259}$Md, based on four measurements.''

\subsection*{$^{260}$Md}\vspace{0.0cm}
In 1989, Hulet et al. described the identification of $^{260}$Md in ``Spontaneous fission properties of $^{258}$Fm, $^{259}$Md, $^{260}$Md, $^{258}$No, and $^{260}$[104]: Bimodal fission'' \cite{1989Hul02}. A $^{254}$Es target was bombarded with $^{18}$O and $^{22}$Ne beams from the Berkeley 88-in.\ cyclotron. Spontaneous fission activity was measured following chemical separation and the isotopic identification was achieved by electromagnetic isotope separation. ``32-d $^{260}$Md: We recently discovered this long-lived isotope of Md in mass-separated samples during the course of investigating the products of transfer reactions originating from heavy-ion bombardments of $^{254}$Es.'' The observation of spontaneous fission of $^{260}$Md from this experiment had been published three years earlier by Hulet et al., however, no details about the $^{260}$Md were included \cite{1986Hul01}. Also, the same group reported a half-life of 31.8(5)~d in a conference proceeding \cite{1986Lou01}.


\section{Discovery of $^{250-260}$No}
The first observation of a nobelium isotope assigned to either $^{251}$No or $^{253}$No by Fields et al. in 1957 \cite{1957Fie01} could not be confirmed later \cite{1958Ghi01,1958Fle01}. Also the assignment of a 3~s half-life to $^{254}$No \cite{1958Ghi01} was incorrect. Eight years later Donets et al.\ \cite{1966Don02} and Zager et al.\ \cite{1966Zag01} simultaneously reported the identification of $^{254}$No. The discovery of nobelium was officially accepted by the IUPAC-IUPAP Transfermium Working Group  in 1993: ``The two 1966 (simultaneously published) Dubna results \cite{1966Don02} and, especially, \cite{1966Zag01}, both submitted in 1965, give conclusive evidence that element 102 had been produced'' \cite{1992Bar01,1993TWG01}. Fields et al.\ had suggest the name nobelium ``We suggest the name nobelium, symbol No, for the new element in recognition of Alfred Nobel's support of scientific research and after the institute where the work was done'' \cite{1957Fie01}. Although this original report was not correct the name was continued to be used and officially accepted in 1997 \cite{1997IUP01,1997IUP02,1997IUP03,1997JCE01}.  Eleven nobelium isotopes have been reported so far.

The assignment of a 54~$\mu$s half-life to $^{249}$No \cite{2003Bel01,2003Yer01,2003Pop01} was later reassigned to $^{250}$No \cite{2006Pet01}. The observation of $^{262}$No was only published in a conference proceeding \cite{1989Lou01} as quoted in \cite{2000Hol01} and in an internal report \cite{1989Hul01}.

\subsection*{$^{250}$No}\vspace{0.0cm}
Oganessian et al.\ identified $^{250}$No in the 2001 article ``Measurements of cross sections for the fusion-evaporation reactions $^{204,206,207,208}$Pb+$^{48}$Ca and $^{207}$Pb+$^{34}$S: Decay properties of the even-even nuclides $^{238}$Cf and $^{250}$No'' \cite{2001Oga03}. Enriched $^{206}$Pb and $^{204}$Pb targets were bombarded with 213.5$-$242.5~MeV $^{48}$Ca beams from the Dubna U400 cyclotron forming $^{250}$No in (4n) and (2n) fusion-evaporation reactions, respectively. Recoil products were separated with the Dubna Gas-filled Recoil Separator and implanted in a position sensitive detector array which also measured subsequent $\alpha$ and spontaneous fission decay. Escaping $\alpha$-particles were also recorded with eight additional detectors arranged in a boxlike configuration. ``The spontaneously fissioning even-even isotope $^{250}$No, with a half-life T$_{1/2}$ = 36~$\mu$s, was identified for the first time in this experiment.'' This half-life of 36$^{+11}_{-6}~\mu$s agrees with the currently adopted value of 46$^{+22}_{-14}~\mu$ for an isomeric state. A spontaneous fission half-life of 250~$\mu$s reported earlier \cite{1975Ter01} was evidently incorrect.

\subsection*{$^{251}$No}\vspace{0.0cm}
$^{251}$No was discovered by Ghiorso et al.\ in ``Isotopes of element 102 with mass 251 to 258'' in 1967 \cite{1967Ghi01}. A $^{244}$Cm target was bombarded with 78$-$90~MeV $^{12}$C beams from the Berkeley heavy-ion linear accelerator (HILAC) to produce $^{251}$No in the (5n) fusion-evaporation reaction. Recoil products were transported by a helium gas stream onto a wheel which rotates in regular intervals to move the activities in front of Au-Si surface-barrier $\alpha$-particle detectors. The results were summarized in a table listing a half-life of 0.8(3)~s for $^{251}$No. A 10~min half-life assigned to either $^{251}$No or $^{253}$No \cite{1957Fie01} was evidently incorrect.

\subsection*{$^{252-253}$No}\vspace{0.0cm}
Mikheev et al.\ identified $^{252}$No and $^{253}$No in 1967 in ``Synthesis of isotopes of element 102 with mass numbers 254, 253, and 252'' \cite{1967Mik02}. A 96~MeV $^{18}$O and a 102~MeV $^{16}$O beam from the Dubna 310~cm heavy-ion cyclotron bombarded a $^{239}$Pu and $^{242}$Pu target forming $^{252}$No and $^{253}$No, respectively, in (5n) fusion evaporation reactions. Recoil products were transported by a helium gas jet onto a metallic catcher which swiveled in front of a silicon surface barrier detector. ``The experimental data obtained confirm the synthesis of the isotope 102$^{252}$ in the reaction Pu$^{239}$(O$^{18}$,5n)102$^{252}$, with T$_{1/2}$ = 4.5$\pm$1.5~sec and E$_\alpha$ = 8.41$\pm$0.03~MeV... Thus, all the experimental data obtained confirm that the half-life of the isotope 102$^{253}$ is 95$\pm$10~sec and that the energy of the most intense group of the $\alpha$-particles is 8.01$\pm$0.03~MeV.'' A 10~min half-life assigned to either $^{251}$No or $^{253}$No \cite{1957Fie01} was evidently incorrect. In addition, a 3~s half-life had previously been incorrectly assigned to $^{254}$No \cite{1958Ghi01}.

\subsection*{$^{254}$No}\vspace{0.0cm}
$^{254}$No was simultaneously discovered in 1966 by Donets et al.\ \cite{1966Don02} and Zager et al.\ \cite{1966Zag01} in two papers with the same title ``The properties of the isotope 102$^{254}$''. Donets et al.\ used a $^{22}$Ne beam from the Dubna 300-cm cyclotron to bombard $^{238}$U producing $^{254}$No in the (6n) fusion-evaporation reaction. Recoil products diffused in a gas onto a disk which was rotated in front of a collection region. The $\alpha$-decay of the $^{250}$Fm daughters was then measured with an $\alpha$-spectrometer following chemical separation. The half-life was measured by varying the disk velocity. ``According to these data the half-life of 102$^{254}$ is 50$\pm$10~sec.'' Zager et al.\ bombarded a $^{243}$Am target with 82$-$84~MeV $^{15}$N beams from the Dubna 150-cm cyclotron to form $^{254}$No in the (4n) fusion-evaporation reaction. Recoils were transported by a helium gas stream onto a metal collector which was periodically transferred to a silicon surface-barrier detector to measure subsequent $\alpha$ decay. ``According to our data, the half-life of 102$^{254}$ is between 20 and 50 sec, and the alpha particle energy is 8.10$\pm$0.05~MeV.'' A 3~s half-life previously assigned to $^{254}$No \cite{1958Ghi01} was evidently incorrect.

\subsection*{$^{255}$No}\vspace{0.0cm}
In the 1967 paper ``Nuclear properties of the isotopes of element 102 with mass numbers 255 and 256'' Druin et al.\ reported the identification of $^{255}$No \cite{1967Dru01}. Natural uranium targets were bombarded with $^{22}$Ne beams of energies up to 177~MeV from the Dubna 310-cm cyclotron forming $^{255}$No in the fusion evaporation reactions $^{238}$U($^{22}$Ne,5n). Alpha-particles emitted from the recoils were measured. No further details about the experimental setup were given referring to a preprint \cite{1966Aka01}. ``By comparing the illustrated excitation functions of reactions leading to the formation of $\alpha$-emitters with 8.08, 8.23, and 8.35~MeV, we see that the reaction reminiscent of U$^{238}$(Ne$^{22}$,5n)102$^{255}$, (as regards the shape and position of the maximum) only gives an $\alpha$-emitter with an energy of 8.08~MeV and a half-life of about 3 min, which may thus be the isotope 102$^{255}$.'' This half-life agrees with the presently adopted value of 3.1(2)~min. A 15~s half-life assigned to $^{255}$No \cite{1961Ghi02} was evidently incorrect.

\subsection*{$^{256}$No}\vspace{0.0cm}
Donets et al.\ described the observation of $^{256}$No in 1963 in ``Synthesis of a new isotope of element 102'' \cite{1963Don01} with details of the experiment published in a subsequent paper \cite{1964Don01}. $^{22}$Ne beams from the Dubna 3~m cyclotron bombarded a uranium target forming $^{256}$No in the fusion-evaporation reaction $^{238}$U($^{22}$Ne,4n). Recoil nuclei were analyzed with a separator following $\alpha$-decay and $^{256}$No was identified by measuring the $\alpha$-decay of the $^{252}$Fm daughter with a high-resolution $\alpha$-spectrometer. ``At the Nuclear Reactions Laboratory of the Joint Institute for Nuclear Research, a new isotope of the element 102, having mass number 256, has been successfully synthesized. It was found that 102$^{256}$ is $\alpha$-active and has a half-life of $\sim$8~sec.'' Although the half-life differs by more than a factor two from the presently adopted value of 2.91(5)~s later measurements credited Donets et al.\ with the discovery \cite{1967Dru01}.

\subsection*{$^{257}$No}\vspace{0.0cm}
$^{257}$No was discovered by Ghiorso et al.\ in ``Isotopes of element 102 with mass 251 to 258'' in 1967 \cite{1967Ghi01}. A $^{248}$Cm target was bombarded with 63$-$68~MeV $^{13}$C beams from the Berkeley heavy-ion linear accelerator (HILAC) to produce $^{257}$No in the (4n) fusion-evaporation reaction. Recoil products were transported by a helium gas stream onto a wheel which rotates in regular intervals to move the activities in front of Au-Si surface-barrier $\alpha$-particle detectors. The results were summarized in a table listing a half-life of 23(2)~s for $^{257}$No. A 15~s half-life had previously been incorrectly assigned to $^{255}$No \cite{1961Ghi02}.

\subsection*{$^{258}$No}\vspace{0.0cm}
In 1989, Hulet et al. described the identification of $^{258}$No in ``Spontaneous fission properties of $^{258}$Fm, $^{259}$Md, $^{260}$Md, $^{258}$No, and $^{260}$[104]: Bimodal fission'' \cite{1989Hul02}. A $^{248}$Cm metal target was bombarded with a 67.6~MeV $^{13}$C beam from the Berkeley 88-in.\ cyclotron and $^{258}$No was produced in the (3n) fusion-evaporation reaction. Spontaneous fission products were measured with the Spinning-Wheel Analyzer for Millisecond Isotopes (SWAMI). ``Assuming a 5\% efficiency for SWAMI, we obtained a 17$—$18~nb cross section for the formation of what we believe to be $^{258}$No.'' A half-life of 1.2(2)~ms was measured for $^{258}$No. This half-life had previously been reported in an unpublished report \cite{1969Nur01}. An even earlier search for $^{258}$No was unsuccessful \cite{1967Ghi01}.

\subsection*{$^{259}$No}\vspace{0.0cm}
In the 1973 paper ``The new nuclide nobelium-259'' Silva et al. described the discovery of $^{259}$No \cite{1973Sil01}. A $^{248}$Cm target was bombarded with 88$-$106~MeV $^{18}$O beams from the Oak Ridge isochronous cyclotron forming $^{259}$No in the ($\alpha$3n) fusion-evaporation reaction. A helium gas-jet system was used to implant the recoil products onto a platinum catcher foil which was pneumatically transferred in front of an $\alpha$-particle detection system. ``We believe the $\alpha$-particle groups at 7.685, 7.605, 7.533, 7.500 and 7.455$\pm$0.010~MeV to be associated with $^{259}$No decay.''

\subsection*{$^{260}$No}\vspace{0.0cm}
Somerville et al.\ identified $^{260}$No in ``Spontaneous fission of rutherfordium isotopes'' in 1985 \cite{1985Som01}. A 99~MeV $^{18}$O beam from the Berkeley 88-in. cyclotron bombarded a $^{254}$Es target forming $^{260}$No in a multi-nucleon transfer reaction. Helium transported the recoils onto a long tape collector which passed one meter of stationary mica track detectors. ``We have found a 106$\pm$8-ms SF activity shown in [the figure] with a production cross section of 1.1$\pm$0.2~$\mu$b in the reaction 99-MeV $^{18}$O + $^{254}$Es... But $^{260}$No is a possible assignment because it is the only even-even nuclide whose production cross section of 1.1~$\mu$b would fit an extrapolation of the yield curve for transfer products from the reaction $^{18}$O + $^{254}$Es... However, a $\sim$l00-ms half-life for $^{260}$No would be surprisingly long, based on an extrapolation of the known nobelium half-lives in [the figure] and a known half-life of only 1~ms for $^{258}$No. Thus, an assignment to $^{260}$No is supported by our cross bombardments but would be surprising in view of the nobelium half-life systematics.''


\section{Discovery of $^{252-260}$Lr}
In 1961, Ghiorso et al.\ reported the discovery of lawrencium with the observation of $^{257}$Lr and suggested the name lawrencium: ``In honor of the late Ernest O. Lawrence, we respectfully suggest that the new element be named lawrencium with the symbol Lw'' \cite{1961Ghi02}. Four years later, Donets et al.\ identified $^{256}$Lr \cite{1965Don01}. The mass assignment of the original observation could later not be confirmed \cite{1968Fle01}. Credit for the discovery was given to both groups in 1993 by the IUPAC-IUPAP Transfermium Working Group: ``In the complicated situation presented by element 103, with several papers of varying degrees of completeness and conviction, none conclusive, and referring to several isotopes, it is impossible to say other than that full confidence was built up over a decade with credit attaching to work in both Berkeley and Dubna'' \cite{1992Bar01,1993TWG01}. The original suggestion of the name lawrencium was adopted but the symbol was later changed to Lr; name and symbol were officially accepted by IUPAC in 1997 \cite{1997IUP01,1997IUP02,1997IUP03,1997JCE01}.  Nine lawrencium isotopes have been reported so far.

The observations of $^{261}$Lr and $^{262}$Lr were only published as internal reports \cite{1989Hul01,1991Hen01} and in a conference proceeding \cite{1990Hul01}.

\subsection*{$^{252}$Lr}\vspace{0.0cm}
He\ss berger et al.\ reported the first observation of $^{252}$Lr in ``Decay properties of neutron-deficient isotopes $^{256,257}$Db, $^{255}$Rf, $^{252,253}$Lr'' in 2001 \cite{2001Hes01}. A $^{209}$Bi target was bombarded with a 5.08~MeV/u $^{50}$Ti beam from the GSI UNILAC accelerator and $^{256}$Db was formed in (3n) fusion-evaporation reactions. Recoil products were separated with the velocity filter SHIP and implanted in a position-sensitive 16-strip PIPS detector which also measured subsequent $\alpha$-decay and spontaneous fission. In addition, escaping $\alpha$-decay and spontaneous fission events were recorded in six silicon detectors located in the backward hemisphere. ``The identification of the isotopes $^{256}$Db and $^{252}$Lr was based on a total of 16 $\alpha$-decay chains, that were followed down to $^{244}$Cf according to the sequences $^{256}$Db$\stackrel{\alpha}{\rightarrow}^{252}$Lr$\stackrel{\alpha}{\rightarrow}^{248}$Md$\stackrel{\alpha}{\rightarrow}^{244}$Es$\stackrel{EC}{\rightarrow}^{244}$Cf$\stackrel{\alpha}{\rightarrow}^{240}$Cm or $^{256}$Db$\stackrel{\alpha}{\rightarrow}^{252}$Lr$\stackrel{\alpha}{\rightarrow}^{248}$Md$\stackrel{EC}{\rightarrow}^{248}$Fm$\stackrel{\alpha}{\rightarrow}^{244}$Cf$\stackrel{\alpha}{\rightarrow}^{240}$Cm.''
Earlier an upper limit for spontaneous fission of $^{252}$Lr was reported \cite{1976Oga01}.

\subsection*{$^{253}$Lr}\vspace{0.0cm}
The discovery of $^{253}$Lr was reported in 1985 in the paper ``The new isotopes $^{258}$105, $^{257}$105, $^{254}$Lr and $^{253}$Lr'' by He\ss berger et al.\ \cite{1985Hes01}. $^{209}$Bi targets were bombarded with 4.65, 4.75, 4.85, and 4.95 MeV/u $^{50}$Ti beams from the GSI UNILAC accelerator forming $^{257}$Db in the (2n) fusion-evaporation reaction. $^{253}$Lr was then populated by $\alpha$-decay. Recoil products were separated with the velocity filter SHIP and implanted in seven position-sensitive surface barrier detectors which also measured subsequent $\alpha$-decay and spontaneous fission. ``Isotope $^{253}$Lr: This isotope was found in the $\alpha$-decay chains of $^{257}$105. Two $\alpha$ lines with mean energies E$_{\alpha 1,2}$ = 8,800, 8,722~keV could be attributed to it. The measured half-life is T$_{1/2}$ = (1.3$^{+0.6}_{-0.3}$)~s.'' Earlier an upper limit for spontaneous fission of $^{253}$Lr was reported \cite{1976Oga01}.

\subsection*{$^{254}$Lr}\vspace{0.0cm}
In the 1981 paper ``Identification of element 107 by $\alpha$ correlation chains'' M\"unzenberg at al.\ described the discovery of $^{254}$Lr \cite{1981Mun02}. A 4.85~MeV/u $^{54}$Cr from the GSI UNILAC linear accelerator bombarded $^{209}$Bi targets forming $^{262}$Bh in the (1n) fusion-evaporation reaction. $^{254}$Lr was then populated by $\alpha$-decay. Recoil products were separated with the velocity filter SHIP and implanted in seven position sensitive surface barrier detectors which also measured the subsequent $\alpha$-decays and spontaneous fission. Three events for the decay of $^{254}$Lr were measured. In addition, $^{254}$Lr was also observed in the fusion evaporation reaction $^{209}$Bi($^{50}$Ti,n) at a beam energy of 4.75~MeV/u: ``$^{258}$105 can be produced in $^{50}$Ti on $^{209}$Bi irradiations by evaporation of one neutron. At 4.75 MeV/u we observed decays of (9,189$\pm$35)~keV and (9,066$\pm$35)~keV with (4.0$^{+1.8}_{-1.6}$)~s half-life and (8,468$\pm$30)~keV with (18$^{+19}_{-6}$)~s half-life corresponding to $^{258}$105 and $^{254}$Lr respectively in good agreement to the data from $^{262}$107 shown in the table.''

\subsection*{$^{255}$Lr}\vspace{0.0cm}
Druin reported the first observation of $^{255}$Lr in the 1971 paper ``Radioactive properties of isotopes of element 103'' \cite{1971Dru02}. A $^{243}$Am target was bombarded with a $^{16}$O beam and $^{255}$Lr was formed in the (4n) fusion-evaporation reaction. Recoil products were swept from the target with a gas stream and collected on a filter where subsequent $\alpha$ decay was measured with two $\alpha$-particle detectors. ``It was shown that the $\alpha$ emitter with E$_\alpha$ = 8.38~MeV and T$_{1/2}$ $\sim$ 20~sec behaves like a product of total fusion of O$^{16}$ and Am$^{243}$ with subsequent evaporation of four neutrons, i.e., like isotope 103$^{255}$.''

\subsection*{$^{256}$Lr}\vspace{0.0cm}
In the 1965 article ``Synthesis of the isotope of element 103 (lawrencium) with mass number 256'' Donets et al.\ described the discovery of $^{256}$Lr \cite{1965Don01}. A $^{18}$O beam with a maximum energy of 96~MeV from the Dubna three-meter, multiply-charged ion cyclotron bombarded a $^{243}$Am target forming $^{256}$Lr in the (5n) fusion-evaporation reaction. Recoil products were diffused by gas on a rotating disc and the half-life of $^{256}$Lr was determined from the distribution of $^{252}$Fm on the collector and the rotation speed of the disc. $^{252}$Fm was identified by its $\alpha$-decay measured in an $\alpha$ spectrometer with surface-barrier Au-Si detectors, following chemical separation. ``The detection and identification of $_{103}$Lw$^{256}$ was made through the isotope $_{100}$Fm$^{252}$, a product of electron capture in $_{101}$Mv$^{252}$ produced by $\alpha$-decay of $_{103}$Lw$^{256}$. The half-life of $_{100}$Lw$^{256}$ is 45~sec.''

\subsection*{$^{257-260}$Lr}\vspace{0.0cm}
Eskola et al.\ identified $^{257}$Lr, $^{258}$Lr, $^{259}$Lr, and $^{260}$Lr in ``Studies of lawrencium isotopes with mass numbers 255 through 260'' in 1971 \cite{1971Esk01}. Boron, nitrogen, and oxygen beams with a maximum energy of 10.4~MeV/u from the Berkeley heavy-ion linear accelerator bombarded $^{249}$Cf, $^{248}$Cm, and $^{249}$Bk targets. Recoil products were swept by rapidly flowing helium gas onto a collection wheel which rotated periodically in front of a series of Si-Au surface-barrier detectors. ``In our bombardments of the $^{249}$Cf target with $^{15}$N ions with the primary goal of making isotopes of element 105, a pronounced 8.87-MeV, 0.6-sec $\alpha$ particle group appeared in the spectra. By producing this activity using three different projectiles, $^{11}$B, $^{14}$N, and $^{15}$N, on the $^{249}$Cf target, we have concluded that the activity must be due to $^{257}$Lr... The excitation functions for the 8.87-MeV, 0.6-sec and the 8.6-MeV, 4.2-sec $\alpha$ activities produced by $^{15}$N ions on $^{249}$Cf are displayed in [the figure]... Such a behavior is in accordance with the assignments of the activities to $^{257}$Lr and $^{258}$Lr,... The 8.45-MeV, 5.4-sec peak has been assigned to $^{259}$Lr... In recent bombardments of a 300-$\mu$g/cm$^2$ $^{249}$Bk target with 95-MeV $^{18}$O ions, we observed an 8.03~MeV, 3-min activity which we assign to $^{260}$Lr.'' Earlier half-life measurements of 8(2)~s \cite{1961Ghi02} and $\sim$35~s \cite{1968Fle01} assigned to $^{257}$Lr \cite{1961Ghi02} were evidently incorrect. The results for $^{257}$Lr and $^{258}$Lr were mentioned by Ghiorso et al.\ about a year earlier \cite{1970Ghi02} referring to the paper by Eskola et al.\ \cite{1971Esk01} as ``to be published''.


\section{Discovery of $^{253-267}$Rf}
In 1969, Rutherfordium was essentially discovered simultaneously in Dubna by Zvara et al.\ \cite{1969Zva01} and in Berkeley by Ghiorso et al.\ \cite{1969Ghi01} as recognized by the IUPAC-IUPAP Transfermium Working Group in 1993: ``The chemical experiments in Dubna \cite{1969Zva01} with \cite{1970Zva01}) and the Berkeley experiments (\cite{1969Ghi01}) were essentially contemporaneous and each show that element 104 had been produced'' \cite{1992Bar01,1993TWG01}. Zvara et al.\ submitted their results on the chemical properties of element 104 on October 14, 1968 and Ghiorso et al.\ reported the observation of the isotopes $^{257}$Rf, $^{258}$Rf, and $^{259}$Rf on May 5, 1969. Later, the Dubna group suggested the name kurchatovium (Ku) while the Berkeley group suggested rutherfordium (Rf). In 1994, the Commission on Nomenclature of Inorganic Chemistry of IUPAC did not accept either suggestion and recommended dubnium instead \cite{1994IUP01}. However, this decision was changed to rutherfordium in 1997 \cite{1997IUP01}. Thirteen rutherfordium isotopes have been reported so far.

$^{266}$Rf was at the end of the isotope chain originating at $^{282}$113, however, the observed spontaneous fission could have been due to either $^{266}$Db or $^{266}$Rf \cite{2007Oga01,2007Oga02}.

\subsection*{$^{253,254}$Rf}\vspace{0.0cm}
In 1997, He\ss berger described the identification of $^{253}$Rf and $^{254}$Rf in ``Spontaneous fission and alpha-decay properties of neutron deficient isotopes $^{257-253}$104 and $^{258}$106'' \cite{1997Hes01}. $^{204}$Pb and $^{206}$Pb targets were bombarded with 4.68~MeV/u and 4.81~MeV/u $^{50}$Ti beams from the GSI UNILAC accelerator forming $^{253}$Rf and $^{254}$Rf in (1n) and (2n) evaporation reactions, respectively. Recoil products were separated with the velocity filter SHIP and implanted in a position sensitive 16-strip silicon wafer which also measured subsequent $\alpha$ decay and spontaneous fission. ``New spontaneous fission activities were identified and assigned to $^{253}$104, $^{254}$104, and $^{258}$106. The half-lives were measured as T$_{1/2}$ = (48$^{+17}_{-10}$)~$\mu$s for $^{253}$104, T$_{1/2}$ = (23$\pm$3)~$\mu$s for $^{254}$104, and T$_{1/2}$ = (2.9$^{+1.3}_{-0.7}$)~ms for $^{258}$106. No indication for $\alpha$-decay of any of these isotopes was found.'' Earlier a 1.8~s half-life was reported for $^{253}$Rf in a conference proceeding \cite{1976Fle01} as quoted in \cite{1981Sch02,1987Hyd01}. For $^{254}$Rf an upper limit of $<$3~ms \cite{1975Oga02} was previously reported and a 0.5(2)~ms \cite{1975Ter01} measurement was considered to be ambiguous \cite{1997Hes01}. The results by He\ss berger were recently confirmed with a new half-life measurement for $^{254}$Rf of 29.6$^{+0.7}_{-0.6}~\mu$s \cite{2008Dra01}.

\subsection*{$^{255,256}$Rf}\vspace{0.0cm}
In the 1975 paper ``Experiments on the synthesis of neutron-deficient isotopes of kurchatovium in reactions with accelerated $^{50}$Ti ions'' Oganessian et al.\ described the observation of $^{255}$Rf and $^{256}$Rf \cite{1975Oga01}. $^{50}$Ti beams with energies up to 260~MeV from the Dubna 310~cm cyclotron bombarded $^{207}$Pb and $^{208}$Pb targets forming $^{255}$Rf and $^{256}$Rf, respectively, in (2n) fusion-evaporation reactions. Spontaneous fission fragments were measured with mica track detectors located around a rotating target. ``The long-lived emitter with half-life about 4~sec, in all probability, is the isotope $^{255}$Ku, which is formed with a maximum cross section in the reaction $^{207}$Pb($^{50}$Ti,2n), and with lower probability in the reaction $^{208}$Pb($^{50}$Ti,3n), and is absent in the reaction $^{206}$Pb($^{50}$Ti,1n)... Thus, analyzing the experimental cross sections of the reactions and the properties of the known isotopes of kurchatovium and lighter elements, it can be assumed that the observed effect is due to decay of the isotope $^{256}$Ku, which is formed in the reaction $^{208}$Pb($^{50}$Ti,2n)$^{256}$Ku.'' The same results were submitted to a different journal less than a month later \cite{1975Oga02}.

\subsection*{$^{257-259}$Rf}\vspace{0.0cm}
Ghiorso et al.\ discovered $^{257}$Rf, $^{258}$Rf, and $^{259}$Rf in 1969 in ``Positive identification of two alpha-particle-emitting isotopes of element 104'' \cite{1969Ghi01}. $^{12}$C and $^{13}$C beams with energies of up to 10.4~MeV/u from the Berkeley heavy ion linear accelerator (Hilac) bombarded $^{249}$Cf targets. $^{257}$Rf was formed in the ($^{12}$C,4n), $^{258}$Rf in both ($^{12}$C,3n) and ($^{13}$C,4n) and $^{259}$Rf in the ($^{13}$C,3n) fusion-evaporation reactions. Recoil products were swept by helium gas to a wheel, which rotated periodically. Alpha-decay and spontaneous fission was recorded with four Si-Au surface-barrier crystal detectors. ``$^{251}$104 is a 4.5-sec alpha-particle activity with a complex spectrum; $^{259}$104 is likewise an alpha emitter with a half-life of 3~sec. $^{258}$104 is tentatively identified as an 11-msec spontaneous-fission activity.''

\subsection*{$^{260}$Rf}\vspace{0.0cm}
Somerville et al.\ identified $^{260}$Rf in ``Spontaneous fission of rutherfordium isotopes'' in 1985 \cite{1985Som01}. Oxygen and nitrogen beams from the Berkeley 88-in. cyclotron were used to form $^{260}$Rf in the reactions $^{249}$Bk($^{15}$N,4n), $^{248}$Cm($^{16}$O,4n), and $^{249}$Cf($^{18}$O,$\alpha$3n) at beam energies of 80, 92, and 96 MeV, respectively. Helium transported the recoils onto a long tape collector which passed one meter of stationary mica track detectors. ``The following tentative assignments are based on several cross bombardments and comparisons between experimental and calculated production cross sections: $^{256}$Rf(9$\pm$2~ms), $^{257}$Rf(3.8$\pm$0.8~s, 14$\pm$9\% SF), $^{258}$Rf(13$\pm$3~ms), $^{259}$Rf(3.4$\pm$1.7~s, 9$\pm$3\% SF), $^{260}$Rf(21$\pm$1~ms), and $^{262}$Rf(47$\pm$5~ms).'' Earlier reports of half-lives of 0.3~s \cite{1964Fle01,1964Fle02}, 0.1~s (assigned to either $^{259}$Rf or $^{260}$Rf) \cite{1970Oga01}, and 80~ms \cite{1976Dru01,1977Dru01} could not be confirmed. A $\sim$20~ms had been observed earlier, however, without a firm mass assignment \cite{1981Nit01}.

\subsection*{$^{261}$Rf}\vspace{0.0cm}
The first observation of $^{261}$Rf was described by Ghiorso et al.\ in the 1970 paper ``$^{261}$Rf; new isotope of element 104'' \cite{1970Ghi03}. A $^{248}$Cm target was bombarded with 90$-$100~MeV $^{18}$O beams from the Berkeley heavy-ion linear accelerator (Hilac) and $^{261}$Rf was populated in the (5n) fusion-evaporation reaction. Recoil products were swept by helium gas to a wheel, which rotated periodically. Alpha-decay and spontaneous fission were recorded with five Si-Au surface-barrier crystal detectors. ``Altogether, as indicated above, the experimental data are consistent with the interpretation of the 8.3 MeV, 65~s $\alpha$ activity being the $\alpha$ precursor of $^{257}$No and thus unambiguously identifying it as $^{261}$Rf.''

\subsection*{$^{262}$Rf}\vspace{0.0cm}
Somerville et al.\ identified $^{262}$Rf in ``Spontaneous fission of rutherfordium isotopes'' in 1985 \cite{1985Som01}. Oxygen and neon beams from the Berkeley 88-in. cyclotron were used to form $^{262}$Rf in the reactions $^{248}$Cm($^{18}$O,4n) and $^{244}$Pu($^{22}$Ne,4n) at beam energies of 89 and 113 MeV, respectively. Helium transported the recoils onto a long tape collector which passed one meter of stationary mica track detectors. ``The following tentative assignments are based on several cross bombardments and comparisons between experimental and calculated production cross sections: $^{256}$Rf(9$\pm$2~ms), $^{257}$Rf(3.8$\pm$0.8~s, l4$\pm$9\% SF), $^{258}$Rf(13$\pm$3~ms), $^{259}$Rf(3.4$\pm$1.7~s, 9$\pm$3\% SF), $^{260}$Rf(2l$\pm$1~ms), and $^{262}$Rf(47$\pm$5~ms).'' Later papers did neither confirm nor reject this measurement reporting half-lives of 1.2~s \cite{1994Laz02} and 2.1(2)~s \cite{1996Lan01}. Recently, it was suggested that these longer half-lives were due to an isomeric state of $^{261}$Rf \cite{2006Dvo01}.

\subsection*{$^{263}$Rf}\vspace{0.0cm}
The first observation of $^{263}$Rf was reported by Kratz et al.\ in the 2003 paper ``An EC-branch in the decay of 27-s $^{263}$Db: Evidence for the isotope $^{263}$Rf'' \cite{2003Kra01}. A 123.1~MeV $^{18}$O beam from the PSI Philips Cyclotron bombarded a $^{249}$Bk target forming $^{263}$Db in the (4n) fusion-evaporation reaction. $^{263}$Rf was then populated by electron capture. Recoil products were transported to a collection site with a helium gas containing KCl aerosols. Alpha-particles and spontenous fission events were measured with sixteen passivated implanted planar silicon detectors. ``Thus, there is growing evidence for a small EC-branch in the decay of $^{263}$Db through which the new isotope $^{263}$Rf is formed. $^{263}$Rf has a relatively long half life of tens of minutes and decays predominantly by SF.'' More recently a spontaneous fission half-life of 8$^{+40}_{-4}$~s was measured for $^{263}$Rf \cite{2008Dvo01,2010Gra01} without referencing the Kratz et al.\ results. This apparent discrepancy has not been resolved.

\subsection*{$^{265}$Rf}\vspace{0.0cm}
Ellison et al.\ described the discovery of $^{265}$Rf in 2010 in ``New superheavy element isotopes: $^{242}$Pu($^{48}$Ca, 5n)$^{285}$114'' \cite{2010Eli01}. $^{242}$PuO$_2$ targets were bombarded with a 247~MeV $^{48}$Ca beams from the Berkeley 88-in. cyclotron and $^{285}$114 was produced in (5n) fusion-evaporation reactions. $^{265}$Rf was populated by subsequent $\alpha$ decay. Residues were separated with the Berkeley Gas-Filled Separator BGS and detected in multiwire proportional counters and silicon strip detectors. Subsequent radioactive decay events were recorded in the strip detectors and additional silicon chips forming a five-sided box. ``The decay chain terminated 152 seconds later with a 208.1~MeV SF-like event...  interpreted as the SF of $^{265}_{104}$Rf.'' A single decay chain was observed. A previously reported observation of $^{265}$Rf \cite{1999Nin01} was later retracted \cite{2002Nin01}.

\subsection*{$^{267}$Rf}\vspace{0.0cm}
In the 2004 paper ``Measurements of cross sections and decay properties of the isotopes of elements 112, 114, and 116 produced in the fusion reactions $^{233,238}$U, $^{242}$Pu, and $^{248}$Cm+$^{48}$Ca'', Oganessian et al.\ identified $^{267}$Rf \cite{2004Oga02}. $^{238}$U and $^{242}$Pu targets were bombarded with $^{48}$Ca beams from the Dubna U400 cyclotron producing $^{283}$Cn and $^{287}$114, respectively. $^{267}$Rf was then populated by $\alpha$ decays. The residues were separated with a gas-filled recoil separator and implanted in a semiconductor detector array. Subsequent $\alpha$ particle decay and spontaneous fission events were recorded in this array and in eight detectors arranged in a box configuration around the implantation detector. ``Data on the decay characteristics of the isotopes $^{286,287}$114, $^{282,283}$112, and $^{279}$110, as well as $^{275}$Hs, $^{271}$Sg, and $^{267}$Rf synthesized in the reactions $^{242}$Pu, $^{238}$U+$^{48}$Ca, are summarized in [the table].'' A single event for $^{267}$Rf was observed, decaying by spontaneous fission with a half-life of 2.3$^{+98}_{-1.7}$~h.


\section{Discovery of $^{256-270}$Db}

Dubnium was essentially discovered simultaneously in Berkeley reporting the discovery of $^{260}$Db by Ghiorso et al.\ on April 17, 1970 \cite{1970Ghi02} and in Dubna describing the observation of $^{261}$Db by Flerov et al.\ on June 30, 1970 \cite{1970Fle01}. Ghiorso et al. suggested the name hahnium: ``In honor of the late Otto Hahn we respectfully suggest that this new element be given the name hahnium with the symbol Ha'' \cite{1970Ghi02} while the Dubna group recommended the name nielsbohrium. These names were in use until the controversy was resolved by IUPAC in 1997. In 1994, the Commission on Nomenclature of Inorganic Chemistry of IUPAC had not accepted either of the suggestions and recommended joliotium instead \cite{1994IUP01}. However, this decision was changed to dubnium in 1997 \cite{1997IUP01} and officially accepted later in the year \cite{1997IUP01,1997IUP02,1997IUP03,1997JCE01}. The discovery of dubnium had officially been accepted by the IUPAC-IUPAP Transfermium Working Group in 1993: ``Independent work reported in 1970 from Berkeley \cite{1970Ghi02} and from Dubna \cite{1971Dru01} was essentially contemporaneous and equally convincing'' \cite{1992Bar01,1993TWG01}. Eleven dubnium isotopes have been reported so far.

Half-lives of 1.5~s and 1.6~s had been reported for $^{255}$Db only in a conference proceeding \cite{1976Fle01} as quoted in references \cite{1987Hyd01,1999Art01} and an internal report \cite{1983Oga01} as quoted in reference \cite{1999Art01}, respectively. $^{266}$Db was at the end of the isotope chain originating at $^{282}$113, however, the observed spontaneous fission could have been due to either $^{266}$Db or $^{266}$Rf \cite{2007Oga01,2007Oga02}.

\subsection*{$^{256}$Db}\vspace{0.0cm}
He\ss berger et al.\ reported the first observation of $^{256}$Db in ``Decay properties of neutron-deficient isotopes $^{256,257}$Db, $^{255}$Rf, $^{252,253}$Lr'' in 2001 \cite{2001Hes01}. A $^{209}$Bi target was bombarded with a 5.08~MeV/u $^{50}$Ti beam from the GSI UNILAC accelerator and $^{256}$Db was formed in (3n) fusion-evaporation reactions. Recoil products were separated with the velocity filter SHIP and implanted in a position-sensitive 16-strip PIPS detector which also measured subsequent $\alpha$-decay and spontaneous fission. In addition, escaping $\alpha$-decay and spontaneous fission events were recorded in six silicon detectors located in the backward hemisphere. ``The identification of the isotopes $^{256}$Db and $^{252}$Lr was based on a total of 16 $\alpha$-decay chains, that were followed down to $^{244}$Cf according to the sequences $^{256}$Db$\stackrel{\alpha}{\rightarrow}^{252}$Lr$\stackrel{\alpha}{\rightarrow}^{248}$Md$\stackrel{\alpha}{\rightarrow}^{244}$Es$\stackrel{EC}{\rightarrow}^{244}$Cf$\stackrel{\alpha}{\rightarrow}^{240}$Cm or $^{256}$Db$\stackrel{\alpha}{\rightarrow}^{252}$Lr$\stackrel{\alpha}{\rightarrow}^{248}$Md$\stackrel{EC}{\rightarrow}^{248}$Fm$\stackrel{\alpha}{\rightarrow}^{244}$Cf$\stackrel{\alpha}{\rightarrow}^{240}$Cm.''

\subsection*{$^{257}$Db}\vspace{0.0cm}
The discovery of $^{257}$Db was reported in 1985 in the paper ``The new isotopes $^{258}$105, $^{257}$105, $^{254}$Lr and $^{253}$Lr'' by He\ss berger et al.\ \cite{1985Hes01}. $^{209}$Bi targets were bombarded with 4.65, 4.75, 4.85, and 4.95 MeV/u $^{50}$Ti beams from the GSI UNILAC accelerator and $^{257}$Db was formed in the (2n) fusion-evaporation reaction. Recoil products were separated with the velocity filter SHIP and implanted in seven position-sensitive surface barrier detectors which also measured subsequent $\alpha$-decay and spontaneous fission. ``Isotope $^{257}$105: This isotope was produced in the reaction $^{209}$Bi($^{50}$Ti,2n)$^{257}$105 and also identified by $\alpha-\alpha$ correlations to its decay products $^{253}$Lr, $^{249}$Md, $^{24S}$Es.'' A previous assignment of a 5~s spontaneous fission half-life to $^{257}$Db \cite{1976Oga01} was later reassigned to $^{258}$Rf and $^{258}$Db \cite{1986Hes01}.

\subsection*{$^{258}$Db}\vspace{0.0cm}
In the 1981 paper ``Identification of element 107 by $\alpha$ correlation chains'' M\"unzenberg at al.\ described the discovery of $^{258}$Db \cite{1981Mun02}. A 4.85~MeV/u $^{54}$Cr from the GSI UNILAC linear accelerator bombarded $^{209}$Bi targets forming $^{262}$Bh in the (1n) fusion-evaporation reaction. $^{258}$Db was then populated by $\alpha$-decay. Recoil products were separated with the velocity filter SHIP and implanted in seven position sensitive surface barrier detectors which also measured the subsequent $\alpha$-decays and spontaneous fission. Four events for the decay of $^{258}$Db were measured. In addition, $^{258}$Db was also formed in the fusion evaporation reaction $^{209}$Bi($^{50}$Ti,n) at a beam energy of 4.75~MeV/u: ``$^{258}$105 can be produced in $^{50}$Ti on $^{209}$Bi irradiations by evaporation of one neutron. At 4.75 MeV/u we observed decays of (9,189$\pm$35)~keV and (9,066$\pm$35)~keV with (4.0$^{+1.8}_{-1.6}$)~s half-life and (8,468$\pm$30)~keV with (18$^{+19}_{-6}$)~s half-life corresponding to $^{258}$105 and $^{254}$Lr respectively in good agreement to the data from $^{262}$107 shown in the table.'' It is interesting to note that four years later the same group published a paper titled ``The New Isotopes $^{258}$105, $^{207}$105, $^{254}$Lr and $^{253}$Lr'' \cite{1985Hes01} describing the formation and decay of $^{258}$Db without discussing the earlier work.

\subsection*{$^{259}$Db}\vspace{0.0cm}
The first observation of $^{259}$Db was reported by Gan et al.\ in the 2001 article ``A new alpha-particle-emitting isotope $^{259}$Db'' \cite{2001Gan01}. The Lanzhou Sector Focus Cyclotron (SFC) was used to bombard a $^{241}$Am target with a 132~MeV $^{22}$Ne beam to populate $^{259}$Db in the $^{241}$Am($^{22}$Ne,4n) fusion-evaporation reaction. Recoil products were transported with helium gas and collected on a rotating wheel which was located in front of four groups of three Si(Au) surface-barrier detectors. ``An obvious $\alpha$-peak with the energy of 9.47~MeV appearing in [the figure] is assigned to $^{259}$Db in the present work. Its half-life is measured to be 0.51$\pm$0.16~s.''

\subsection*{$^{260}$Db}\vspace{0.0cm}
Ghiorso et al.\ discovered $^{260}$Db as described in ``New element hahnium, atomic number 105'' in 1970 \cite{1970Ghi02}. A $^{249}$Cf target was bombarded with a 85~MeV $^{15}$N beam from the Berkeley heavy-ion linear accelerator (HILAC) forming $^{260}$Db in the (4n) fusion-evaporation reaction. Recoil products were removed from the target with a helium jet and implanted on a wheel which was periodically rotated in front of a series of solid-state Si-Au surface barrier detectors to measure $\alpha$-spectra. ``In the inset above the sum spectrum in [the figure] there is shown an alpha spectrum of 30-sec $^{256}$Lr produced by the reaction $^{249}$Cf($^{11}$B,4n)$^{256}$Lr. Because of the similarity of the sum spectrum with the spectrum in the inset, and the good agreement of the half-lives, the daughter activity is assigned to $^{256}$Lr and therefore the 9.1-MeV mother activity has to be $^{260}$105.'' Ghiorso et al.\ could not confirm earlier results by Flerov et al.\ which were only published in a conference proceeding \cite{1968Fle03} and an internal report \cite{1968Fle02}. Bemis et al.\ confirmed the identification of $^{260}$Db by measuring L-series X-rays of lawrencium: ``Our results for $^{260}$105 completely corroborate and extend the earlier experiments of Ghiorso et al. The unique identification provided for element 105 in our present experiments unequivocally supports the discovery claims for element 105 proffered by Ghiorso et al.'' \cite{1977Bem01}.

\subsection*{$^{261}$Db}\vspace{0.0cm}
The first observation of $^{261}$Db was reported in 1970 by Flerov et al.\ in ``The synthesis of element 105'' \cite{1970Fle01}. A 114~MeV $^{22}$Ne beam bombarded a $^{243}$Am target forming $^{261}$Db in the (4n) fusion-evaporation reactions. Recoil products were implanted on a nickel ribbon moving at a constant speed passing by 105 phosphate glass fission fragment detectors. ``Considering the data obtained altogether, we arrive at the conclusion that the product experiencing spontaneous fission with a half-life of $\sim$2~sec observed in the reaction of Am$^{243}$ + Ne$^{22}$ is an isotope of element 105... The most probable mass number of the isotope of the new element is 261.'' The same results were submitted to Nuclear Physics a month later \cite{1971Fle01}. A month earlier, $\alpha$-decay with a half-life of 1.4~s was assigned to either $^{260}$Db or $^{261}$Db \cite{1971Dru01}.

\subsection*{$^{262}$Db}\vspace{0.0cm}
$^{262}$Db was identified by Ghiorso et al.\ in the 1971 paper ``Two new alpha-particle emitting isotopes of element 105, $^{261}$Ha and $^{262}$Ha'' \cite{1971Ghi01}. A $^{249}$Bk target was bombarded with 92$-$97~MeV MeV $^{18}$O beams from the Berkeley heavy-ion linear accelerator (HILAC) and $^{262}$Db was formed in the (5n) fusion-evaporation reaction. Recoil produced were transported by a He jet onto a wheel which was periodically rotated in front of seven Au-Si surface-barrier detectors. ``The new 40$\pm$10-sec activity which is assigned to $^{262}$Ha has a complex $\alpha$-particle spectrum with the most prominent peaks at 8.45 and 8.66 MeV.''

\subsection*{$^{263}$Db}\vspace{0.0cm}
In the 1992 paper ``New nuclide $^{263}$Ha'' Kratz et al.\ reported the discovery of $^{263}$Db \cite{1992Kra01}. A 93~MeV $^{18}$O beam from the Berkeley 88-in.\ cyclotron bombarded a $^{249}$Bk target and $^{263}$Db was populated in the (4n) fusion-evaporation reaction. Recoil products were removed from the target with a helium gas system containing KCl aerosols. At a collection station $\alpha$-decay and spontaneous fission events were recorded with silicon detectors on-line, and subsequently analyzed with the automated rapid chemistry apparatus ARCA II. ``After chemical separation, $^{263}$Ha was found to decay by spontaneous fission (57$^{+13}_{-15}$\%) and by $\alpha$ emission (E$_\alpha$ = 8.35~MeV, 43\%) with a half-life of 27$^{+10}_{-7}$s.''

\subsection*{$^{267,268}$Db}\vspace{0.0cm}
$^{267}$Db and $^{268}$Db were first observed by Oganessian et al.\ in 2004 as reported in ``Experiments on the synthesis of element 115 in the reaction $^{243}$Am($^{48}$Ca,xn)$^{291-x}$115'' \cite{2004Oga01}. The Dubna U400 cyclotron was used to bombard an AmO$_2$ target enriched in $^{243}$Am with 253~MeV and 248~MeV $^{48}$Ca beams to form $^{287}$115 and $^{288}$115 in (4n) and (3n) fusion evaporation reactions, respectively. $^{267}$Db and $^{268}$Db were populated by subsequent $\alpha$-decays. The residues were separated with a gas-filled recoil separator and implanted in a semiconductor detector array. Subsequent $\alpha$ particle decay and spontaneous fission events were recorded in this array and in eight detectors arranged in a box configuration around the implantation detector. ``The experimental decay scheme for $^{287}$115 is also supported by the agreement of the observed decay properties of the other nuclides in the decay chain with the expectations of theory. This means that the SF occurs directly in the decay of $^{267}$Db since the calculated $\alpha$-decay energies and EC-decay energies for this isotope are rather low (Q$_\alpha$ =7.41 MeV, Q$_{EC}$ = 1 MeV) and their expected partial half-lives significantly exceed the observed time interval of 106 min... In the decay chains shown in [the figure], we assigned SF events to the isotope $^{268}$Db following five consecutive $\alpha$ decays.'' One decay chain involving $^{267}$Db and three chains involving $^{268}$Db were observed.

\subsection*{$^{270}$Db}\vspace{0.0cm}
In the 2010 paper ``Synthesis of a new element with atomic number Z = 117'', Oganessian et al.\ reported the first observation of $^{270}$Db \cite{2010Oga01}. A $^{249}$Bk target was bombarded with a 252~MeV $^{48}$Ca beam to form $^{294}$117 in the (3n) evaporation reaction. $^{270}$Db was populated by subsequent $\alpha$-decay. The residues were separated with a gas-filled recoil separator and implanted in a semiconductor detector array. Subsequent $\alpha$ particle decay and spontaneous fission events were recorded in this array and in eight detectors arranged in a box configuration around the implantation detector. ``The data are consistent with the observation of two isotopes of element 117, with atomic masses 293 and 294. These isotopes undergo $\alpha$ decay with E$_\alpha$ = 11.03(8)~MeV and 10.81(10)~MeV and half-lives 14(+11,$-$4) and 78(+370,$-$36)~ms, respectively, giving rise to sequential $\alpha$-decay chains ending in spontaneous fission of $^{281}$Rg (T$_{SF} \sim$ 26~s) and $^{270}$Db (T$_{SF} \sim$ 1~d), respectively.'' A single decay chain beginning at $^{294}$117 and ending with the spontaneous fission of  $^{270}$Db was observed.


\section{Discovery of $^{258-271}$Sg}
Ghiorso et al.\ reported the discovery of $^{263}$Sg on September 9, 1974 \cite{1974Ghi01} and only three days later, Oganessian et al.\ tentatively assigned spontaneous fission events to $^{259}$Sg \cite{1974Oga01}. The two groups from Berkeley and Dubna had communicated their results at conferences and in internal reports and Ghiorso et al.\ stated: ``In view of the simultaneity of the experiments at the Dubna and Lawrence
laboratories, and their very different nature, we shall postpone suggesting a name for element 106 until the situation has been clarified'' \cite{1974Ghi01}. In 1993, the IUPAC-IUPAP Transfermium Working Group gave the credit for the discovery of seaborgium to Ghiorso et al.: ``Independent work reported in 1974 from Berkeley-Livermore \cite{1974Ghi01} and from Dubna \cite{1974Oga01} was essentially contemporaneous. The Dubna work is highly important for later developments but does not demonstrate the formation of a new element with adequate conviction, whereas that from Berkeley-Livermore does'' \cite{1992Bar01,1993TWG01}. The suggestion for the name seaborgium was not accepted by the 1994 Commission on Nomenclature of Inorganic Chemistry of IUPAC recommending the name rutherfordium instead \cite{1994IUP01}. This decision was changed in 1997 when the name seaborgium was officially accepted \cite{1997IUP01,1997IUP02,1997IUP03,1997JCE01}. Twelve seaborgium isotopes have been reported so far.

\subsection*{$^{258}$Sg}\vspace{0.0cm}
In 1997, He\ss berger et al.\ described the identification of $^{258}$Sg in ``Spontaneous fission and alpha-decay properties of neutron deficient isotopes $^{257-253}$104 and $^{258}$106'' \cite{1997Hes01}. $^{209}$Bi targets were bombarded with 4.77, 4.91, and 4.99~MeV/u $^{51}$V beams from the GSI UNILAC accelerator and $^{258}$Sg was produced in (2n) evaporation reactions.  Recoil products were separated with the velocity filter SHIP and implanted in a position sensitive 16-strip silicon wafer which also measured subsequent $\alpha$ decay and spontaneous fission. ``The spontaneous fission activity was attributed to $^{258}$106, the 2n deexcitation channel, since its maximum production rate was found to be close to the E¤ value where the measured excitation function for the similar reaction $^{50}$Ti + $^{208}$Pb $\rightarrow ^{258}$104 showed the maximum of the 2n deexcitation channel.'' A total of eleven spontaneous fission events of $^{258}$Sg were observed.

\subsection*{$^{259}$Sg}\vspace{0.0cm}
$^{259}$Sg was identified in ``The isotopes $^{259}$106 ,$^{260}$106, and $^{261}$106'' by M\"unzenberg et al.\ in 1985 \cite{1985Mun01}. $^{207}$Pb targets were bombarded with a 262~MeV $^{54}$Cr beam from the GSI UNILAC heavy-ion accelerator forming $^{259}$Sg in the (2n) fusion-evaporation reaction. Recoil products were separated with the velocity filter SHIP and implanted in an array of position sensitive surface barrier detectors which also measured subsequent $\alpha$ decay and spontaneous fission. ``In an irradiation of $^{207}$Pb with $^{54}$Cr at 4.90 MeV/u, the optimum energy for the 2n channel, we produced the isotope $^{259}$106.'' Seven $\alpha$-decay events of $^{259}$Sg were observed. In 1974, Oganessian et al.\ had assumed that spontaneous fission events produced in reactions of $^{54}$Cr on $^{207}$Pb and $^{208}$Pb originated from $^{259}$Sg \cite{1974Oga01}.

\subsection*{$^{260}$Sg}\vspace{0.0cm}
Demin et al.\ described the identification of $^{260}$Sg in ``On the properties of the element 106 isotopes produced in the reactions Pb + $^{54}$Cr'' in 1984 \cite{1984Dem01}. Enriched $^{206,207,208}$Pb targets were bombarded with a 290~MeV $^{54}$Cr beam from the Dubna U400 cyclotron. No experimental details were given referring to earlier publications. ``Thus from the relation given above it follows that the 6~ms activity yield corresponds to the total yield from the reaction $^{208}$Pb($^{54}$Cr,2n)$^{260}$106.''

\subsection*{$^{261}$Sg}\vspace{0.0cm}
M\"unzenberg et al.\ discovered $^{261}$Sg in the 1984 paper ``The identification of element 108'' \cite{1984Mun01}. A 5.02~MeV/u $^{58}$Fe beam from the GSI heavy ion accelerator UNILAC bombarded an enriched $^{208}$Pb target and $^{265}$Hs was formed in the (1n) fusion-evaporation reaction. $^{261}$Sg was then populated by $\alpha$-decay. Recoil products were separated with the velocity filter SHIP and implanted in an array of seven position sensitive surface barrier detectors which also measured the subsequent $\alpha$-decay and spontaneous fission. ``In particular, the decay of the daughter was seen with full energy in the second chain. The observed energy of (9.57$\pm$0.03) MeV is in excellent agreement with that of the isotope $^{261}$106 - which has a prominent transition of (9.56$\pm$0.03) MeV - unambiguously identified in 8 events by correlation to the daughter $^{257}$104 in a companion experiment using the reaction $^{208}$Pb($^{54}$Cr,1n)$^{261}$106. The half-life of the three daughter decays of (0.11$^{+0.14}_{-0.04}$)~s overlaps with the {0.26$^{+0.11}_{-0.06}$)~s half-life obtained for $^{261}$105.'' The results of the companion experiment were published a year later \cite{1985Mun01}.

\subsection*{$^{262}$Sg}\vspace{0.0cm}
The first observation of $^{262}$Sg was reported in 2001 in ``The new isotope $^{270}$110 and its decay products $^{266}$Hs and $^{262}$Sg'' by Hofmann et al.\ \cite{2001Hof01}. A 317 MeV $^{64}$Ni beam accelerated by the GSI UNILAC bombarded an enriched $^{207}$Pb target producing $^{270}$Ds in the (1n) fusion evaporation reaction. $^{270}$Ds and the subsequent $\alpha$-decay daughters $^{266}$Hs and $^{262}$Sg were identified with a detector system at the velocity filter SHIP. ``The nucleus $^{262}$Sg decays by fission with a half-life of (6.9$^{+3.8}_{-1.8}$)~ms and a total kinetic energy of the fission fragments of (222$\pm$10) MeV.''

\subsection*{$^{263}$Sg}\vspace{0.0cm}
Ghiorso et al.\ discovered $^{263}$Sg in 1974 as described in the paper ``Element 106'' \cite{1974Ghi01}. A 95~MeV $^{18}$O beam from the Berkeley SuperHILAC bombarded a $^{249}$Cf target and $^{263}$Sg was formed in the (4n) fusion-evaporation reaction. Recoil products were swept to a series of detection stations with a helium flow containing NaCl aerosol. Alpha-particles and spontaneous fission fragments were measured with Si(Au) surface barrier detectors. ``The new nuclide $^{263}$106, produced by the ($^{18}$O,4n) reaction, is shown to decay by $\alpha$ emission with a half-life of 0.9$\pm$0.2~sec and a principal $\alpha$ energy of 9.06$\pm$0.04~MeV to the known nuclide $^{259}$Rf, which in turn is shown to decay to the known nuclide $^{255}$No.'' The experimental results were confirmed for the first time twenty years later by Gregorich et al.\ \cite{1994Gre01}.

\subsection*{$^{264}$Sg}\vspace{0.0cm}
$^{264}$Sg was first observed by Gregorich et al.\ in ``New isotope $^{264}$Sg and decay properties of $^{262-264}$Sg'' in 2006 \cite{2006Gre01}. The Berkeley 88-in. cyclotron was used to accelerate $^{30}$Si beams to 5.2$-$6.0~MeV/nucleon and bombard $^{238}$UF$_4$ targets. $^{264}$Sg was populated in the (4n) fusion-evaporation reaction and separated with the Berkeley Gas-filled Separator (BGS). A Si-strip detector array measured the implanted recoil products and the subsequent $\alpha$-decay and spontaneous fission. ``Five SF events assigned to new isotope $^{264}$Sg, produced by the $^{238}$U($^{30}$Si,4n)$^{264}$Sg reaction, were observed at the lowest $^{30}$Si energy.'' The measured half-life was 37$^{+27}_{-11}$~ms. Less than a month later Nishio et al.\ independently reported the observation of three spontaneous fission events for $^{264}$Sg \cite{2006Nis01}.

\subsection*{$^{265}$Sg}\vspace{0.0cm}
The identification of $^{265}$Sg was reported by Lazarev et al.\ in the 1994 paper ``Discovery of enhanced nuclear stability near the deformed shells N = 162 and Z = 108'' \cite{1994Laz02}. A $^{248}$Cm target was bombarded with a 121 MeV $^{22}$Ne beam from the Dubna U400 cyclotron and $^{265}$Sg was formed in the (5n) fusion-evaporation reaction. Recoil products were separated with a gas-filled recoil separator and implanted in a position sensitive surface barrier detector array which also measured subsequent $\alpha$-decay and spontaneous fission. ``We assigned the four $\alpha-\alpha-(\alpha)$ correlations at 121 MeV with E$_{\alpha_1}$ = 8.71 to 8.91~MeV to the decay chain $^{265}$106 $\rightarrow ^{261}$104 (T$_{1/2}$ = 65~s, E$_\alpha \sim$ 8.29~MeV) $\rightarrow ^{257}$102 (T$_{1/2}$ = 26~s, E$_\alpha \sim$ 8.22, 8.27, 8.32~MeV) for which we measured a production cross section of 260~pb.''

\subsection*{$^{266}$Sg}\vspace{0.0cm}
Dvorak et al.\ described the identification of $^{266}$Sg in the 2006 paper ``Doubly magic nucleus $^{270}_{108}$Hs$_{162}$'' \cite{2006Dvo01}. A $^{248}$Cm target was bombarded with 185 and 193~MeV $^{26}$Mg beams from the GSI UNILAC accelerator forming $^{270}$Hs in the (4n) fusion-evaporation reaction. $^{266}$Sg was then populated by $\alpha$-decay. Alpha-particles and spontaneous fission events were detected with 2 $\times$ 32 PIPS detectors following rapid chemical separation of hassium. Four chains terminating with spontaneous fission of $^{266}$Sg were observed: ``Three out of four chains were detected at the lower beam energy at the expected maximum of the 4n evaporation channel. Therefore, we assign these four chains to the decay of the new isotope $^{270}$Hs and its daughter $^{266}$Sg.'' The earlier reported $\alpha$-decay of $^{266}$Sg \cite{1994Laz02,1998Tur01} could not be confirmed.

\subsection*{$^{267}$Sg}\vspace{0.0cm}
The observation of $^{267}$Sg was reported in 2008 by Dvorak et al.\ in ``Observation of the 3n evaporation channel in the complete hot-fusion reaction $^{26}$Mg + $^{248}$Cm leading to the new superheavy nuclide $^{271}$Hs'' \cite{2008Dvo01}. $^{26}$Mg beams accelerated by the GSI linear accelerator UNILAC to 130 and 140~MeV bombarded a $^{248}$Cm target to form $^{271}$Hs in the (3n) fusion-evaporation reaction. $^{271}$Sg was populated by subsequent $\alpha$ decay. Alpha-decay chains were measured with the online chemical separation and detection system COMPACT. ``A relatively long half-life was measured for $^{267}$Sg, which is consistent with the observed low $\alpha$-particle energy, causing $\alpha$/SF branching in this nucleus.'' In a table a half-life of 80$^{+60}_{-20}$~s is quoted for $^{267}$Sg.

\subsection*{$^{269}$Sg}\vspace{0.0cm}
Ellison et al.\ described the discovery of $^{269}$Sg in 2010 in ``New superheavy element isotopes; $^{242}$Pu($^{48}$Ca, 5n)$^{285}$114'' \cite{2010Eli01}. $^{242}$PuO$_2$ targets were bombarded with a 247~MeV $^{48}$Ca beams from the Berkeley 88-in. cyclotron and $^{285}$114 was produced in (5n) fusion-evaporation reactions. $^{269}$Sg was populated by subsequent $\alpha$ decay. Residues were separated with the Berkeley Gas-Filled Separator BGS and detected in multiwire proportional counters and silicon strip detectors. Subsequent radioactive decay events were recorded in the strip detectors and additional silicon chips forming a five-sided box. ``The chain continued with four subsequent $\alpha$-like events... after 140 ms, 8.21 ms, 346 ms, and 185 s with energies of 10.31, 10.57, 9.59, and 8.57 MeV, which are interpreted as the successive $\alpha$ decays of $^{281}_{112}$Cn, $^{277}_{110}$Ds, $^{273}_{108}$Hs, and $^{269}_{106}$Sg, respectively.'' A single decay chain was observed. A previously reported observation of $^{269}$Sg \cite{1999Nin01} was later retracted \cite{2002Nin01}.

\subsection*{$^{271}$Sg}\vspace{0.0cm}
In the 2004 paper ``Measurements of cross sections and decay properties of the isotopes of elements 112, 114, and 116 produced in the fusion reactions $^{233,238}$U, $^{242}$Pu, and $^{248}$Cm+$^{48}$Ca'', Oganessian et al.\ identified $^{271}$Sg \cite{2004Oga02}. $^{238}$U and $^{242}$Pu targets were bombarded with $^{48}$Ca beams from the Dubna U400 cyclotron producing $^{283}$Cn and $^{287}$114, respectively. $^{271}$Sg was then populated by $\alpha$ decays. The residues were separated with a gas-filled recoil separator and implanted in a semiconductor detector array. Subsequent $\alpha$ particle decay and spontaneous fission events were recorded in this array and in eight detectors arranged in a box configuration around the implantation detector. ``Data on the decay characteristics of the isotopes $^{286,287}$114, $^{282,283}$112, and $^{279}$110, as well as $^{275}$Hs, $^{271}$Sg, and $^{267}$Rf synthesized in the reactions $^{242}$Pu, $^{238}$U+$^{48}$Ca, are summarized in [the table].'' 2 events for $^{271}$Sg were observed, one decaying by $\alpha$-emission the other one by spontaneous fission with a half-life of 2.4$^{+4.3}_{-1.0}$~min.


\section{Discovery of $^{260-274}$Bh}
Bohrium was discovered in 1981 by M\"unzenberg et al.\ with the observation of $^{262}$Bh \cite{1981Mun02}. An earlier observation of spontaneous fission of $^{261}$Bh \cite{1976Oga01} could later not be confirmed. The observation of $^{262}$Bh by M\"unzenberg et al.\ was officially accepted by the IUPAC-IUPAP Transfermium Working Group in 1993: ``The Darmstadt work \cite{1981Mun02} provides convincing evidence for the formation of element 107'' \cite{1992Bar01,1993TWG01}. The name bohrium was officially accepted in 1997 \cite{1997IUP01,1997IUP02,1997IUP03,1997JCE01}. Originally, the name nielsbohrium (Ns) had been suggested \cite{1994IUP01}. Ten bohrium isotopes have been reported so far.

Although $^{271}$Bh was part of the decay chain originating in $^{287}$115, the decay of $^{271}$Bh itself was not observed \cite{2004Oga01}.

\subsection*{$^{260}$Bh}\vspace{0.0cm}
The first observation of $^{260}$Bh was reported by Nelson et al.\ in ``Lightest isotope of Bh produced via the $^{209}$Bi($^{52}$Cr,n)$^{260}$Bh reaction'' in 2008 \cite{2008Nel01}. $^{209}$Bi targets were bombarded with a 257.0~MeV $^{52}$Cr beam from the Berkeley 88-inch cyclotron. Evaporation residues were separated with the Berkeley Gas-filled Separator BGS and implanted in a silicon strip detector array which also measured subsequent $\alpha$ decays. ``[The figure] contains the eight observed decay chains attributed to the decay of $^{260}$Bh... Using the eight alpha decay lifetimes, the half-life of $^{260}$Bh was found to be 35$^{+19}_{-9}$~ms.''

\subsection*{$^{261}$Bh}\vspace{0.0cm}
In 1989, M\"unzenberg et al.\ identified $^{261}$Bh in the paper ``Element 107'' \cite{1989Mun01}. A $^{209}$Bi target was bombarded with $^{54}$Cr beams with energies between 4.87 and 5.07~MeV/u from the GSI UNILAC accelerator forming $^{261}$Bh in (2n) fusion-evaporation reactions. Recoil products were separated with the velocity filter SHIP and implanted in seven position sensitive silicon surface-barrier detectors which also detected the subsequent $\alpha$-decay and spontaneous fission. ``We deduce from 10 events observed for decay of $^{261}$107, and no fission event with t $<$ 100~ms that the fission branching ratio is smaller than about 10\%, corresponding to a half-life for spontaneous fission of larger than 0.12~s.'' An earlier observation of spontaneous fission of $^{261}$Bh \cite{1976Oga01} could not be confirmed.

\subsection*{$^{262}$Bh}\vspace{0.0cm}
In the 1981 paper ``Identification of element 107 by $\alpha$ correlation chains'' M\"unzenberg et al.\ described the discovery of $^{262}$Bh \cite{1981Mun02}. A 4.85~MeV/u $^{54}$Cr from the GSI UNILAC linear accelerator bombarded $^{209}$Bi targets forming $^{262}$Bh in the (1n) fusion-evaporation reaction. Recoil products were separated with the velocity filter SHIP and implanted in seven position sensitive surface barrier detectors which also measured the subsequent $\alpha$-decays and spontaneous fission. ``Our results show the discovery of the $\alpha$ decay of element 107. The $\alpha$ chains end in known transitions of $^{250}$Fm and $^{250}$Md, respectively, indicating the observation of the isotope $^{262}$107 formed by the 1n channel from the compound nucleus $^{263}$107.'' Five events for the decay of $^{262}$Bh were measured.

\subsection*{$^{264}$Bh}\vspace{0.0cm}
Hofmann et al.\ discovered $^{264}$Bh in 1995 as reported in ``The new element 111'' \cite{1995Hof02}. Bismuth targets were bombarded with 318 and 320 MeV $^{64}$Ni beams from the GSI UNILAC. $^{272}$Rg was formed in the (1n) fusion-evaporation reaction and $^{264}$Bh was populated by subsequent $\alpha$-decays. Reaction residues were separated with the velocity filter SHIP and $\alpha$ decays were recorded in a position sensitive silicon detector. ``The transitions $\alpha$2 and $\alpha$3 are consequently assigned to the new isotopes $^{268}$109 and $^{264}$107.'' A half-life of 440$^{+600}_{-160}$~ms was reported.

\subsection*{$^{265}$Bh}\vspace{0.0cm}
The first observation of $^{265}$Bh was reported in 2004 by Gan et al.\ in ``New isotope $^{265}$Bh'' \cite{2004Gan01}. An americium-oxide target was bombarded with a 168~MeV $^{26}$Mg beam from the Lanzhou Sector Focus Cyclotron (SFC) forming $^{265}$Bh in the $^{243}$Am($^{26}$Mg,4n) fusion-evaporation reaction. Recoil products were collected with a helium transport system and deposited on a rotating wheel in front of four pairs of PIPS detectors. ``A total of 8 correlated decay events of $^{265}$Bh and 4 decay events of $^{264}$Bh were observed. $^{265}$Bh decays with a 0.94$^{+0.70}_{-0.31}$~s half-life by emission of $\alpha$-particles with an average energy of 9.24$\pm$0.05~MeV.''

\subsection*{$^{266,267}$Bh}\vspace{0.0cm}
The discovery of $^{266}$Bh and $^{267}$Bh was reported by Wilk et al.\ in the 2000 paper ``Evidence for new isotopes of element 107: $^{266}$Bh and $^{267}$Bh'' \cite{2000Wil01}. A $^{249}$Bk target was bombarded with 117~MeV and 123~MeV $^{22}$Ne beams from the Berkeley 88-in.\ cyclotron and $^{266}$Bh and $^{267}$Bh were formed in (5n) and (4n) fusion-evaporation reactions, respectively. Recoil products were swept with helium gas containing KCl aerosols onto a merry-go-round rotating wheel system. Alpha-decays were then recorded with six pairs of passivated, ion-implanted planar silicon detectors. ``Five atoms of $^{267}$Bh, E$_\alpha$ ranging from 8.73 to 8.87~MeV and one atom of $^{266}$Bh with an E$_\alpha$ of 9.29~MeV were identified during the experiment.'' The single event of $^{266}$Bh was observed at 123~MeV beam energy, while for $^{267}$Bh two events were measured at 123~MeV and three events at 117~MeV.

\subsection*{$^{270}$Bh}\vspace{0.0cm}
Oganessian et al.\ reported the observation of $^{270}$Bh in 2007 in ``Synthesis of the isotope $^{282}$113 in the $^{237}$Np+$^{48}$Ca fusion reaction'' \cite{2007Oga01}. A 244 MeV $^{48}$Ca beam from the Dubna U400 cyclotron bombarded a $^{237}$Np target and $^{282}$113 was populated in the (3n) fusion evaporation reaction. $^{270}$Bh was populated by subsequent $\alpha$ decays. The residues were separated with a gas-filled recoil separator and implanted in a semiconductor detector array. Alpha particle decay and spontaneous fission events were recorded in this array and in eight detectors arranged in a box configuration around the implantation detector. Only one of the two decay chains included the decay of $^{270}$Bh: ``For the last $\alpha$ decay observed in the first decay chain of $^{282}$113, the $\alpha$-particle energy as well as half-life are in agreement with those expected for $^{270}$Bh (E$_\alpha$ = 8.93$\pm$0.08 MeV, T$_{1/2}$ = 61$^{+292}_{-28}$~s, T$_{calc}$ = 5~s).''

\subsection*{$^{272}$Bh}\vspace{0.0cm}
$^{272}$Bh was first observed by Oganessian et al.\ in 2004 as reported in ``Experiments on the synthesis of element 115 in the reaction $^{243}$Am($^{48}$Ca,xn)$^{291-x}$115'' \cite{2004Oga01}. The Dubna U400 cyclotron was used to bombard an AmO$_2$ target enriched in $^{243}$Am with a 248~MeV $^{48}$Ca beam to form $^{288}$115 in (3n) fusion evaporation reactions. $^{272}$Bh was populated by subsequent $\alpha$-decays. The residues were separated with a gas-filled recoil separator and implanted in a semiconductor detector array. Alpha particle decay and spontaneous fission events were recorded in this array and in eight detectors arranged in a box configuration around the implantation detector. ``The $\alpha$-decay energies attributed to the isotopes of Mt and Bh coincide well with theoretical values'' Three decay chains involving $^{272}$Bh were observed and a half-life of 9.8$^{+11.7}_{-3.5}$~s was reported.

\subsection*{$^{274}$Bh}\vspace{0.0cm}
In the 2010 paper ``Synthesis of a new element with atomic number Z = 117'', Oganessian et al.\ reported the first observation of $^{274}$Bh \cite{2010Oga01}. A $^{249}$Bk target was bombarded with a 252~MeV $^{48}$Ca beam from the Dubna U400 cyclotron to form $^{294}$117 in the (3n) evaporation reaction. $^{274}$Bh was populated by subsequent $\alpha$-decay. The residues were separated with a gas-filled recoil separator and implanted in a semiconductor detector array.  Alpha particle decay and spontaneous fission events were recorded in this array and in eight detectors arranged in a box configuration around the implantation detector. $^{274}$Bh is not specifically mentioned in the text but an $\alpha$-energy of 9.55(19)~MeV with a lifetime of 1.3~min is quoted in the figure displaying the single observed decay chain.


\section{Discovery of $^{263-277}$Hs}
On April 14, 1984, M\"unzenberg et al.\ submitted the identification of $^{265}$Hs for publication followed by the submission by Oganessian et al.\ on June 14, 1984 reporting the observation of $^{263-265}$Hs \cite{1984Oga01}. The Dubna group did not observe the $\alpha$ decay of the hassium isotopes directly and inferred their formation from known decays of the granddaughters ($^{263,264}$Hs) or the great-great-great-granddaughter ($^{265}$Hs). The discovery of hassium was officially accepted by the IUPAC-IUPAP Transfermium Working Group in 1993: ``The formation of element 108 was established by simultaneous and independent work in Darmstadt \cite{1984Mun01} and Dubna \cite{1984Oga01}'' \cite{1992Bar01,1993TWG01}. The name hassium was officially accepted in 1997 \cite{1997IUP01,1997IUP02,1997IUP03,1997JCE01}. Previously, the name hahnium had been recommended \cite{1994IUP01}. Twelve hassium isotopes have been reported so far.

\subsection*{$^{263}$Hs}\vspace{0.0cm}
$^{263}$Hs was discovered by Dragojevi\'c et al.\ in 2009 as described in ``New isotope $^{263}$Hs \cite{2009Dra01}. A 280 MeV $^{56}$Fe beam from the Berkeley 88-in.\ cyclotron bombarded an enriched $^{208}$Pb target and $^{263}$Hs was formed in the (1n) fusion-evaporation reaction. Recoil products were separated with the Berkeley gas-filled separator (BGS) and implanted into a Si-strip focal plane detector array which also recorded the subsequent $\alpha$-decay and spontaneous fission. ``$^{263}$Hs was identified by observing an `EVR-like event' followed by a `$^{263}$Hs-like event' within 10~ms, and then by (i) at least two of the $^{259}$Sg, $^{255}$Rf, and $^{251}$No daughters... within 15~s, or (ii) SF (E $>$ 90 MeV), within 10~s.'' Six decay chains from $^{263}$Hs were observed. In 1984 Oganessian et al.\ reported evidence for the formation of $^{263}$Hs by identifying the decay of daughter nuclei, however, no direct evidence for the observation of $^{263}$Hs was measured \cite{1984Oga01}.

\subsection*{$^{264}$Hs}\vspace{0.0cm}
The first identification of $^{264}$Hs was reported by M\"unzenberg et al.\ in the 1986 paper ``Evidence for $^{264}$108, the heaviest known even-even isotope'' \cite{1986Mun01}. An enriched $^{207}$Pb target was bombarded with a 5.04~MeV/u beam from the GSI UNILAC accelerator and $^{264}$Hs was populated in the (1n) fusion-evaporation reaction. Recoil products were separated with the velocity filter SHIP and implanted in position sensitive surface barrier detectors which also measured subsequent $\alpha$-decays and spontaneous fission. ``We have observed the decay of one atom of the doubly even isotope 108. we observed $\alpha$-decay with a half-life of (76$^{+364}_{-36}$)~$\mu$s.'' In 1984 Oganessian et al.\ reported evidence for the formation of $^{264}$Hs by identifying the decay of daughter nuclei, however, no direct evidence for the observation of $^{264}$Hs was measured \cite{1984Oga01}.

\subsection*{$^{265}$Hs}\vspace{0.0cm}
M\"unzenberg et al.\ discovered $^{265}$Hs in the 1984 paper ``The identification of element 108'' \cite{1984Mun01}. A 5.02~MeV/u $^{58}$Fe beam from the GSI heavy ion accelerator UNILAC bombarded an enriched $^{208}$Pb target and $^{265}$Hs was formed in the (1n) fusion-evaporation reaction. Recoil products were separated with the velocity filter SHIP and implanted in an array of seven position sensitive surface barrier detectors which also measured the subsequent $\alpha$-decay and spontaneous fission. ``Our interpretation that this first transition is due to the alpha decay of the isotope $^{265}$108, rests primarily on the fact that the remaining four full-energy alpha signals can all be assigned to known transitions in nuclei belonging to the alpha decay chain that starts with the isotope $^{265}$108.'' Three decay chains were observed. In 1984 Oganessian et al.\ reported evidence for the formation of $^{265}$Hs by identifying the decay of daughter nuclei, however, no direct evidence for the observation of $^{265}$Hs was measured \cite{1984Oga01}.

\subsection*{$^{266}$Hs}\vspace{0.0cm}
The first observation of $^{266}$Hs was reported in 2001 in ``The new isotope $^{270}$110 and its decay products $^{266}$Hs and $^{262}$Sg'' by Hofmann et al.\ \cite{2001Hof01}. A 317 MeV $^{64}$Ni beam accelerated by the GSI UNILAC bombarded an enriched $^{207}$Pb target producing $^{270}$Ds in the (1n) fusion evaporation reaction. $^{270}$Ds and the subsequent $\alpha$-decay daughters $^{266}$Hs and $^{263}$Sg were identified with a detector system at the velocity filter SHIP. ``The nucleus $^{266}$Hs decays by $\alpha$ emission with an energy of (10.18$\pm$0.02) MeV and a half-life of (2.3$^{+1.3}_{-0.6})$~ms.''

\subsection*{$^{267}$Hs}\vspace{0.0cm}
Lazarev et al.\ identified $^{267}$Hs in 1995 in ``New nuclide $^{267}$108 produced by the $^{238}$U + $^{34}$S reaction'' \cite{1995Laz02}. A $^{238}$U target was bombarded with a 186~MeV $^{34}$S beam from the Dubna U400 cyclotron to form $^{267}$Hs in the (5n) fusion-evaporation reaction. Recoil products were separated with the Dubna Gas-filled Recoil Separator and implanted in a position sensitive detector array which also recorded subsequent $\alpha$-decays and spontaneous fission events. ``The above observations and arguments provide strong evidence for the identification of $^{267}$108. From measured time intervals between implantation and $\alpha$ decay events of the $^{267}$108 nuclides, we calculate a maximum likelihood half-life value of 19$^{+29}_{-10}$~ms.'' Three chains originating at $^{267}$Hs were recorded.

\subsection*{$^{268}$Hs}\vspace{0.0cm}
$^{268}$Hs was first observed by Nishio et al.\ as described in the 2010 paper ``Nuclear orientation in the reaction $^{34}$S + $^{238}$U and synthesis of the new isotope $^{268}$Hs'' \cite{2010Nis01}. A 5.16~MeV/u $^{34}$S beam from the GSI linear accelerator UNILAC bombarded a $^{238}$U target and $^{268}$Hs was formed in the (4n) fusion-evaporation reaction. Recoil products as well as subsequent $\alpha$-particle emission and spontaneous fission events were measured with the detector setup of the velocity filter SHIP. ``At 152.0~MeV one decay of the new isotope $^{268}$Hs was observed. It decays with a half-life of 0.38$^{+1.8}_{-0.17}$~s by 9479$\pm$16 keV $\alpha$-particle emission.''

\subsection*{$^{269}$Hs}\vspace{0.0cm}
In the 1996 paper ``The new element 112'', Hofmann et al.\ reported the identification of $^{269}$Hs \cite{1996Hof02}. A 344 MeV $^{70}$Zn beam from the GSI UNILAC bombarded enriched $^{208}$Pb targets and $^{277}$Cn was populated in the single neutron fusion-evaporation reaction. $^{269}$Hs was populated by subsequent $\alpha$-decays. Reaction residues were separated with the velocity filter SHIP and the $\alpha$ decays were recorded in a position sensitive silicon detector. ``Therefore, the observed chain must be assigned to the isotope with mass number A = 277 of element Z = 112, produced by fusion of $^{70}$Zn and $^{208}$Pb and emission of one neutron.'' Two chains were observed, however, the first chain was later retracted \cite{2002Hof01}. Within the second chain $^{269}$Hs decayed with an $\alpha$ energy of 9.23~MeV within 19.7~s. Earlier, Lazarev et al.\ had reported the observation of several decay chains beginning at $^{273}$Ds, however, only one included values for the decay of $^{269}$Hs and in the text $^{269}$Hs is always referred to as an ``unknown nucleus'' \cite{1996Laz01}.

\subsection*{$^{270}$Hs}\vspace{0.0cm}
Dvorak et al.\ described the first observation of $^{270}$Hs in the 2006 paper ``Doubly magic nucleus $^{270}_{108}$Hs$_{162}$'' \cite{2006Dvo01}. A $^{248}$Cm target was bombarded with 185 and 193~MeV $^{26}$Mg beams from the GSI UNILAC accelerator forming $^{270}$Hs in the (4n) fusion-evaporation reaction. Alpha-particles and spontaneous fission events were detected with 2 $\times$ 32 PIPS detectors following rapid chemical separation of hassium. Four chains originating in $^{270}$Hs were observed: ``Three out of four chains were detected at the lower beam energy at the expected maximum of the 4n evaporation channel. Therefore, we assign these four chains to the decay of the new isotope $^{270}$Hs and its daughter $^{266}$Sg.'' Previously two events had tentatively been assigned to the decay of $^{270}$Hs \cite{2002Dul01,2003Tur01}, however, this assignment was based on decay properties of $^{266}$Sg \cite{1994Laz02,1998Tur01} which were not confirmed.

\subsection*{$^{271}$Hs}\vspace{0.0cm}
$^{271}$Hs was discovered in 2008 by Dvorak et al.\ as reported in ``Observation of the 3n evaporation channel in the complete hot-fusion reaction $^{26}$Mg + $^{248}$Cm leading to the new superheavy nuclide $^{271}$Hs'' \cite{2008Dvo01}. $^{26}$Mg beams accelerated by the GSI linear accelerator UNILAC to 130 and 140~MeV bombarded a $^{248}$Cm target to form $^{271}$Hs in the (3n) fusion-evaporation reaction. Alpha-decay chains were measured with the online chemical separation and detection system COMPACT. ``Increased stability, as evidenced by long partial SF and $\alpha$ decay half-lives is expected when approaching the closed shell at N = 162 and is most pronounced in odd-A nuclei due to the well-known hindrance effect associated with the odd neutron. Considering the beam energy and the decay properties of the chain members, these chains are attributed to the decay of the new isotope $^{271}$Hs.'' Six chains originating in $^{271}$Hs were measured. Previously, the same group reported one event tentatively assigned to $^{271}$Hs \cite{2006Dvo01}.

\subsection*{$^{273}$Hs}\vspace{0.0cm}
Ellison et al.\ described the discovery of $^{273}$Hs in 2010 in ``New superheavy element isotopes; $^{242}$Pu($^{48}$Ca, 5n)$^{285}$114'' \cite{2010Eli01}. $^{242}$PuO$_2$ targets were bombarded with a 247~MeV $^{48}$Ca beams from the Berkeley 88-in. cyclotron and $^{285}$114 was produced in (5n) fusion-evaporation reactions. $^{273}$Hs was populated by subsequent $\alpha$ decay. Residues were separated with the Berkeley Gas-Filled Separator BGS and detected in multiwire proportional counters and silicon strip detectors. Subsequent radioactive decay events were recorded in the strip detectors and additional silicon chips forming a five-sided box. ``The chain continued with four subsequent $\alpha$-like events... after 140 ms, 8.21 ms, 346 ms, and 185 s with energies of 10.31, 10.57, 9.59, and 8.57 MeV, which are interpreted as the successive $\alpha$ decays of $^{281}_{112}$Cn, $^{277}_{110}$Ds, $^{273}_{108}$Hs, and $^{269}_{106}$Sg, respectively.'' A single decay chain was observed. A previously reported observation of $^{273}$Hs \cite{1999Nin01} was later retracted \cite{2002Nin01}.

\subsection*{$^{275}$Hs}\vspace{0.0cm}
In the 2004 paper ``Measurements of cross sections and decay properties of the isotopes of elements 112, 114, and 116 produced in the fusion reactions $^{233,238}$U, $^{242}$Pu, and $^{248}$Cm+$^{48}$Ca'', Oganessian et al.\ identified $^{275}$Hs \cite{2004Oga02}. $^{238}$U and $^{242}$Pu targets were bombarded with $^{48}$Ca beams from the Dubna U400 cyclotron producing $^{283}$Cn and $^{287}$114, respectively. $^{275}$Hs was then populated by $\alpha$ decays. The residues were separated with a gas-filled recoil separator and implanted in a semiconductor detector array. Subsequent $\alpha$ particle decay and spontaneous fission events were recorded in this array and in eight detectors arranged in a box configuration around the implantation detector. ``Data on the decay characteristics of the isotopes $^{286,287}$114, $^{282,283}$112, and $^{279}$110, as well as $^{275}$Hs, $^{271}$Sg, and $^{267}$Rf synthesized in the reactions $^{242}$Pu, $^{238}$U+$^{48}$Ca, are summarized in [the table].'' 2 decay chains were observed quoting a half-life of 0.15$^{+0.27}_{-0.06}$~s for $^{275}$Hs.

\subsection*{$^{277}$Hs}\vspace{0.0cm}
The discovery of $^{277}$Hs was reported in 2010 by D\"ullmann et al.\ in ``Production and decay of element 114: High cross sections and the new nucleus $^{277}$Hs'' \cite{2010Dul01}. The GSI Universal Linear Accelerator (UNILAC) was used to bombard a $^{244}$Pu target with 236.4$-$241.0~MeV $^{48}$Ca beams to form $^{289}$114 in the (3n) fusion-evaporation reaction. Reaction products as well as $\alpha$-emission and spontaneous fission decays were measured with the detection system of the gas-filled recoil separator TASCA. $^{277}$Hs was then populated in subsequent $\alpha$-emission. One decay event of $^{277}$Hs was observed: ``The chain was then terminated 4.5~ms later by SF of the $\alpha$-decay daughter of $^{281}$Ds, i.e., the new nucleus $^{277}$Hs with Z = 108 and N = 169.'' An earlier reported observation of the spontaneous fission of $^{277}$Hs \cite{1999Oga02} could not be reproduced.


\section{Discovery of $^{266-278}$Mt}

The discovery of meitnerium was reported in 1982 by M\"unzenberg et al.\ with the observation of $^{266}$Mt \cite{1982Mun01} and officially accepted by the IUPAC-IUPAP Transfermium Working Group in 1993: ``The Darmstadt work \cite{1982Mun01} gives confidence that element 109 has been observed'' \cite{1992Bar01,1993TWG01}. The name meitnerium was officially accepted in 1997 \cite{1997IUP01,1997IUP02,1997IUP03,1997JCE01}. Seven meitnerium isotopes have been reported so far.

\subsection*{$^{266}$Mt}\vspace{0.0cm}
The first identification of $^{266}$Mt was reported in ``Observation of one correlated $\alpha$-decay in the reaction $^{58}$Fe on $^{209}$Bi$\rightarrow ^{267}$109'' \cite{1982Mun01}. A 5.15~MeV/nucleon $^{58}$Fe beam from the GSI UNILAC heavy ion accelerator bombarded a bismuth target. $^{266}$Mt was produced in the (1n) fusion-evaporation reaction and separated with the velocity filter SHIP. The residues and subsequent $\alpha$ and spontaneous fission decays were recorded in seven position sensitive surface barrier detectors. ``In an irradiation of $^{209}$Bi targets with accelerated $^{58}$Fe ions we found a decay chain consisting of two consecutive alpha disintegrations followed by fission. This decay chain most probably originates from the isotope $^{266}$109.'' The time between the implantation and the decay of the daughter $^{262}$Bh was 5.0~ms.

\subsection*{$^{268}$Mt}\vspace{0.0cm}
Hofmann et al.\ discovered $^{268}$Mt in 1995 as reported in ``The new element 111'' \cite{1995Hof02}. Bismuth targets were bombarded with 318 and 320 MeV $^{64}$Ni beams from the GSI UNILAC. $^{272}$Rg was formed in the (1n) fusion-evaporation reaction and $^{268}$Mt was populated by $\alpha$-decay. Reaction residues were separated with the velocity filter SHIP and subsequent $\alpha$ decays were recorded in a position sensitive silicon detector. ``The transitions $\alpha$2 and $\alpha$3 are consequently assigned to the new isotopes $^{268}$109 and $^{264}$107.'' A half-life of 70$^{+100}_{-30}$~ms was reported.

\subsection*{$^{270}$Mt}\vspace{0.0cm}
The first identification of $^{270}$Mt was reported by Morita et al.\ in ``Experiment on the synthesis of element 113 in the reaction $^{209}$Bi($^{70}$Zn,n)$^{278}$113'' in 2004 \cite{2004Mor01}. Bismuth targets were bombarded with a 352.6~MeV $^{70}$Zn beam from the RIKEN linear accelerator facility RILAC and $^{270}$Mt was populated by $\alpha$-decays from $^{278}$113. Recoil products were separated with the gas-filled recoil ion separator GARIS and detected with micro-channel plates and a silicon strip detector. Spontaneous fission and $\alpha$-decay events were recorded with a silicon semiconductor detector box consisting of the central detector plus four additional silicon strip detectors forming a box. ``In conclusion, the reaction product, followed by the decay chain observed in our experiment, was considered to be most probably due to the $^{209}$Bi($^{70}$Zn,n)$^{278}$113 reaction. As a result, the members of the decay chain were consequently assigned as $^{278}$113, $^{274}$111, $^{270}$Mt, $^{266}$Bh, and $^{262}$Db.'' A single decay chain was observed and the observed decay time measured between $^{270}$Mt and $^{266}$Bh was 7.16~ms.

\subsection*{$^{274}$Mt}\vspace{0.0cm}
Oganessian et al.\ reported the observation of $^{274}$Mt in 2007 in ``Synthesis of the isotope $^{282}$113 in the $^{237}$Np+$^{48}$Ca fusion reaction'' \cite{2007Oga01}. A 244 MeV $^{48}$Ca beam from the Dubna U400 cyclotron bombarded a $^{237}$Np target and $^{282}$113 was populated in the (3n) fusion evaporation reaction. $^{274}$Mt was populated by subsequent $\alpha$ decays. The residues were separated with a gas-filled recoil separator and implanted in a semiconductor detector array. Alpha particle decay and spontaneous fission events were recorded in this array and in eight detectors arranged in a box configuration around the implantation detector. $^{274}$Mt is not specifically mentioned in the text but a figure of the two decay chains shows that $^{274}$Mt decayed within 87.98~s and 97.02~s with $\alpha$ decay energies of 8.93(8)~MeV and 8.52(10)~MeV.

\subsection*{$^{275,276}$Mt}\vspace{0.0cm}
$^{275}$Mt and $^{276}$Mt were first observed by Oganessian et al.\ in 2004 as reported in ``Experiments on the synthesis of element 115 in the reaction $^{243}$Am($^{48}$Ca,xn)$^{291-x}$115'' \cite{2004Oga01}. The Dubna U400 cyclotron was used to bombard an AmO$_2$ target enriched in $^{243}$Am with 253~MeV and 248~MeV $^{48}$Ca beams to form $^{287}$115 and $^{288}$115 in (4n) and (3n) fusion evaporation reactions, respectively. $^{275}$Mt and $^{276}$Mt were populated by subsequent $\alpha$-decays. The residues were separated with a gas-filled recoil separator and implanted in a semiconductor detector array. Alpha particle decay and spontaneous fission events were recorded in this array and in eight detectors arranged in a box configuration around the implantation detector. ``The $\alpha$-decay energies attributed to the isotopes of Mt and Bh coincide well with theoretical values.'' One decay chain involving $^{275}$Mt and three chains involving $^{276}$Mt were observed.

\subsection*{$^{278}$Mt}\vspace{0.0cm}
In the 2010 paper ``Synthesis of a new element with atomic number Z = 117'', Oganessian et al.\ reported the first observation of $^{278}$Mt \cite{2010Oga01}. A $^{249}$Bk target was bombarded with a 252~MeV $^{48}$Ca beam from the Dubna U400 cyclotron to form $^{294}$117 in the (3n) evaporation reaction. $^{278}$Mt was populated by subsequent $\alpha$-decay. The residues were separated with a gas-filled recoil separator and implanted in a semiconductor detector array. Alpha particle decay and spontaneous fission events were recorded in this array and in eight detectors arranged in a box configuration around the implantation detector. $^{278}$Mt is not specifically mentioned in the text but an $\alpha$-energy of 9.00(10)~MeV with a lifetime of 11.0~s is quoted in the figure displaying the single observed decay chain.


\section{Discovery of $^{267-281}$Ds}

Darmstadtium was discovered by Hofmann et al.\ in 1995 with the first observation of $^{269}$Ds reported in a paper submitted on November 14, 1994 \cite{1995Hof01}. Only eight days later, on November 22, 1994 Ghiorso et al.\ submitted their ``possible synthesis of element 110'' describing the observation of a single event of $^{267}$Ds \cite{1995Ghi01}. The discovery of darmstadtium was officially accepted by the IUPAC/IUPAP Joint Working Party (JWP) in 2001: ``In accordance with the criteria for the discovery of elements, previously established by the 1992 IUPAC/IUPAP Transfermium Working Group, it was determined that the claim by the Hofmann et al.\ research collaboration for the discovery of element 110 at GSI has fulfilled those criteria'' \cite{2001Kar01}. The name darmstadtium was officially accepted in 2003 \cite{2003Cor01}. Eight darmstadtium isotopes have been reported so far.

\subsection*{$^{267}$Ds}\vspace{0.0cm}
$^{267}$Ds was first reported by Ghiorso et al.\ in ``Evidence for the possible synthesis of element 110 produced by the $^{59}$Co+$^{209}$Bi reaction'' in 1995 \cite{1995Ghi01}. A 5.1 MeV/nucleon $^{59}$Co beam from the Berkeley SuperHILAC accelerator bombarded a bismuth target and $^{267}$Ds was formed in the (1n) fusion-evaporation reaction. Recoil products were separated with the gas-filled magnetic spectrometer SASSY2. The recoils and subsequent $\alpha$ decays were recorded in five position sensitive silicon wafers. ``One event with many of the expected characteristics of a successful synthesis of $^{267}$110 was observed.'' The same results were also published in a conference proceeding in the same year \cite{1995Ghi02}.

\subsection*{$^{269}$Ds}\vspace{0.0cm}
Hofmann et al.\ discovered $^{269}$Ds in 1995 as reported in ``Production and decay of $^{269}$110'' \cite{1995Hof01}. An enriched $^{208}$Pb target was bombarded with a 311 MeV $^{62}$Ni beam from the GSI UNILAC. $^{269}$Ds was formed in the single neutron evaporation reaction and separated with the velocity filter SHIP. A detector system consisting of two time-of-flight detectors, seven 16-strip silicon wafers, and three germanium detectors measured the heavy-ion, $\alpha$-, X- and $\gamma$-rays. ``We therefore, assign the observed decay chain to the $\alpha$-decay of $^{269}$110. The half-life is (270$^{+1300}_{-120})~\mu$s.'' In a note added in proof it was mentioned that three additional chains had been observed. The observation of one of the decay chains was later retracted \cite{2002Hof01}.

\subsection*{$^{270}$Ds}\vspace{0.0cm}
The first observation of $^{270}$Ds was reported in 2001 in ``The new isotope $^{270}$110 and its decay products $^{266}$Hs and $^{262}$Sg'' by Hofmann et al.\ \cite{2001Hof01}. A 317 MeV $^{64}$Ni beam accelerated by the GSI UNILAC bombarded an enriched $^{207}$Pb target. $^{270}$Ds was produced in the (1n) fusion evaporation reaction and identified with a detector system at the velocity filter SHIP. ``The ground state of $^{270}$110 decays by $\alpha$ emission with an energy of (11.03$\pm$0.05)~MeV and a half life of (100$^{+140}_{-40}$)~$\mu$s.''

\subsection*{$^{271}$Ds}\vspace{0.0cm}
The observation of $^{271}$Ds was first reported in a review article by Hofmann in 1998: ``New elements approaching Z = 114'' \cite{1998Hof01}. An enriched $^{208}$Pb target was bombarded with 311.7, 313, and 315.5~MeV $^{64}$Ni beams from the GSI UNILAC. $^{271}$Ds was produced in the (1n) fusion evaporation reaction and identified with a detector system at the velocity filter SHIP. ``The measured $\alpha$-decays of $^{271}$110 can be subdivided into three groups; five events decay with an average energy of 10.738~MeV, two with 10.682~MeV, plus one escape event decay with the same lifetime $\tau$ = (1.6$^{+0.9}_{-0.4}$)~ms or a half-life T$_{1/2}$ = (1.1$^{+0.6}_{-0.3}$)~ms.''

\subsection*{$^{273}$Ds}\vspace{0.0cm}
In the 1996 paper ``$\alpha$ decay of $^{273}$110: Shell closure at N = 162'', Lazarev et al.\ reported the discovery of $^{273}$Ds \cite{1996Laz01}. A 190 MeV $^{34}$S beam from the Dubna U400 cyclotron bombarded enriched $^{244}$Pu targets. $^{273}$Ds was formed in the (5n) fusion-evaporation reaction. Reaction residues were separated with the Dubna Gas-filled Recoil Separator and subsequent $\alpha$ and spontaneous fission decays were recorded in a position sensitive silicon detector. ``As a result of the above-described selection, 14 candidate chains of the $^{273}$110 type were observed in detector strips 1–6, and one four-member sequence, with E$_{\alpha 1}$ = 11.35~MeV, was detected in strip 7''. A half-life of 0.3$^{+1.3}_{-0.2}$~ms was quoted. About a month later Hofmann et al.\ independently reported the observation of $^{273}$Ds in the $\alpha$-decay of $^{277}$Cn \cite{1996Hof02}.

\subsection*{$^{277}$Ds}\vspace{0.0cm}
Ellison et al.\ described the discovery of $^{277}$Ds in 2010 in ``New superheavy element isotopes; $^{242}$Pu($^{48}$Ca, 5n)$^{285}$114'' \cite{2010Eli01}. $^{242}$PuO$_2$ targets were bombarded with a 247~MeV $^{48}$Ca beams from the Berkeley 88-in. cyclotron and $^{285}$114 was produced in (5n) fusion-evaporation reactions. $^{277}$Ds was populated by subsequent $\alpha$ decay. Residues were separated with the Berkeley Gas-Filled Separator BGS and detected in multiwire proportional counters and silicon strip detectors. Subsequent radioactive decay events were recorded in the strip detectors and additional silicon chips forming a five-sided box. ``The chain continued with four subsequent $\alpha$-like events... after 140 ms, 8.21 ms, 346 ms, and 185 s with energies of 10.31, 10.57, 9.59, and 8.57 MeV, which are interpreted as the successive $\alpha$ decays of $^{281}_{112}$Cn, $^{277}_{110}$Ds, $^{273}_{108}$Hs, and $^{269}_{106}$Sg, respectively.'' A single decay chain was observed. A previously reported observation of $^{277}$Ds \cite{1999Nin01} was later retracted \cite{2002Nin01}.

\subsection*{$^{279}$Ds}\vspace{0.0cm}
$^{279}$Ds was first identified by Oganessian et al.\ in ``Measurements of cross sections for the fusion-evaporation reactions $^{244}$Pu($^{48}$Ca,xn)$^{292-x}$114 and $^{245}$Cm($^{48}$Ca,xn)$^{293-x}$116'' in 2004 \cite{2004Oga03}. $^{48}$Ca beams of 243 and 257~MeV from the Dubna U400 cyclotron bombarded a PuO$_2$ target enriched $^{244}$Pu and a CmO$_2$ target enriched in $^{245}$Cm. $^{279}$Ds was populated by $\alpha$ decays from $^{291}$116 and $^{287}$114 which were formed in (2n) and (5n) evaporation reactions on the CmO$_2$ and PuO$_2$ targets, respectively. The residues were separated with a gas-filled recoil separator and implanted in a semiconductor detector array. Subsequent $\alpha$ particle decay and spontaneous fission events were recorded in this array and in eight detectors arranged in a box configuration around the implantation detector. The observation of $^{279}$Ds was not specifically mentioned in the text but a table listed the spontaneous fission half-life to be 0.29$^{+0.35}_{-0.10}$~s. Three decay chains ending in $^{279}$Ds were reported.

\subsection*{$^{281}$Ds}\vspace{0.0cm}
$^{281}$Ds was first identified by Oganessian et al.\ in ``Measurements of cross sections for the fusion-evaporation reactions $^{244}$Pu($^{48}$Ca,xn)$^{292-x}$114 and $^{245}$Cm($^{48}$Ca,xn)$^{293-x}$116'' in 2004 \cite{2004Oga03}. $^{48}$Ca beams of 243 and 250~MeV from the Dubna U400 cyclotron bombarded a PuO$_2$ target enriched $^{244}$Pu. $^{281}$Ds was populated by $\alpha$ decays from $^{289}$114 which was formed in the (3n) evaporation reaction. The residues were separated with a gas-filled recoil separator and implanted in a semiconductor detector array. Subsequent $\alpha$ particle decay and spontaneous fission events were recorded in this array and in eight detectors arranged in a box configuration around the implantation detector. The observation of $^{281}$Ds was not specifically mentioned in the text but a table listed the spontaneous fission half-life to be 9.6$^{+5.0}_{-2.5}$~s. Eight decay chains ending in $^{281}$Ds were reported.


\section{Discovery of $^{272-282}$Rg}
Hofmann et al.\ discovered the first isotope of roentgenium ($^{272}$Rg) in 1995 \cite{1995Hof02}. This discovery was officially accepted by the IUPAC/IUPAP Joint Working Party (JWP) in 2003: ``In concordance with the criteria established for validating claims, the JWP has agreed that the priority of the Hofmann et al.\ collaboration's discovery of element 111 at GSI is acknowledged'' \cite{2003Kar01}. The name roentgenium was officially accepted in 2003 \cite{2004Cor01}. Seven roentgenium isotopes have been reported.

\subsection*{$^{272}$Rg}\vspace{0.0cm}
Hofmann et al.\ discovered $^{272}$Rg in 1995 as reported in ``The new element 111'' \cite{1995Hof02}. Bismuth targets were bombarded with 318 and 320 MeV $^{64}$Ni beams from the GSI UNILAC and $^{272}$Rg was formed in the (1n) fusion-evaporation reaction. Reaction residues were separated with the velocity filter SHIP and subsequent $\alpha$ decays were recorded in a position sensitive silicon detector. ``We assign the three measured decay chains to the previously unknown isotope $^{272}$111. This nucleus is the first one observed of the new element Z = 111.'' A half-life of 1.5$^{+2.0}_{-0.5}$~ms was reported.

\subsection*{$^{274}$Rg}\vspace{0.0cm}
The first identification of $^{274}$Rg was reported by Morita et al.\ in ``Experiment on the synthesis of element 113 in the reaction $^{209}$Bi($^{70}$Zn,n)$^{278}$113'' in 2004 \cite{2004Mor01}. Bismuth targets were bombarded with a 352.6~MeV $^{70}$Zn beam from the RIKEN linear accelerator facility RILAC and $^{274}$Rg was populated by $\alpha$-decay from $^{278}$113. Recoil products were separated with the gas-filled recoil ion separator GARIS and detected with micro-channel plates and a silicon strip detector. Spontaneous fission and $\alpha$-decay events were recorded with a silicon semiconductor detector box consisting of the central detector plus four additional silicon strip detectors forming a box. ``In conclusion, the reaction product, followed by the decay chain observed in our experiment, was considered to be most probably due to the $^{209}$Bi($^{70}$Zn,n)$^{278}$113 reaction. As a result, the members of the decay chain were consequently assigned as $^{278}$113, $^{274}$111, $^{270}$Mt, $^{266}$Bh, and $^{262}$Db.'' A single decay chain was observed.

\subsection*{$^{278}$Rg}\vspace{0.0cm}
Oganessian et al.\ reported the observation of $^{278}$Rg in 2007 in ``Synthesis of the isotope $^{282}$113 in the $^{237}$Np+$^{48}$Ca fusion reaction'' \cite{2007Oga01}. A 244 MeV $^{48}$Ca beam from the Dubna U400 cyclotron bombarded a $^{237}$Np target and $^{282}$113 was populated in the (3n) fusion evaporation reaction. $^{278}$Rg was populated by subsequent $\alpha$ decay. The residues were separated with a gas-filled recoil separator and implanted in a semiconductor detector array. Alpha particle decay and spontaneous fission events were recorded in this array and in eight detectors arranged in a box configuration around the implantation detector. ``The $\alpha$-decay energies attributed to the isotopes $^{282}$113 and $^{278}$Rg agree well with expected values resulting from the trend of the Q$_\alpha$(N) systematics measured for the neighboring isotopes $^{278,283,284}$113 and $^{274,279,280}$Rg.'' Two decay chains were observed.

\subsection*{$^{279,280}$Rg}\vspace{0.0cm}
$^{279}$Rg and $^{280}$Rg were first observed by Oganessian et al.\ in 2004 as reported in ``Experiments on the synthesis of element 115 in the reaction $^{243}$Am($^{48}$Ca,xn)$^{291-x}$115'' \cite{2004Oga01}. The Dubna U400 cyclotron was used to bombard an AmO$_2$ target enriched in $^{243}$Am with 253~MeV and 248~MeV $^{48}$Ca beams to form $^{287}$115 and $^{288}$115 in (4n) and (3n) fusion evaporation reactions, respectively. $^{279}$Rg and $^{280}$Rg were populated by subsequent $\alpha$-decays. The residues were separated with a gas-filled recoil separator and implanted in a semiconductor detector array. Alpha particle decay and spontaneous fission events were recorded in this array and in eight detectors arranged in a box configuration around the implantation detector. ``The $\alpha$-decay energies attributed to the isotopes of Mt and Bh coincide well with theoretical values. For the isotopes $^{279,280}$111 and $^{283,284}$113 the difference between theoretical and experimental Q$_\alpha$ values amounts to 0.6-0.9 MeV.'' One decay chain involving $^{279}$Rg and three chains involving $^{280}$Rg were observed.

\subsection*{$^{281,282}$Rg}\vspace{0.0cm}
In the 2010 paper ``Synthesis of a new element with atomic number Z = 117'', Oganessian et al.\ reported the first observation of $^{281}$Rg and $^{282}$Rg \cite{2010Oga01}. A $^{249}$Bk target was bombarded with 252~MeV and 247~MeV $^{48}$Ca beam from the Dubna U400 cyclotron to form $^{293}$117 and $^{294}$117 in (4n) and (3n) evaporation reactions, respectively. $^{281}$Rg and $^{282}$Rg were populated by subsequent $\alpha$-decays. The residues were separated with a gas-filled recoil separator and implanted in a semiconductor detector array.  Alpha particle decay and spontaneous fission events were recorded in this array and in eight detectors arranged in a box configuration around the implantation detector. ``Despite the strong hindrance resulting in the relatively long half-life, SF is a principal decay mode of the odd-even nucleus $^{281}$111. On the other hand, the heavier isotope $^{282}$111 undergoes $\alpha$ decay.'' Five decay chains involving $^{281}$Rg and one chain involving $^{282}$Rg were observed.


\section{Discovery of $^{277-285}$Cn}
Copernicium was discovered in 1996 with the identification of $^{277}$Cn by Hofmann et al.\ \cite{1996Hof02}. The IUPAC/IUPAP Joint Working Party (JWP) official accepted this discovery in 2009: ``In concordance with the criteria established for validating claims, the JWP has agreed that the priority of the Hofmann et al. 1996 \cite{1996Hof02} and 2002 \cite{2002Hof01} collaborations' discovery of the element with atomic number 112 at GSI is acknowledged'' \cite{2009Bar01}. Claims for the observation of element 112 in tungsten targets bombarded with 24 GeV protons at CERN \cite{1984Mar01} were not credible \cite{2001Kar01}. The name copernicium was officially accepted in 2010 \cite{2010Tat01}. Six copernicium isotopes have been reported so far.

\subsection*{$^{277}$Cn}\vspace{0.0cm}
In the 1996 paper ``The new element 112'', Hofmann et al.\ reported the discovery of $^{277}$Cn \cite{1996Hof02}. A 344 MeV $^{70}$Zn beam from the GSI UNILAC bombarded enriched $^{208}$Pb targets and $^{277}$Cn was populated in the single neutron fusion-evaporation reaction. Reaction residues were separated with the velocity filter SHIP and subsequent $\alpha$ decays were recorded in a position sensitive silicon detector. ``Therefore, the observed chain must be assigned to the isotope with mass number A = 277 of element Z = 112, produced by fusion of $^{70}$Zn and $^{208}$Pb and emission of one neutron. This chain represents the first unambiguous identification of the new element Z = 112.'' Two chains were observed, however, the first chain was later retracted \cite{2002Hof01}.

\subsection*{$^{281}$Cn}\vspace{0.0cm}
Ellison et al.\ described the discovery of $^{281}$Cn in 2010 in ``New superheavy element isotopes; $^{242}$Pu($^{48}$Ca, 5n)$^{285}$114'' \cite{2010Eli01}. $^{242}$PuO$_2$ targets were bombarded with a 247~MeV $^{48}$Ca beams from the Berkeley 88-in. cyclotron and $^{285}$114 was produced in (5n) fusion-evaporation reactions. $^{281}$Cn was populated by subsequent $\alpha$ decay. Residues were separated with the Berkeley Gas-Filled Separator BGS and detected in multiwire proportional counters and silicon strip detectors. Subsequent radioactive decay events were recorded in the strip detectors and additional silicon chips forming a five-sided box. ``The chain continued with four subsequent $\alpha$-like events... after 140 ms, 8.21 ms, 346 ms, and 185 s with energies of 10.31, 10.57, 9.59, and 8.57 MeV, which are interpreted as the successive $\alpha$ decays of $^{281}_{112}$Cn, $^{277}_{110}$Ds, $^{273}_{108}$Hs, and $^{269}_{106}$Sg, respectively.'' A single decay chain was observed. A previously reported observation of $^{281}$Cn \cite{1999Nin01} was later retracted \cite{2002Nin01}.

\subsection*{$^{282-285}$Cn}\vspace{0.0cm}
$^{282}$Cn, $^{283}$Cn, $^{284}$Cn, and $^{285}$Cn were first identified by Oganessian et al.\ in ``Measurements of cross sections for the fusion-evaporation reactions $^{244}$Pu($^{48}$Ca,xn)$^{292-x}$114 and $^{245}$Cm($^{48}$Ca,xn)$^{293-x}$116'' in 2004 \cite{2004Oga03}. $^{48}$Ca beams of 243, 250, and 257~MeV from the Dubna U400 cyclotron bombarded a PuO$_2$ target enriched in $^{244}$Pu and a CmO$_2$ target enriched in $^{245}$Cm. $^{282}$Cn and $^{283}$Cn were populated by $\alpha$ decays from $^{290}$116 and $^{291}$116 which were formed of the 243 MeV beam on the CmO$_2$ target in (3n) and (2n) evaporation reaction, respectively. $^{283}$Cn was also populated by $\alpha$ decay following the (5n) reaction forming $^{287}$114 on the PuO$_2$ target at 257~MeV.  $^{284}$Cn and $^{285}$Cn were populated by $\alpha$ decay following (4n) and (3n) reactions forming $^{289}$114 and $^{290}$114, respectively, on the PuO$_2$ target. $^{284}$Cn was observed at 243, 250, and 257~MeV, and $^{285}$Cn at 243 and 250~MeV. The residues were separated with a gas-filled recoil separator and implanted in a semiconductor detector array. Subsequent $\alpha$ particle decay and spontaneous fission events were recorded in this array and in eight detectors arranged in a box configuration around the implantation detector. Only $^{284}$Cn is specifically mentioned in the text: ``The isotope $^{284}$112 decays via SF with a half-life of $\sim$98~ms,... ''. The decay properties are listed in a table. One decay chain ended with $^{282}$Cn decaying by spontaneous fission with a half-life of 1.0$^{+4.8}_{-0.5}$~ms. Three $\alpha$ decays of $^{283}$Cn with a half-life of 6.1$^{+7.2}_{-2.2}$~s were observed. A spontaneous fission half-life of 98$^{+41}_{-23}$~s was extracted from eleven decay chains for $^{284}$Cn, and eight $\alpha$ decays were recorded for $^{285}$Cn with a half-life of 34$^{+17}_{-9}$~s. Based on these results the previous assignment for the observation of $^{284}$Cn \cite{2000Oga01,2000Oga02} had to be changed to $^{285}$Cn. Earlier reports of $^{283}$Cn \cite{1999Oga01,1999Oga03,2004Oga05} and $^{285}$Cn \cite{1999Oga01,1999Oga02} could not be confirmed. A specific search for $^{283}$Cn in 2002 by Loveland et al.\ did not result in any events \cite{2002Lov01}. Finally, the 2004 results for $^{283}$Cn could not be reproduced by Gregorich et al.\ \cite{2005Gre01} but were confirmed by Hofmann et al.\ \cite{2007Hof01}. A comprehensive review of the discovery of these isotopes is presented in reference  \cite{2007Oga02}.


\section{Discovery of $^{278-286}$113}

The discovery of element 113 has not yet been accepted by the IUPAC/IUPAP Joint Working Party: ``The results are encouraging but do not meet the criteria for discovery because of the paucity of events, the lack of connections to known nuclides, and the absence of cross-bombardments.'' \cite{2011Bar01}. Six isotopes of element 113 have been observed.

\subsection*{$^{278}$113}\vspace{0.0cm}
The first identification of $^{278}$113 was reported by Morita et al.\ in ``Experiment on the synthesis of element 113 in the reaction $^{209}$Bi($^{70}$Zn,n)$^{278}$113'' in 2004 \cite{2004Mor01}. Bismuth targets were bombarded with a 352.6~MeV $^{70}$Zn beam from the RIKEN linear accelerator facility RILAC. Recoil products were separated with the gas-filled recoil ion separator GARIS and detected with micro-channel plates and a silicon strip detector. Spontaneous fission and $\alpha$-decay events were recorded with a silicon semiconductor detector box consisting of the central detector plus four additional silicon strip detectors forming a box. ``In conclusion, the reaction product, followed by the decay chain observed in our experiment, was considered to be most probably due to the $^{209}$Bi($^{70}$Zn,n)$^{278}$113 reaction. As a result, the members of the decay chain were consequently assigned as $^{278}$113, $^{274}$111, $^{270}$Mt, $^{266}$Bh, and $^{262}$Db.'' A single decay chain was observed.

\subsection*{$^{282}$113}\vspace{0.0cm}
Oganessian et al.\ reported the observation of $^{282}$113 in 2007 in ``Synthesis of the isotope $^{282}$113 in the $^{237}$Np+$^{48}$Ca fusion reaction'' \cite{2007Oga01}. A 244 MeV $^{48}$Ca beam from the Dubna U400 cyclotron bombarded a $^{237}$Np target and $^{282}$113 was populated in the (3n) fusion evaporation reaction. The residues were separated with a gas-filled recoil separator and implanted in a semiconductor detector array. Subsequent $\alpha$ particle decay and spontaneous fission events were recorded in this array and in eight detectors arranged in a box configuration around the implantation detector. Two decay chains were observed: ``Based on the similar $\alpha$-particle energies and decay times of the first three $\alpha$ transitions, we assign both decay chains to the same parent nucleus, namely $^{282}$113 produced in the $^{237}$Np($^{48}$Ca,3n) reaction.''

\subsection*{$^{283,284}$113}\vspace{0.0cm}
$^{283}$113 and $^{284}$113 were first observed by Oganessian et al.\ in 2004 as reported in ``Experiments on the synthesis of element 115 in the reaction $^{243}$Am($^{48}$Ca,xn)$^{291-x}$115'' \cite{2004Oga01}. The Dubna U400 cyclotron was used to bombard an AmO$_2$ target enriched in $^{243}$Am with 253~MeV and 248~MeV $^{48}$Ca beams to form $^{287}$115 and $^{288}$115 in (4n) and (3n) fusion evaporation reactions, respectively. $^{283}$113 and $^{284}$113 were populated by subsequent $\alpha$-decay. The residues were separated with a gas-filled recoil separator and implanted in a semiconductor detector array. Alpha particle decay and spontaneous fission events were recorded in this array and in eight detectors arranged in a box configuration around the implantation detector. ``The $\alpha$-decay energies attributed to the isotopes of Mt and Bh coincide well with theoretical values. For the isotopes $^{279,280}$111 and $^{283,284}$113 the difference between theoretical and experimental Q$_\alpha$ values amounts to 0.6-0.9 MeV.'' One decay chain involving $^{283}$113 and three chains involving $^{284}$113 were observed.

\subsection*{$^{285,286}$113}\vspace{0.0cm}
In the 2010 paper ``Synthesis of a new element with atomic number Z = 117'', Oganessian et al.\ reported the first observation of $^{285}$113 and $^{286}$113 \cite{2010Oga01}. A $^{249}$Bk target was bombarded with 252~MeV and 247~MeV $^{48}$Ca beam from the Dubna U400 cyclotron to form $^{293}$117 and $^{294}$117 in (4n) and (3n) evaporation reactions, respectively. $^{285}$113 and $^{286}$113 were populated by subsequent $\alpha$-decays. The residues were separated with a gas-filled recoil separator and implanted in a semiconductor detector array. Alpha particle decay and spontaneous fission events were recorded in this array and in eight detectors arranged in a box configuration around the implantation detector. ``The decay properties of the neighboring isotopes $^{293}$117 and $^{294}$117, their daughters $^{289}$115 and $^{290}$115, as well as granddaughters $^{285}$113 and $^{286}$113, do not display substantial differences.'' Five decay chains involving $^{285}$113 and one chain involving $^{286}$113 were observed.


\section{Discovery of $^{285-289}$114}
Isotopes of element 114 with mass numbers 286$-$289 were reported by Oganessian in 2004 \cite{2004Oga03}. Independent confirmation of the formation of element 114 were recently reported with the observation of $^{285}$114 in Berkeley \cite{2010Eli01} and $^{288,289}$116 at GSI \cite{2010Dul01,2011Gat01}. The discovery of element 114 was officially accepted by the IUPAC/IUPAP Joint Working Party in 2011: ``For the elements Z = 114 and 116, the establishment of the identity of the isotope $^{283}$Cn by a large number of decaying chains, originating from a variety of production pathways essentially triangulating its A,Z character enables that nuclide's use in unequivocally recognizing higher-Z isotopes that are observed to decay through it. The JWP notes that the internal redundancy and extended decay chain sequence for identification of Z = $^{287}$114 from $^{48}$Ca + $^{242}$Pu fusion by the 2004 Dubna-Livermore collaborations \cite{2004Oga05,2004Oga02} and recommends that the Dubna-Livermore collaboration be credited with discovery of this new element.'' \cite{2011Bar01}.  Five isotopes of element 114 have been reported.

\subsection*{$^{285}$114}\vspace{0.0cm}
Ellison et al.\ described the discovery of $^{285}$114 in 2010 in ``New superheavy element isotopes; $^{242}$Pu($^{48}$Ca, 5n)$^{285}$114'' \cite{2010Eli01}. $^{242}$PuO$_2$ targets were bombarded with a 247~MeV $^{48}$Ca beams from the Berkeley 88-in. cyclotron and $^{285}$114 was produced in (5n) fusion-evaporation reactions. Residues were separated with the Berkeley Gas-Filled Separator BGS and detected in multiwire proportional counters and silicon strip detectors. Subsequent radioactive decay events were recorded in the strip detectors and additional silicon chips forming a five-sided box. ``Element-114 atoms were identified by detecting time- and position-correlated events corresponding to their implantation and subsequent radioactive decay chain, terminating with the detection of a SF event. [The table] contains the times, energies, and positions of the two correlated decay chains observed in the experiment. Based on a comparison with predicted decay properties, the first event was assigned to the decay of $^{285}$114 and its daughters.'' A single decay chain was observed. A previously reported observation of $^{285}$114 \cite{1999Nin01} was later retracted \cite{2002Nin01}.

\subsection*{$^{286-289}$114}\vspace{0.0cm}
$^{286}$114, $^{287}$114, $^{288}$114, and $^{289}$114 were first identified by Oganessian et al.\ in ``Measurements of cross sections for the fusion-evaporation reactions $^{244}$Pu($^{48}$Ca,xn)$^{292-x}$114 and $^{245}$Cm($^{48}$Ca,xn)$^{293-x}$116'' in 2004 \cite{2004Oga03}. $^{48}$Ca beams of 243, 250, and 257~MeV from the Dubna U400 cyclotron bombarded a PuO$_2$ target enriched in $^{244}$Pu and a CmO$_2$ target enriched in $^{245}$Cm. $^{286}$114 and $^{287}$114 were populated by $\alpha$-decay from $^{290}$116 and $^{291}$116 which were formed in (3n) and (2n) evaporation reaction of the 243 MeV beam on the CmO$_2$ target, respectively. $^{287}$114 was also formed in the (5n) reaction on the PuO$_2$ target at 257~MeV.  $^{288}$114 and $^{289}$114 were produced in (4n) and (3n) reactions, respectively, on the PuO$_2$ target. The residues were separated with a gas-filled recoil separator and implanted in a semiconductor detector array. Subsequent $\alpha$ particle decay and spontaneous fission events were recorded in this array and in eight detectors arranged in a box configuration around the implantation detector. ``Then, the shorter chains should be assigned to the decay of even-even $^{290}$116, the product of 3n-evaporation... For the daughter nucleus $^{286}$114, in one decay chain we observed a decay and SF was registered in two other cases... The properties of the daughter nucleus of one of the element 116 isotopes produced in the $^{245}$Cm+$^{48}$Ca reaction (two events) essentially reproduces the characteristics of the 10-s (ER$-\alpha-\alpha$-SF) chain arising in the $^{244}$Pu+$^{48}$Ca reaction that we assign to the decay of isotope $^{287}$114... At E$^*$=41 MeV, 47 and 53 MeV, we observed 12 events of the decay of a new nuclide that undergoes sequential ER$-\alpha-$SF decay over the span of about 1 second. The maximum yield of this nuclide, $^{288}$114, corresponds to E$^*<$43 MeV and a peak production cross section of 5.3$^{+3.6}_{-2.1}$~pb... At E$^*$=41 MeV and 47 MeV, three chains of sequential ER$-\alpha-\alpha-$SF decays were observed. These chains are identical to those detected in previous $^{244}$Pu+$^{48}$Ca experiments at 236 MeV (E$^*$=35 MeV) and to those produced as decay products of the Z=116 nucleus observed in the $^{248}$Cm+$^{48}$Ca reaction. The maximum yield of this nuclide, $^{289}$114, is observed at E$^*$=41 MeV with a peak production cross section of 1.7$^{+2.5}_{-1.1}$~pb.'' Based on these results the previous assignment for the observation of $^{288}$114 \cite{2000Oga01,2000Oga02} was changed to $^{289}$114. Earlier reports of $^{287}$114 \cite{1999Oga01} and $^{289}$114 \cite{1999Oga01,1999Oga02} could not be confirmed. A comprehensive review of the discovery of these isotopes is presented in reference  \cite{2007Oga02}.


\section{Discovery of $^{287-290}$115}

The discovery of element 115 has not yet been accepted by the IUPAC/IUPAP Joint Working Party. Although 23 events were assigned to $^{288}$115 as described below they are not connected to any known nuclei and the chemical analysis cannot ``distinguish the properties of Groups 4 and 5 elements in this region with confidence'' \cite{2011Bar01}. Four isotopes of element 115 have been observed.

\subsection*{$^{287,288}$115}\vspace{0.0cm}
$^{287}$115 and $^{288}$115 were first observed by Oganessian et al.\ in 2004 as reported in ``Experiments on the synthesis of element 115 in the reaction $^{243}$Am($^{48}$Ca,xn)$^{291-x}$115'' \cite{2004Oga01}. The Dubna U400 cyclotron was used to bombard an AmO$_2$ target enriched in $^{243}$Am with 253~MeV and 248~MeV $^{48}$Ca beams to form $^{287}$115 and $^{288}$115 in (4n) and (3n) fusion evaporation reactions, respectively. The residues were separated with a gas-filled recoil separator and implanted in a semiconductor detector array. Subsequent $\alpha$ particle decay and spontaneous fission events were recorded in this array and in eight detectors arranged in a box configuration around the implantation detector. ``The decay properties of these synthesized nuclei are consistent with consecutive $\alpha$ decays originating from the parent isotopes of the new element 115, $^{288}$115 and $^{287}$115, produced in the 3n- and 4n-evaporation channels with cross sections of about 3 pb and 1 pb, respectively.'' One decay chain for $^{287}$115 and three chains for $^{288}$115 were observed.

\subsection*{$^{289,290}$115}\vspace{0.0cm}
In the 2010 paper ``Synthesis of a new element with atomic number Z = 117'', Oganessian et al.\ reported the first observation of $^{289}$115 and $^{290}$115 \cite{2010Oga01}. A $^{249}$Bk target was bombarded with 252~MeV and 247~MeV $^{48}$Ca beam from the Dubna U400 cyclotron to form $^{293}$117 and $^{294}$117 in (4n) and (3n) evaporation reactions, respectively. $^{289}$115 and $^{290}$115 were populated by subsequent $\alpha$-decay. The residues were separated with a gas-filled recoil separator and implanted in a semiconductor detector array. Alpha particle decay and spontaneous fission events were recorded in this array and in eight detectors arranged in a box configuration around the implantation detector. ``The decay properties of the neighboring isotopes $^{293}$117 and $^{294}$117, their daughters $^{289}$115 and $^{290}$115, as well as granddaughters $^{285}$113 and $^{286}$113, do not display substantial differences.'' Five decay chains involving $^{289}$115 and one chain involving $^{290}$115 were observed.


\section{Discovery of $^{290-293}$116}

The discovery of element 116 was officially accepted by the IUPAC/IUPAP Joint Working Party in 2011: ``For the elements Z = 114 and 116, the establishment of the identity of the isotope $^{283}$Cn by a large number of decaying chains, originating from a variety of production pathways essentially triangulating its A,Z character enables that nuclide's use in unequivocally recognizing higher-Z isotopes that are observed to decay through it... The Dubna-Livermore collaboration \cite{2004Oga03} should be credited with the discovery of the new element with Z = 116.'' \cite{2011Bar01}. So far, four isotopes of element 116 have been reported. The observation of $^{289}$116 \cite{1999Nin01} was later retracted \cite{2002Nin01}.

\subsection*{$^{290,291}$116}\vspace{0.0cm}
$^{290}$116 and $^{291}$116 were first identified by Oganessian et al.\ in ``Measurements of cross sections for the fusion-evaporation reactions $^{244}$Pu($^{48}$Ca,xn)$^{292-x}$114 and $^{245}$Cm($^{48}$Ca,xn)$^{293-x}$116'' in 2004 \cite{2004Oga03}. A 243~MeV $^{48}$Ca beam from the Dubna U400 cyclotron bombarded a CmO$_2$ target enriched in $^{245}$Cm. $^{290}$116 and $^{291}$116 populated in (3n) and (2n) fusion-evaporation reactions, respectively. The residues were separated with a gas-filled recoil separator and implanted in a semiconductor detector array. Subsequent $\alpha$ particle decay and spontaneous fission events were recorded in this array and in eight detectors arranged in a box configuration around the implantation detector. ``As a result, the longer ER-$\alpha-\alpha-\alpha$-SF chains observed in the $^{245}$Cm+$^{48}$Ca reaction must arise from the decay of $^{291}$116 produced via the 2n-evaporation channel. Then, the shorter chains should be assigned to the decay of even-even $^{290}$116, the product of 3n-evaporation.'' Two decay chains for each of the isotopes were observed.

\subsection*{$^{292}$116}\vspace{0.0cm}
In the 2004 paper ``Measurements of cross sections and decay properties of the isotopes of elements 112, 114, and 116 produced in the fusion reactions $^{233,238}$U, $^{242}$Pu, and $^{248}$Cm+$^{48}$Ca'', Oganessian et al.\ identified $^{292}$116 \cite{2004Oga02}. A $^{248}$Cm target was bombarded with a 247~MeV $^{48}$Ca beam from the Dubna U400 cyclotron and $^{292}$116 was produced in the (3n) fusion evaporation reaction. The residues were separated with a gas-filled recoil separator and implanted in a semiconductor detector array. Subsequent $\alpha$ particle decay and spontaneous fission events were recorded in this array and in eight detectors arranged in a box configuration around the implantation detector. ``We observed the new nuclide $^{292}$116 (T$_\alpha$ = 18$^{+16}_{-6}$~ms, E$_\alpha$ = 10.66$\pm$0.07~MeV) in the irradiation of the $^{248}$Cm target at a higher energy than in previous experiments.'' 6 decay chains were observed. Previous assignments of  $^{292}$116 \cite{2001Oga01,2001Oga02,2002Oga02} were reassigned to  $^{293}$116 \cite{2004Oga03}.

\subsection*{$^{293}$116}\vspace{0.0cm}
$^{293}$116 was identified by Oganessian et al.\ in ``Measurements of cross sections for the fusion-evaporation reactions $^{244}$Pu($^{48}$Ca,xn)$^{292-x}$114 and $^{245}$Cm($^{48}$Ca,xn)$^{293-x}$116'' in 2004 \cite{2004Oga03}. A 257~MeV $^{48}$Ca beam from the Dubna U400 cyclotron bombarded a PuO$_2$ target enriched in $^{244}$Pu. The residues were separated with a gas-filled recoil separator and implanted in a semiconductor detector array. Subsequent $\alpha$ particle decay and spontaneous fission events were recorded in this array and in eight detectors arranged in a box configuration around the implantation detector. The earlier assignment of the isotope $^{292}$116 was changed to $^{293}$116 based on the reassignment of three Z = 114 decay chains from A = 288 to A = 289: ``Note, in this interpretation of the data, the previously observed decay of the parent nuclei discovered in the reactions $^{244}$Pu+$^{48}$Ca and $^{248}$Cm+$^{48}$Ca originated from the isotopes $^{289}$114 and $^{293}$116.'' This reassignment affected one decay chain published in \cite{2001Oga01,2001Oga02} and two additional decay chains in reference \cite{2002Oga02}. The latter two chains were also mentioned in a note added in proof in reference \cite{2001Oga02}.


\section{Discovery of $^{293-294}$117}

The element 117 has not been considered to be accepted as a new element by the IUPAC/IUPAP Joint working party. So far only two isotopes of element 117 have been reported.

\subsection*{$^{293,294}$117}\vspace{0.0cm}
In the 2010 paper ``Synthesis of a new element with atomic number Z = 117'', Oganessian et al.\ reported the first observation of $^{293}$117 and $^{294}$117 \cite{2010Oga01}. A $^{249}$Bk target was bombarded with 252~MeV and 247~MeV $^{48}$Ca beam from the Dubna U400 cyclotron to form $^{293}$117 and $^{294}$117 in (4n) and (3n) evaporation reactions, respectively. The residues were separated with a gas-filled recoil separator and implanted in a semiconductor detector array. Subsequent $\alpha$ particle decay and spontaneous fission events were recorded in this array and in eight detectors arranged in a box configuration around the implantation detector. ``The data are consistent with the observation of two isotopes of element 117, with atomic masses 293 and 294. These isotopes undergo $\alpha$ decay with E$_\alpha$ = 11.03(8)~MeV and 10.81(10)~MeV and half-lives 14(+11,$-$4) and 78(+370,$-$36)~ms, respectively, giving rise to sequential $\alpha$-decay chains ending in spontaneous fission of $^{281}$Rg (T$_{SF} \sim$ 26~s) and $^{270}$Db (T$_{SF} \sim$ 1~d), respectively.'' Five decay chains for $^{293}$117 and one chain for $^{294}$117 were observed. Further experimental details were included in a subsequent publication \cite{2011Oga01}.


\section{Discovery of $^{294}$118}

Only one isotope has been reported for element 118. The discovery of this element has not yet been accepted by the IUPAC/IUPAP Joint Working Party. The observation of the three decay chains attributed to $^{294}$118 as described below does not satisfy the criteria for discovery of element 118 because they are not connected to any known nuclide \cite{2011Bar01}. The observation of $^{293}$118 \cite{1999Nin01} was later retracted \cite{2002Nin01}.

\subsection*{$^{294}$118}\vspace{0.0cm}
Oganessian et al.\ reported the first identification of $^{294}$118 in the 2006 paper ``Synthesis of the isotopes of elements 118 and 116 in the $^{249}$Cf and $^{245}$Cm+$^{48}$Ca fusion reactions'' \cite{2006Oga01}. A 251~MeV $^{48}$Ca beam from the Dubna U400 cyclotron bombarded an enriched $^{249}$Cf target and $^{294}$118 was formed in the (3n) evaporation reaction. The residues were separated with a gas-filled recoil separator and implanted in a semiconductor detector array. Subsequent $\alpha$ particle decay and spontaneous fission events were recorded in this array and in eight detectors arranged in a box configuration around the implantation detector. ``From the comparison of the decay properties of the nuclei synthesized in the two experiments with targets of $^{249}$Cf and $^{245}$Cm, it follows that in the $^{249}$Cf+$^{48}$Ca reaction an isotope of the new element with Z = 118 and A = 294 was observed.'' One of the three measured decay chains had been mentioned in a previous publication by the same group \cite{2004Oga02} referring to internal reports \cite{2002Oga01}. In another publication it was speculated that two events could have resulted from either $^{294}$118 or $^{295}$118 \cite{2004Oga04}.


\section{Summary}
A large fraction ($>$42\%) of the 159 isotopes discovered so far for the elements with Z $\ge$ 100 has been discovered during the last ten years as shown in Figure \ref{f:year-z100}. In many cases the discovery is based on a single or a few events which still have to be independently confirmed. Thus, some of the present assignments might change in the future. As can also be seen in the figure there are still many intermediate isotopes that have not been observed yet. In a conservative estimate counting only the missing isotopes within the envelope of the discovered isotopes at least 100 additional isotopes are yet to be discovered.

The majority of the isotopes ($>$90\%) have been observed in fusion-evaporation reaction which is at the present time the only mechanism available to produce isotopes with Z $>$ 102. In the near future discoveries therefore have to rely on improvements of the production and detection techniques. In the longer term high intensity radioactive beams might open up new opportunities to populate heavy elements that are currently out of reach.

\begin{figure}
	\centering
	\includegraphics[scale=0.9]{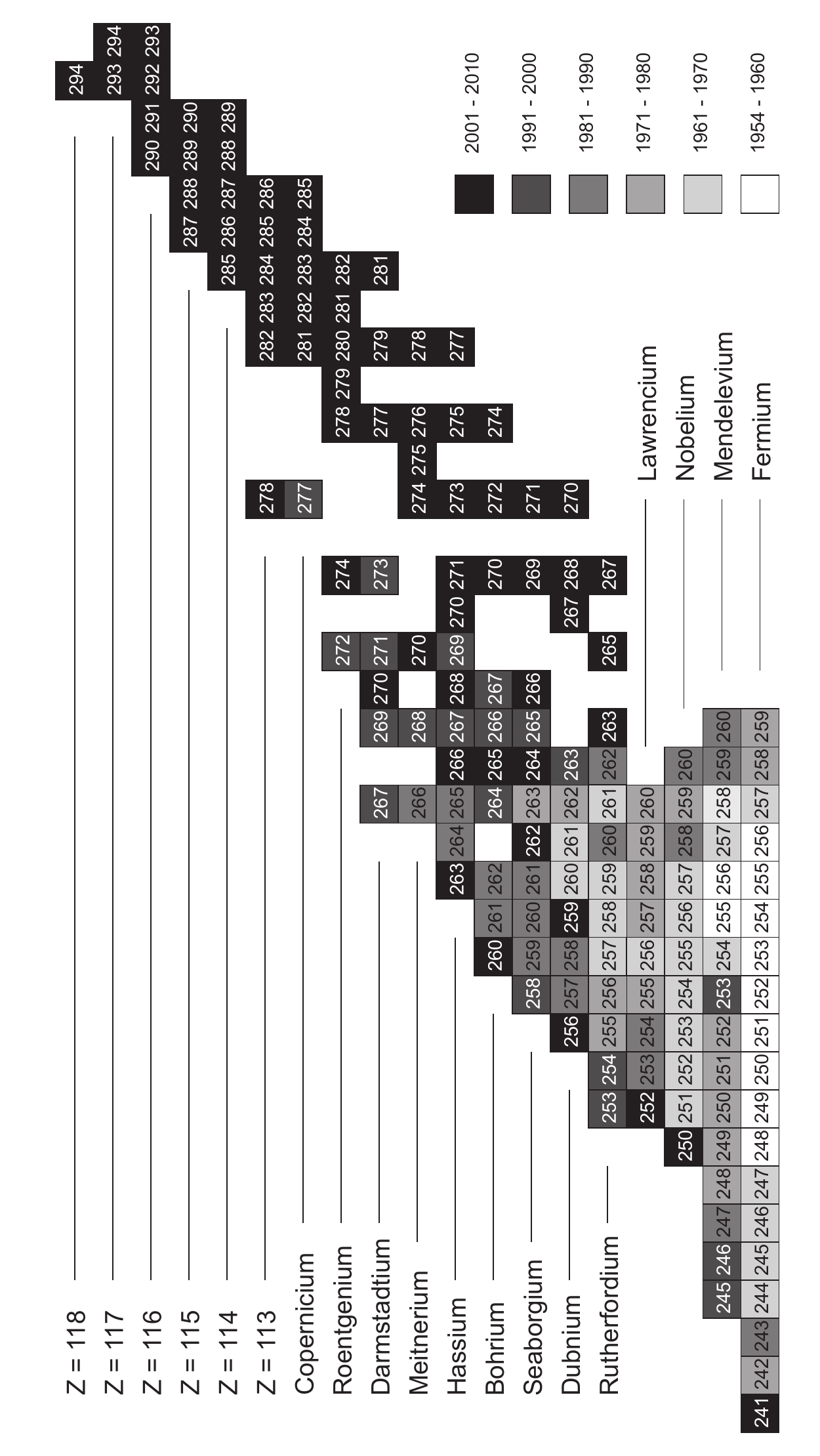}
	\caption{Discovery of the isotopes of elements with Z $\ge$ 100. The shading of the boxes corresponds to the year of discovery as indicated in the figure.}
\label{f:year-z100}
\end{figure}

The assignment of elements of Z = 114 and above can still be considered as uncertain. The observed decay chains have not yet been linked to known isotopes and thus there is still the possibility that the Z and A assignment might be incorrect. Figure \ref{f:cross-z100} indicates the isotopes that have been initially populated in the fusion-evaporation reactions (black squares) and the isotopes populated by subsequent $\alpha$-decay (light squares). The figure clearly shows the separation of the linked isotopes with Z = 109$-$113 populated by ``cold fusion'', reactions where only one neutron is evaporated and the isolated isotopes with Z $\le$ 113 which were populated by ``hot fusion'', where mostly 3$-$5 neutrons are evaporated (2 in the case of $^{291}$116). The reaction parameters (beams, targets, beam energy, excitation energy, reaction channel and cross sections) for the fusion-evaporation reactions leading to the formation of isotopes with Z $\le$ 103 are listed in Table 2. Only the values included in the original publication are listed.

\begin{figure}
	\centering
	\includegraphics[scale=0.9]{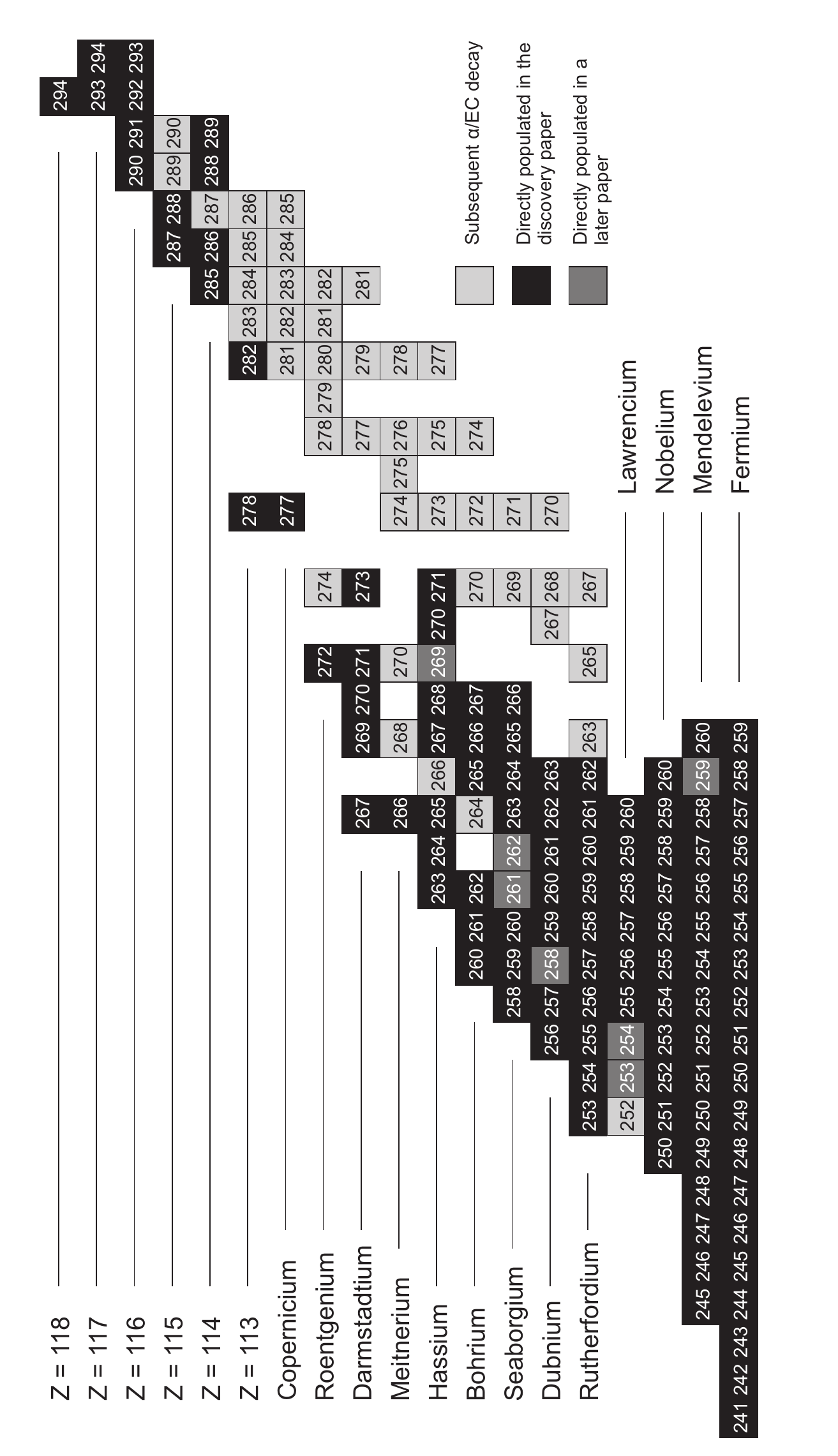}
	\caption{Discovery of the isotopes of elements with Z $\ge$ 100. The shading of the boxes corresponds to the year of discovery as indicated in the figure.}
\label{f:cross-z100}
\end{figure}

\ack

I would like to thank Walter Loveland, Mark Stoyer, and the referee for their many productive comments and suggestions, and Cathleen Fry for carefully proofreading the manuscript. This work was supported by the National Science Foundation under grant No. PHY06-06007.

\bibliography{../isotope-discovery-references}

\newpage

\newpage

\TableExplanation

\bigskip
\renewcommand{\arraystretch}{1.0}

\section{Table 1.\label{tbl1te} Discovery of isotopes of elements with Z $\ge$ 100 }
\begin{tabular*}{0.95\textwidth}{@{}@{\extracolsep{\fill}}lp{5.5in}@{}}
\multicolumn{2}{p{0.95\textwidth}}{ }\\

Isotope & Isotopes of elements with Z $\ge$ 100 \\
First author & First author of refereed publication \\
Journal & Journal of publication \\
Ref. & Reference \\
Method & Production method used in the discovery: \\

  & FE: fusion evaporation \\
  & NC: Neutron capture reactions \\
  & LP: light-particle reactions (including neutrons) \\
  & DI: deep inelastic reactions \\

Laboratory & Laboratory where the experiment was performed\\
Country & Country of laboratory\\
Year & Year of discovery \\
\end{tabular*}
\label{tableI}

\section{Table 2.\label{tbl2te} Production cross sections of isotopes of elements with Z $\ge$ 103 }
\begin{tabular*}{0.95\textwidth}{@{}@{\extracolsep{\fill}}lp{5.5in}@{}}
\multicolumn{2}{p{0.95\textwidth}}{ }\\

Isotope & Isotopes of elements with Z $\ge$ 103 \\
First author & First author of refereed publication \\
Journal & Journal of publication \\
Beam & Projectile isotope \\
Target & Target isotope \\
Energy & Beam Energy \\
E* (MeV) & Excitation Energy in MeV\\
Evaporation channel & Number of neutrons evaporated\\
Cross section & Evaporation residue population cross section\\

\end{tabular*}
\label{tableII}

\datatables 



\setlength{\LTleft}{0pt}
\setlength{\LTright}{0pt}


\setlength{\tabcolsep}{0.5\tabcolsep}

\renewcommand{\arraystretch}{1.0}

\footnotesize 

\begin{longtable}{@{\extracolsep\fill}llllllll@{}}
\caption{Discovery of isotopes of elements with Z $\ge$ 100. See page\ \pageref{tbl1te} for Explanation of Tables}
Isotope & First Author & Journal & Ref. & Method & Laboratory & Country & Year\\
\hline\\
\endfirsthead\\
\caption[]{(continued)}
Isotope & First author & Journal & Ref. & Method & Laboratory & Country & Year\\
\hline\\
\endhead
$^{241}$Fm & J. Khuyagbaatar & Eur. Phys. J. A &\cite{2008Khu01}& FE & Darmstadt & Germany &2008\\
$^{242}$Fm & G.M. Ter-Akopyan & Nucl. Phys. A &\cite{1975Ter01}& FE & Dubna & USSR &1975\\
$^{243}$Fm & G. M\"unzenberg & Z. Phys. A &\cite{1981Mun01}& FE & Darmstadt & W. Germany &1981\\
$^{244}$Fm & M. Nurmia & Phys. Lett. B &\cite{1967Nur01}& FE & Berkeley & USA &1967\\
$^{245}$Fm & M. Nurmia & Phys. Lett. B &\cite{1967Nur01}& FE & Berkeley & USA &1967\\
$^{246}$Fm & G.N. Akapev & Sov. At. Energy &\cite{1966Aka01}& FE & Dubna & USSR &1966\\
$^{247}$Fm & G.N. Flerov & Sov. At. Energy &\cite{1967Fle01}& FE & Dubna & USSR &1967\\
$^{248}$Fm & A. Ghiorso & Phys. Rev. Lett. &\cite{1958Ghi01}& FE & Berkeley & USA &1958\\
$^{249}$Fm & V.P. Perelygin & Sov. Phys. JETP &\cite{1960Per01}& FE & Moscow & USSR &1960\\
$^{250}$Fm & H. Atterling & Phys. Rev. &\cite{1954Att01}& FE & Stockholm & Sweden &1954\\
$^{251}$Fm & S. Amiel & Phys. Rev. &\cite{1957Ami01}& LP & Berkeley & USA &1957\\
$^{252}$Fm & A.M. Friedman & Phys. Rev. &\cite{1956Fri01}& LP & Argonne & USA &1956\\
$^{253}$Fm & S. Amiel & Phys. Rev. &\cite{1957Ami02}& LP & Berkeley & USA &1957\\
$^{254}$Fm & B.G. Harvey & Phys. Rev. &\cite{1954Har01}& NC & Berkeley & USA &1954\\
$^{255}$Fm & G.R. Choppin & Phys. Rev. &\cite{1954Cho01}& NC & Berkeley & USA &1954\\
$^{256}$Fm & G.R. Choppin & Phys. Rev. &\cite{1955Cho01}& NC & Berkeley & USA &1955\\
$^{257}$Fm & E.K. Hulet & Phys. Rev. Lett. &\cite{1964Hul01}& NC & Berkeley & USA &1964\\
$^{258}$Fm & E.K. Hulet & Phys. Rev. Lett. &\cite{1971Hul01}& LP & Berkeley & USA &1971\\
$^{259}$Fm & E.K. Hulet & Phys. Rev. C &\cite{1980Hul01}& LP & Los Alamos & USA &1980\\
 & & & & & &  \\
 & & & & & &  \\
$^{245}$Md & V. Ninov & Z. Phys. A &\cite{1996Nin01}& FE & Darmstadt & Germany &1996 \\
$^{246}$Md & V. Ninov & Z. Phys. A &\cite{1996Nin01}& FE & Darmstadt & Germany &1996 \\
$^{247}$Md & G. M\"unzenberg & Z. Phys. A &\cite{1981Mun01}& FE & Darmstadt & W. Germany &1981 \\
$^{248}$Md & P. Eskola & Phys. Rev. C &\cite{1973Esk02}& FE & Berkeley & USA &1973 \\
$^{249}$Md & P. Eskola & Phys. Rev. C &\cite{1973Esk02}& FE & Berkeley & USA &1973 \\
$^{250}$Md & P. Eskola & Phys. Rev. C &\cite{1973Esk02}& FE & Berkeley & USA &1973 \\
$^{251}$Md & P. Eskola & Phys. Rev. C &\cite{1973Esk02}& FE & Berkeley & USA &1973 \\
$^{252}$Md & P. Eskola & Phys. Rev. C &\cite{1973Esk02}& FE & Berkeley & USA &1973 \\
$^{253}$Md & B. Kadkhodayan & Radiochim. Acta &\cite{1992Kad01}& FE & Berkeley & USA &1992 \\
$^{254}$Md & P.R. Fields & Nucl. Phys. A &\cite{1970Fie01}& LP & Argonne & USA &1970 \\
$^{255}$Md & L. Phillips & Phys. Rev. Lett. &\cite{1958Phi01}& LP & Berkeley & USA &1958 \\
$^{256}$Md & A. Ghiorso & Phys. Rev. &\cite{1955Ghi02}& LP & Berkeley & USA &1955 \\
$^{257}$Md & T. Sikkeland & Phys. Rev. &\cite{1965Sik01}& FE & Berkeley & USA &1965 \\
$^{258}$Md & P.R. Fields & Nucl. Phys. A &\cite{1970Fie01}& LP & Argonne & USA &1970 \\
$^{259}$Md & J.F. Wild & Phys. Rev. C &\cite{1982Wil01}& FE & Berkeley & USA &1982 \\
$^{260}$Md & E.K. Hulet & Phys. Rev. C &\cite{1989Hul02}& DI & Berkeley & USA &1989 \\
 & & & & & &  \\
 & & & & & &  \\
$^{250}$No & Yu.T. Oganessian & Phys. Rev. C &\cite{2001Oga03}& FE & Dubna & Russia &2003 \\
$^{251}$No & A. Ghiorso & Phys. Rev. Lett. &\cite{1967Ghi01} & FE & Berkeley & USA &1967 \\
$^{252}$No & V.L. Mikheev & Sov. At. Energy &\cite{1967Mik02}& FE & Dubna & USSR &1967 \\
$^{253}$No & V.L. Mikheev & Sov. At. Energy &\cite{1967Mik02}& FE & Dubna & USSR &1967 \\
$^{254}$No & E.D. Donets & Sov. At. Energy &\cite{1966Don02}& FE & Dubna & USSR &1966 \\
           & B.A. Zager & Sov. At. Energy &\cite{1966Zag01}& FE & Dubna & USSR &1966 \\
$^{255}$No & V.A. Druin & Sov. At. Energy &\cite{1967Dru01}& FE & Dubna & USSR &1967 \\
$^{256}$No & E.D. Donets  & Sov. At. Energy &\cite{1963Don01}& FE & Dubna & USSR &1963 \\
$^{257}$No & A. Ghiorso & Phys. Rev. Lett. &\cite{1967Ghi01} & FE & Berkeley & USA &1967 \\
$^{258}$No & E.K. Hulet & Phys. Rev. C &\cite{1989Hul02}& FE & Berkeley & USA &1989 \\
$^{259}$No & R.J. Silva & Nucl. Phys. A &\cite{1973Sil01}& FE & Oak Ridge & USA &1973 \\
$^{260}$No & L.P. Somerville & Phys. Rev. C &\cite{1985Som01} & DI & Berkeley & USA &1985 \\
 & & & & & &  \\
 & & & & & &  \\
$^{252}$Lr & F.P. He\ss berger & Eur. Phys. J. A &\cite{2001Hes01}& FE & Darmstadt & Germany &2001 \\
$^{253}$Lr & F.P. He\ss berger & Z. Phys. A &\cite{1985Hes01}& FE & Darmstadt & W. Germany &1985 \\
$^{254}$Lr & G. M\"unzenberg & Z. Phys. A &\cite{1981Mun02}& FE & Darmstadt & W. Germany &1981 \\
$^{255}$Lr & V.A. Druin & Sov. J. Nucl. Phys. &\cite{1971Dru02}& FE & Dubna & USSR &1971 \\
$^{256}$Lr & E.D. Donets & Sov. At. Energy &\cite{1965Don01}& FE & Dubna & USSR &1965 \\
$^{257}$Lr & K. Eskola & Phys. Rev. C &\cite{1971Esk01}& FE & Berkeley & USA &1971 \\
$^{258}$Lr & K. Eskola & Phys. Rev. C &\cite{1971Esk01}& FE & Berkeley & USA &1971 \\
$^{259}$Lr & K. Eskola & Phys. Rev. C &\cite{1971Esk01}& FE & Berkeley & USA &1971 \\
$^{260}$Lr & K. Eskola & Phys. Rev. C &\cite{1971Esk01}& FE & Berkeley & USA &1971 \\
 & & & & & &  \\
 & & & & & &  \\
$^{253}$Rf & F.P. He\ss berger & Z. Phys. A &\cite{1997Hes01}& FE & Darmstadt & Germany &1997 \\
$^{254}$Rf & F.P. He\ss berger & Z. Phys. A &\cite{1997Hes01}& FE & Darmstadt & Germany &1997 \\
$^{255}$Rf & Yu.T. Oganessian & Sov. At. Energy &\cite{1975Oga01}& FE & Dubna & USSR &1975 \\
$^{256}$Rf & Yu.T. Oganessian & Sov. At. Energy &\cite{1975Oga01}& FE & Dubna & USSR &1975 \\
$^{257}$Rf & A. Ghiorso & Phys. Rev. Lett. &\cite{1969Ghi01}& FE & Berkeley & USA &1969 \\
$^{258}$Rf & A. Ghiorso & Phys. Rev. Lett. &\cite{1969Ghi01}& FE & Berkeley & USA &1969 \\
$^{259}$Rf & A. Ghiorso & Phys. Rev. Lett. &\cite{1969Ghi01}& FE & Berkeley & USA &1969 \\
$^{260}$Rf & L.P. Somerville & Phys. Rev. C &\cite{1985Som01} & FE & Berkeley & USA &1985 \\
$^{261}$Rf & A. Ghiorso & Phys. Lett. B &\cite{1970Ghi03}& FE & Berkeley & USA &1970 \\
$^{262}$Rf & L.P. Somerville & Phys. Rev. C &\cite{1985Som01} & FE & Berkeley & USA &1985 \\
$^{263}$Rf & J.V. Kratz & Radiochim. Acta &\cite{2003Kra01}& FE & Villigen & Switzerland &2003 \\
$^{264}$Rf &&&&&&& \\
$^{265}$Rf & P.A. Ellison & Phys. Rev. Lett. &\cite{2010Eli01}& FE & Berkeley & USA &2010 \\
$^{266}$Rf &&&&&&& \\
$^{267}$Rf & Yu. T. Oganessian & Phys. Rev. C &\cite{2004Oga02}& FE & Dubna & Russia &2004 \\
 & & & & & &  \\
 & & & & & &  \\
$^{256}$Db & F.P. He\ss berger & Eur. Phys. J. A &\cite{2001Hes01}& FE & Darmstadt & Germany &2001 \\
$^{257}$Db & F.P. He\ss berger & Z. Phys. A &\cite{1985Hes01}& FE & Darmstadt & W. Germany &1985 \\
$^{258}$Db & G. M\"unzenberg & Z. Phys. A &\cite{1981Mun02}& FE & Darmstadt & W. Germany &1981 \\
$^{259}$Db & Z.G. Gan & Eur. Phys. J. A &\cite{2001Gan01}& FE & Lanzhou & China & 2001 \\
$^{260}$Db & A. Ghiorso & Phys. Rev. Lett. &\cite{1970Ghi02}& FE & Berkeley & USA &1970 \\
$^{261}$Db & G.N. Flerov & Sov. At. Energy &\cite{1970Fle01}& FE & Dubna & USSR &1970 \\
$^{262}$Db & A. Ghiorso & Phys. Rev. C &\cite{1971Ghi01}& FE & Berkeley & USA &1971 \\
$^{263}$Db & J.V. Kratz & Phys. Rev. C &\cite{1992Kra01}& FE & Berkeley & USA &1992 \\
$^{264}$Db &&&&&&& \\
$^{265}$Db &&&&&&& \\
$^{266}$Db &&&&&&& \\
$^{267}$Db & Yu.T. Oganessian & Phys. Rev. C &\cite{2004Oga01}& FE & Dubna & Russia &2004 \\
$^{268}$Db & Yu.T. Oganessian & Phys. Rev. C &\cite{2004Oga01}& FE & Dubna & Russia &2004 \\
$^{269}$Db &&&&&&& \\
$^{270}$Db & Yu.T. Oganessian & Phys. Rev. Lett. &\cite{2010Oga01}& FE & Dubna & Russia &2010 \\
 & & & & & &  \\
 & & & & & &  \\
$^{258}$Sg & F.P. Hessberger & Z. Phys. A &\cite{1997Hes01}& FE & Darmstadt & Germany &1997\\
$^{259}$Sg & G. M\"unzenberg & Z. Phys. A &\cite{1985Mun01}& FE & Darmstadt & W. Germany &1985\\
$^{260}$Sg & A.G. Demin & Z. Phys. A &\cite{1984Dem01}& FE & Dubna & Russia &1984\\
$^{261}$Sg & G. M\"unzenberg & Z. Phys. A &\cite{1984Mun01}& FE & Darmstadt & W. Germany &1984\\
$^{262}$Sg & S. Hofmann & Eur. Phys. J. A &\cite{2001Hof01}& FE & Darmstadt & Germany &2001\\
$^{263}$Sg & A. Ghiorso & Phys. Rev. Lett. &\cite{1974Ghi01}& FE & Berkeley & USA &1974\\
$^{264}$Sg & K.E. Gregorich & Phys. Rev. C &\cite{2006Gre01}& FE & Berkeley & USA &2006\\
$^{265}$Sg & Yu.A. Lazarev & Phys. Rev. Lett. &\cite{1994Laz02}& FE & Dubna & Russia &1994\\
$^{266}$Sg & J. Dvorak & Phys. Rev. Lett. &\cite{2006Dvo01}& FE & Darmstadt & Germany &2006\\
$^{267}$Sg & J. Dvorak & Phys. Rev. Lett. &\cite{2008Dvo01}& FE & Darmstadt & Germany &2008\\
$^{268}$Sg &&&&&&&\\
$^{269}$Sg & P.A. Ellison & Phys. Rev. Lett. &\cite{2010Eli01}& FE & Berkeley & USA &2010\\
$^{270}$Sg &&&&&&&\\
$^{271}$Sg & Yu.T. Oganessian & Phys. Rev. C &\cite{2004Oga02}& FE & Dubna & Russia &2004\\
 & & & & & &  \\
 & & & & & &  \\
$^{260}$Bh & S. L. Nelson & Phys. Rev. Lett. &\cite{2008Nel01}& FE & Berkeley & USA &2008\\
$^{261}$Bh & G. M\"unzenberg & Z. Phys. A &\cite{1989Mun01}& FE & Darmstadt & W. Germany &1989\\
$^{262}$Bh & G. M\"unzenberg & Z. Phys. A &\cite{1981Mun02}& FE & Darmstadt & W. Germany &1981\\
$^{263}$Bh &&&&&&&\\
$^{264}$Bh & S. Hofmann & Z. Phys. A &\cite{1995Hof02}& FE & Darmstadt & Germany &1995\\
$^{265}$Bh & Z. G. Gan & Eur. Phys. J. A& \cite{2004Gan01}& FE & Lanzhou & China &2004\\
$^{266}$Bh & P.A. Wilk & Phys. Rev. Lett. &\cite{2000Wil01}& FE & Berkeley & USA &2000\\
$^{267}$Bh & P.A. Wilk & Phys. Rev. Lett. &\cite{2000Wil01}& FE & Berkeley & USA &2000\\
$^{268}$Bh &&&&&&&\\
$^{269}$Bh &&&&&&&\\
$^{270}$Bh & Yu.T. Oganessian & Phys. Rev. C &\cite{2007Oga01}& FE & Dubna & Russia &2007\\
$^{271}$Bh &&&&&&&\\
$^{272}$Bh & Yu.T. Oganessian & Phys. Rev. C &\cite{2004Oga01}& FE & Dubna & Russia &2004\\
$^{273}$Bh &&&&&&&\\
$^{274}$Bh & Yu.T. Oganessian & Phys. Rev. Lett. &\cite{2010Oga01}& FE & Dubna & Russia &2010\\
 & & & & & &  \\
 & & & & & &  \\
$^{263}$Hs & I. Dragojevic & Phys. Rev. C &\cite{2009Dra01}& FE & Berkeley & USA &2009\\
$^{264}$Hs & G. M\"unzenberg & Z. Phys. A &\cite{1986Mun01}& FE & Darmstadt & W. Germany &1986\\
$^{265}$Hs & G. M\"unzenberg & Z. Phys. A &\cite{1984Mun01}& FE & Darmstadt & W. Germany &1984\\
$^{266}$Hs & S. Hofmann & Eur. Phys. J. A &\cite{2001Hof01}& FE & Darmstadt & Germany &2001\\
$^{267}$Hs & Yu.A. Lazarev & Phys. Rev. Lett. &\cite{1995Laz02}& FE & Dubna & Russia &1995\\
$^{268}$Hs & K. Nishio & Phys. Rev. C &\cite{2010Nis01}& FE & Darmstadt & Germany &2010\\
$^{269}$Hs & S. Hofmann & Z. Phys. A &\cite{1996Hof02}& FE & Darmstadt & Germany &1996\\
$^{270}$Hs & J. Dvorak & Phys. Rev. Lett. &\cite{2006Dvo01}& FE & Darmstadt & Germany &2006\\
$^{271}$Hs & J. Dvorak & Phys. Rev. Lett. &\cite{2008Dvo01}& FE & Darmstadt & Germany &2008\\
$^{272}$Hs &&&&&&&\\
$^{273}$Hs & P.A. Ellison & Phys. Rev. Lett. &\cite{2010Eli01}& FE & Berkeley & USA &2010\\
$^{274}$Hs &&&&&&&\\
$^{275}$Hs & Yu.T. Oganessian & Phys. Rev. C &\cite{2004Oga02}& FE & Dubna & Russia &2004\\
$^{276}$Hs &&&&&&&\\
$^{277}$Hs & Ch.E. D\"ullmann & Phys. Rev. Lett. &\cite{2010Dul01}& FE & Darmstadt & Germany &2010\\
 & & & & & &  \\
 & & & & & &  \\
$^{266}$Mt & G. M\"unzenberg & Z. Phys. A &\cite{1982Mun01}& FE & Darmstadt & W. Germany &1982\\
$^{267}$Mt &&&&&&&\\
$^{268}$Mt & S. Hofmann & Z. Phys. A &\cite{1995Hof02}& FE & Darmstadt & Germany &1995\\
$^{269}$Mt &&&&&&&\\
$^{270}$Mt & K. Morita & J. Phys. Soc. Japan &\cite{2004Mor01}& FE & RIKEN & Japan &2004\\
$^{271}$Mt &&&&&&&\\
$^{272}$Mt &&&&&&&\\
$^{273}$Mt &&&&&&&\\
$^{274}$Mt & Yu.T. Oganessian & Phys. Rev. C &\cite{2007Oga01}& FE & Dubna & Russia &2007\\
$^{275}$Mt & Yu.T. Oganessian & Phys. Rev. C &\cite{2004Oga01}& FE & Dubna & Russia &2004\\
$^{276}$Mt & Yu.T. Oganessian & Phys. Rev. C &\cite{2004Oga01}& FE & Dubna & Russia &2004\\
$^{277}$Mt &&&&&&&\\
$^{278}$Mt & Yu.T. Oganessian & Phys. Rev. Lett. &\cite{2010Oga01}& FE & Dubna & Russia &2010\\
 & & & & & &  \\
 & & & & & &  \\
$^{267}$Ds& A. Ghiorso & Phys. Rev. C &\cite{1995Ghi01}& FE & Berkeley & USA &1995\\
$^{268}$Ds&&&&&&&\\
$^{269}$Ds& S. Hofmann & Z. Phys. A &\cite{1995Hof01}& FE & Darmstadt & Germany &1995\\
$^{270}$Ds& S. Hofmann & Eur. Phys. J. A &\cite{2001Hof01}& FE & Darmstadt & Germany &2001\\
$^{271}$Ds& S. Hofmann & Rep. Prog. Phys. &\cite{1998Hof01}& FE & Darmstadt & Germany &1998\\
$^{272}$Ds&&&&&&&\\
$^{273}$Ds& Yu.A. Lazarev & Phys. Rev. C &\cite{1996Laz01}& FE & Dubna & Russia &1996\\
$^{274}$Ds&&&&&&&\\
$^{275}$Ds&&&&&&&\\
$^{276}$Ds&&&&&&&\\
$^{277}$Ds& P.A. Ellison & Phys. Rev. Lett. &\cite{2010Eli01}& FE & Berkeley & USA &2010\\
$^{278}$Ds&&&&&&&\\
$^{279}$Ds& Yu.T. Oganessian & Phys. Rev. C &\cite{2004Oga03}& FE & Dubna & Russia &2004\\
$^{280}$Ds&&&&&&&\\
$^{281}$Ds& Yu.T. Oganessian & Phys. Rev. C &\cite{2004Oga03}& FE & Dubna & Russia &2004\\
 & & & & & &  \\
 & & & & & &  \\
$^{272}$Rg& S. Hofmann & Z. Phys. A &\cite{1995Hof02}& FE & Darmstadt & Germany &1995\\
$^{273}$Rg&&&&&&&\\
$^{274}$Rg& K. Morita & J. Phys. Soc. Japan &\cite{2004Mor01}& FE & RIKEN & Japan &2004\\
$^{275}$Rg&&&&&&&\\
$^{276}$Rg&&&&&&&\\
$^{277}$Rg&&&&&&&\\
$^{278}$Rg& Yu.T. Oganessian & Phys. Rev. C &\cite{2007Oga01}& FE & Dubna & Russia &2007\\
$^{279}$Rg& Yu.T. Oganessian & Phys. Rev. C &\cite{2004Oga01}& FE & Dubna & Russia &2004\\
$^{280}$Rg& Yu.T. Oganessian & Phys. Rev. C &\cite{2004Oga01}& FE & Dubna & Russia &2004\\
$^{281}$Rg& Yu.T. Oganessian & Phys. Rev. Lett. &\cite{2010Oga01}& FE & Dubna & Russia &2010\\
$^{282}$Rg& Yu.T. Oganessian & Phys. Rev. Lett. &\cite{2010Oga01}& FE & Dubna & Russia &2010\\
 & & & & & &  \\
 & & & & & &  \\
$^{277}$Cn& S. Hofmann & Z. Phys. A &\cite{1996Hof02}& FE & Darmstadt & Germany &1996\\
$^{278}$Cn&&&&&&&\\
$^{279}$Cn&&&&&&&\\
$^{280}$Cn&&&&&&&\\
$^{281}$Cn& P.A. Ellison & Phys. Rev. Lett. &\cite{2010Eli01}& FE & Berkeley & USA &2010\\
$^{282}$Cn& Yu.T. Oganessian & Phys. Rev. C &\cite{2004Oga03}& FE & Dubna & Russia &2004\\
$^{283}$Cn& Yu.T. Oganessian & Phys. Rev. C &\cite{2004Oga03}& FE & Dubna & Russia &2004\\
$^{284}$Cn& Yu.T. Oganessian & Phys. Rev. C &\cite{2004Oga03}& FE & Dubna & Russia &2004\\
$^{285}$Cn& Yu.T. Oganessian & Phys. Rev. C &\cite{2004Oga03}& FE & Dubna & Russia &2004\\
 & & & & & &  \\
 & & & & & &  \\
$^{278}$113& K. Morita & J. Phys. Soc. Japan &\cite{2004Mor01}& FE & RIKEN & Japan &2004\\
$^{279}$113&&&&&&&\\
$^{280}$113&&&&&&&\\
$^{281}$113&&&&&&&\\
$^{282}$113& Yu.T. Oganessian & Phys. Rev. C &\cite{2007Oga01}& FE & Dubna & Russia &2007\\
$^{283}$113& Yu.T. Oganessian & Phys. Rev. C &\cite{2004Oga01}& FE & Dubna & Russia &2004\\
$^{284}$113& Yu.T. Oganessian & Phys. Rev. C &\cite{2004Oga01}& FE & Dubna & Russia &2004\\
$^{285}$113& Yu.T. Oganessian & Phys. Rev. Lett. &\cite{2010Oga01}& FE & Dubna & Russia &2010\\
$^{286}$113& Yu.T. Oganessian & Phys. Rev. Lett. &\cite{2010Oga01}& FE & Dubna & Russia &2010\\
 & & & & & &  \\
 & & & & & &  \\
$^{285}$114& P.A. Ellison & Phys. Rev. Lett. &\cite{2010Eli01}& FE & Berkeley & USA &2010\\
$^{286}$114& Yu.T. Oganessian & Phys. Rev. C &\cite{2004Oga03}& FE & Dubna & Russia &2004\\
$^{287}$114& Yu.T. Oganessian & Phys. Rev. C &\cite{2004Oga03}& FE & Dubna & Russia &2004\\
$^{288}$114& Yu.T. Oganessian & Phys. Rev. C &\cite{2004Oga03}& FE & Dubna & Russia &2004\\
$^{289}$114& Yu.T. Oganessian & Phys. Rev. C &\cite{2004Oga03}& FE & Dubna & Russia &2004\\
 & & & & & &  \\
 & & & & & &  \\
$^{287}$115& Yu.T. Oganessian & Phys. Rev. C &\cite{2004Oga01}& FE & Dubna & Russia &2004\\
$^{288}$115& Yu.T. Oganessian & Phys. Rev. C &\cite{2004Oga01}& FE & Dubna & Russia &2004\\
$^{289}$115& Yu.T. Oganessian & Phys. Rev. Lett. &\cite{2010Oga01}& FE & Dubna & Russia &2010\\
$^{290}$115& Yu.T. Oganessian & Phys. Rev. Lett. &\cite{2010Oga01}& FE & Dubna & Russia &2010\\
 & & & & & &  \\
 & & & & & &  \\
$^{290}$116& Yu.T. Oganessian & Phys. Rev. C &\cite{2004Oga03}& FE & Dubna & Russia &2004\\
$^{291}$116& Yu.T. Oganessian & Phys. Rev. C &\cite{2004Oga03}& FE & Dubna & Russia &2004\\
$^{292}$116& Yu.T. Oganessian & Phys. Rev. C &\cite{2004Oga02}& FE & Dubna & Russia &2004\\
$^{293}$116& Yu.T. Oganessian & Phys. Rev. C &\cite{2004Oga03}& FE & Dubna & Russia &2004\\
 & & & & & &  \\
 & & & & & &  \\
$^{293}$117& Yu.T. Oganessian & Phys. Rev. Lett. &\cite{2010Oga01}& FE & Dubna & Russia &2010\\
$^{294}$117& Yu.T. Oganessian & Phys. Rev. Lett. &\cite{2010Oga01}& FE & Dubna & Russia &2010\\
 & & & & & &  \\
 & & & & & &  \\
$^{294}$118& Yu.T. Oganessian & Phys. Rev. C &\cite{2006Oga01}& FE & Dubna & Russia &2006\\
 \\
\end{longtable}

\setlength{\LTleft}{0pt}
\setlength{\LTright}{0pt}


\setlength{\tabcolsep}{0.5\tabcolsep}

\renewcommand{\arraystretch}{1.0}

\footnotesize 

\begin{longtable}{@{\extracolsep\fill}lllcccccc@{}}
\caption{Production cross sections of isotopes of elements with Z $\ge$ 103. Only the values included in the original publication are listed. See page\ \pageref{tbl2te} for Explanation of Tables}
Isotope & Author & Journal & Beam & Target & Energy & E* (MeV) & Evaporation channel & Cross section \\
\hline\\
\endfirsthead\\
\caption[]{(continued)}
Isotope & Author & Journal & Beam & Target & Energy & E* (MeV) & Evaporation channel & Cross section \\
\hline\\
\endhead
$^{252}$Lr & F.P. He\ss berger &\cite{2001Hes01}& \multicolumn{6}{l}{ Not directly populated } \\
$^{253}$Lr & A.V. Yeremin\footnote{Subsequent publication, not the original discovery paper}&\cite{1997Yer01}&$^{27}$Al&$^{232}$Th& &60&6n&1.3$\pm$0.5~nb \\
$^{254}$Lr & A.V. Yeremin\footnotemark[1]&\cite{1997Yer01}&$^{27}$Al&$^{232}$Th& &53&5n&1.7$\pm$0.5~nb \\
$^{255}$Lr & V.A. Druin &\cite{1971Dru02}&$^{16}$O&$^{243}$Am& & &4n&  \\
$^{256}$Lr & E.D. Donets &\cite{1965Don01}&$^{18}$O&$^{243}$Am&$\sim$96~MeV& &5n&60~nb \\
$^{257}$Lr & K. Eskola &\cite{1971Esk01}&$^{11}$B&$^{249}$Cf&$\sim$65~MeV& &3n&$\sim$20~nb \\
$^{258}$Lr & K. Eskola &\cite{1971Esk01}&$^{15}$N&$^{248}$Cm&$\sim$85~MeV& &5n&$\sim$200~nb \\
$^{259}$Lr & K. Eskola &\cite{1971Esk01}&$^{15}$N&$^{248}$Cm&$\sim$80~MeV& &4n&$\sim$50~nb \\
$^{260}$Lr & K. Eskola &\cite{1971Esk01}&$^{15}$N&$^{248}$Cm&78~MeV& &3n&$\sim$2~nb \\
&&&&&&&& \\
&&&&&&&& \\
$^{253}$Rf & F.P. He\ss berger &\cite{1997Hes01}&$^{50}$Ti&$^{204}$Pb&4.68~AMeV&15.6&1n&0.11$\pm$0.04~nb \\
$^{254}$Rf & F.P. He\ss berger &\cite{1997Hes01}&$^{50}$Ti&$^{206}$Pb&4.81~AMeV&21.5&2n&2.4$\pm$0.2~nb \\
$^{255}$Rf & Yu.T. Oganessian &\cite{1975Oga01}&$^{50}$Ti&$^{207}$Pb&260~MeV& &2n&3~nb \\
$^{256}$Rf & Yu.T. Oganessian &\cite{1975Oga01}&$^{50}$Ti&$^{208}$Pb&260~MeV& &2n&6~nb \\
$^{257}$Rf & A. Ghiorso &\cite{1969Ghi01}&$^{12}$C&$^{249}$Cf& & &4n&  \\
$^{258}$Rf & A. Ghiorso &\cite{1969Ghi01}&$^{12}$C&$^{249}$Cf& & &3n&  \\
 & & &$^{13}$C&$^{249}$Cf& & &4n&  \\
$^{259}$Rf & A. Ghiorso &\cite{1969Ghi01}&$^{13}$C&$^{249}$Cf& & &3n&  \\
$^{260}$Rf & L.P. Somerville &\cite{1985Som01} &$^{15}$N&$^{249}$Bk&80~MeV& &4n&10~nb \\
$^{261}$Rf & A. Ghiorso &\cite{1970Ghi03}&$^{18}$O&$^{248}$Cm&90-100~MeV& &5n&5~nb \\
$^{262}$Rf & L.P. Somerville &\cite{1985Som01} &$^{18}$O&$^{248}$Cm&89~MeV& &4n&5~nb \\
$^{263}$Rf & J.V. Kratz &\cite{2003Kra01}&\multicolumn{6}{l}{ Not directly populated } \\
$^{264}$Rf & &&&&&&& \\
$^{265}$Rf & P.A. Ellison &\cite{2010Eli01}&\multicolumn{6}{l}{ Not directly populated } \\
$^{266}$Rf & &&&&&&& \\
$^{267}$Rf & Yu.T. Oganessian &\cite{2004Oga02}&\multicolumn{6}{l}{ Not directly populated } \\
&&&&&&&& \\
&&&&&&&& \\
$^{256}$Db & F.P. He\ss berger &\cite{2001Hes01}&$^{50}$Ti&$^{209}$Bi&5.08~AMeV&$\sim$31&3n&0.2~nb \\
$^{257}$Db & F.P. He\ss berger &\cite{1985Hes01}&$^{50}$Ti&$^{209}$Bi&4.75-4.95~MeV/u& &2n&2.1$\pm$0.8~nb \\
$^{258}$Db & F.P. He\ss berger\footnotemark[1] &\cite{1985Hes01}&$^{50}$Ti&$^{209}$Bi&4.75-4.95~MeV/u& &1n&2.9$\pm$0.3~nb \\
$^{259}$Db & Z.G. Gan &\cite{2001Gan01}& $^{22}$Ne & $^{241}$Am & 118~MeV& &4n&1.6$\pm$1.2~nb \\
$^{260}$Db & A. Ghiorso &\cite{1970Ghi02}&$^{15}$N&$^{249}$Cf&85~MeV& & 4n &$\sim$3~nb \\
$^{261}$Db & G.N. Flerov &\cite{1970Fle01}&$^{22}$Ne&$^{243}$Am&$\sim$115~MeV& &4n&$\sim$7~nb\\
$^{262}$Db & A. Ghiorso &\cite{1971Ghi01}&$^{18}$O&$^{249}$Bk& & &5n&  \\
$^{263}$Db & J.V. Kratz &\cite{1992Kra01}&$^{18}$O&$^{249}$Bk&93~MeV& &4n&10$\pm$6~nb \\
$^{264}$Db & &&&&&&& \\
$^{265}$Db & &&&&&&& \\
$^{266}$Db & &&&&&&& \\
$^{267}$Db & Yu.T. Oganessian &\cite{2004Oga01}&\multicolumn{6}{l}{ Not directly populated } \\
$^{268}$Db & Yu.T. Oganessian &\cite{2004Oga01}&\multicolumn{6}{l}{ Not directly populated } \\
$^{269}$Db & &&&&&&& \\
$^{270}$Db & Yu.T. Oganessian &\cite{2010Oga01}&\multicolumn{6}{l}{ Not directly populated } \\
&&&&&&&& \\
&&&&&&&& \\
$^{258}$Sg & F.P. He\ss berger &\cite{1997Hes01}&$^{51}$V&$^{209}$Bi&4.91~AMeV&21.5&2n&38$\pm$13~pb \\

$^{259}$Sg & G. M\"unzenberg &\cite{1985Mun01}&$^{54}$Cr&$^{207}$Pb&262$\pm$3~MeV& &2n&320$^{+250}_{-120}$~pb \\
$^{260}$Sg & A.G. Demin &\cite{1984Dem01}&$^{54}$Cr&$^{207}$Pb&290~MeV& &1n&300~pb \\
 & & &$^{54}$Cr&$^{208}$Pb&290~MeV& &2n&400~pb \\
$^{261}$Sg & G. M\"unzenberg\footnotemark[1] &\cite{1985Mun01}&$^{54}$Cr&$^{208}$Pb&257$\pm$3~MeV& &1n&500$\pm$140~pb \\
$^{262}$Sg & K. Nishio\footnotemark[1]&\cite{2006Nis01}&$^{30}$Si&$^{238}$U&163.5~MeV&50.6&6n&22$^{+51}_{-18}$~pb \\
 & K.E. Gregorich\footnotemark[1]&\cite{2006Gre01}&$^{30}$Si&$^{238}$U&165.1$\pm$2.2~MeV&53.7$\pm$2.0&6n&$\sim$25~pb \\
$^{263}$Sg & A. Ghiorso &\cite{1974Ghi01}&$^{18}$O&$^{249}$Cf&95~MeV& &4n&$\sim$0.3~nb \\
$^{264}$Sg & K. Gregorich &\cite{2006Gre01}&$^{30}$Si&$^{238}$U&148.7$\pm$2.2~MeV&39.3$\pm$2.0&4n&9$^{+6}_{-4}$~pb \\
$^{265}$Sg & Yu.A. Lazarev &\cite{1994Laz02}&$^{22}$Ne&$^{248}$Cm&121~MeV& &5n&260~pb \\
$^{266}$Sg & J. Dvorak &\cite{2006Dvo01}&$^{26}$Mg&$^{248}$Cm&134.6-136.6~MeV& &4n&3~pb \\
$^{267}$Sg & J. Dvorak &\cite{2008Dvo01}&\multicolumn{6}{l}{ Not directly populated } \\
$^{268}$Sg & &&&&&&& \\
$^{269}$Sg & P.A. Ellison &\cite{2010Eli01}&\multicolumn{6}{l}{ Not directly populated } \\
$^{270}$Sg & &&&&&&& \\
$^{271}$Sg & Yu.T. Oganessian &\cite{2004Oga02}&\multicolumn{6}{l}{ Not directly populated } \\
&&&&&&&& \\
&&&&&&&& \\
$^{260}$Bh & S.L. Nelson &\cite{2008Nel01}   &$^{52}$Cr&$^{209}$Bi&257.0~MeV   &	15.0 & 1n & 35$^{+29}_{-20}$~pb \\
$^{261}$Bh & G. M\"unzenberg &\cite{1989Mun01}&$^{54}$Cr&$^{209}$Bi&4.92~MeV/u, 5.00~MeV/u&24$\pm$2&2n&36$^{+22}_{-14}$~pb \\
$^{262}$Bh & G. M\"unzenberg &\cite{1981Mun02}&$^{54}$Cr&$^{209}$Bi&4.85~MeV/u, 4.95~MeV/u&20$\pm$2\footnote{Value from \cite{1989Mun01}}&1n&163$\pm$34~pb\footnotemark[2] \\
$^{263}$Bh & &&&&&&& \\
$^{264}$Bh & S. Hofmann &\cite{1995Hof02}&\multicolumn{6}{l}{ Not directly populated } \\
$^{265}$Bh & Z.G. Gan   &\cite{2004Gan01}&$^{26}$Mg&$^{243}$Am&168~MeV& &4n&   \\
$^{266}$Bh & P.A. Wilk &\cite{2000Wil01}&$^{22}$Ne&$^{249}$Bk&123~MeV& &5n&25-250~pb \\
$^{267}$Bh & P.A. Wilk &\cite{2000Wil01}&$^{22}$Ne&$^{249}$Bk&117~MeV& &4n&96$^{+55}_{-25}$~pb \\
$^{268}$Bh & &&&&&&& \\
$^{269}$Bh & &&&&&&& \\
$^{270}$Bh & Yu.T. Oganessian &\cite{2007Oga01}&\multicolumn{6}{l}{ Not directly populated } \\
$^{271}$Bh & &&&&&&& \\
$^{272}$Bh & Yu.T. Oganessian &\cite{2004Oga01}&\multicolumn{6}{l}{ Not directly populated } \\
$^{273}$Bh & &&&&&&& \\
$^{274}$Bh & Yu.T. Oganessian &\cite{2010Oga01}&\multicolumn{6}{l}{ Not directly populated } \\
&&&&&&&& \\
&&&&&&&& \\
$^{263}$Hs & I. Dragojevic &\cite{2009Dra01}&$^{56}$Fe&$^{208}$Pb&276.4~MeV&15.2&1n&21$^{+13}_{-8.4}$~pb \\
$^{264}$Hs & G. M\"unzenberg &\cite{1986Mun01}&$^{58}$Fe&$^{207}$Pb&5.04~MeV/u&19$\pm$2&1n&3.2$^{+6.1}_{-2.6}$~pb \\
$^{265}$Hs & G. M\"unzenberg &\cite{1984Mun01}&$^{58}$Fe&$^{208}$Pb&5.02~MeV/u&18$\pm$2&1n&19$^{+18}_{-11}$~pb \\
$^{266}$Hs & S. Hofmann &\cite{2001Hof01}&\multicolumn{6}{l}{ Not directly populated } \\
$^{267}$Hs & Yu.A. Lazarev &\cite{1995Laz02}&$^{34}$S&$^{238}$U&186~MeV&$\sim$50&5n&2.5~pb \\
$^{268}$Hs & K. Nishio &\cite{2010Nis01}&$^{34}$S&$^{238}$U&152.0~MeV&40&4n&0.54$^{+1.3}_{-0.45}$~pb \\
$^{269}$Hs & A. T\"urler\footnotemark[1]&\cite{2003Tur01}&$^{26}$Mg&$^{248}$Cm&143.7-146.8~MeV& &5n&$\sim$6~pb \\
$^{270}$Hs & J. Dvorak &\cite{2006Dvo01}&$^{26}$Mg&$^{248}$Cm&136~MeV&40&4n&7~pb \\
$^{271}$Hs & J. Dvorak &\cite{2008Dvo01}&$^{26}$Mg&$^{248}$Cm&130~MeV&35&3n&$\sim$2-3~pb \\
$^{272}$Hs & &&&&&&& \\
$^{273}$Hs & P.A. Ellison &\cite{2010Eli01}&\multicolumn{6}{l}{ Not directly populated } \\
$^{274}$Hs & &&&&&&& \\
$^{275}$Hs & Yu.T. Oganessian &\cite{2004Oga02}&\multicolumn{6}{l}{ Not directly populated } \\
$^{276}$Hs & &&&&&&& \\
$^{277}$Hs & Ch.E. D\"ullmann &\cite{2010Dul01}&\multicolumn{6}{l}{ Not directly populated } \\
&&&&&&&& \\
&&&&&&&& \\
$^{266}$Mt & G. M\"unzenberg &\cite{1982Mun01}&$^{58}$Fe&$^{209}$Bi&5.15~MeV/u&20-26&1n&$\sim$10~pb \\
$^{267}$Mt & &&&&&&& \\
$^{268}$Mt & S. Hofmann &\cite{1995Hof02}&\multicolumn{6}{l}{ Not directly populated } \\
$^{269}$Mt & &&&&&&& \\
$^{270}$Mt & K. Morita &\cite{2004Mor01}&\multicolumn{6}{l}{ Not directly populated } \\
$^{271}$Mt & &&&&&&& \\
$^{272}$Mt & &&&&&&& \\
$^{273}$Mt & &&&&&&& \\
$^{274}$Mt & Yu.T. Oganessian &\cite{2007Oga01}&\multicolumn{6}{l}{ Not directly populated } \\
$^{275}$Mt & Yu.T. Oganessian &\cite{2004Oga01}&\multicolumn{6}{l}{ Not directly populated } \\
$^{276}$Mt & Yu.T. Oganessian &\cite{2004Oga01}&\multicolumn{6}{l}{ Not directly populated } \\
$^{277}$Mt & &&&&&&& \\
$^{278}$Mt & Yu.T. Oganessian &\cite{2010Oga01}&\multicolumn{6}{l}{ Not directly populated } \\
&&&&&&&& \\
&&&&&&&& \\
$^{267}$Ds& A. Ghiorso &\cite{1995Ghi01}&$^{59}$Co&$^{209}$Bi&290-310~MeV&11-27&1n&  \\
$^{268}$Ds& &&&&&&& \\
$^{269}$Ds& S. Hofmann &\cite{1995Hof01}&$^{62}$Ni&$^{208}$Pb&311~MeV&12.3&1n&3.3$^{+6.2}_{-2.7}$~pb \\
$^{270}$Ds& S. Hofmann &\cite{2001Hof01}&$^{64}$Ni&$^{207}$Pb&317~MeV&14.0&1n&13$\pm$5~pb \\
$^{271}$Ds& S. Hofmann &\cite{1998Hof01}&$^{64}$Ni&$^{208}$Pb&313.0~MeV&9.85&1n&15$^{+9}_{-6}$~pb \\
$^{272}$Ds& &&&&&&& \\
$^{273}$Ds& Yu.A. Lazarev &\cite{1996Laz01}&$^{34}$S&$^{244}$Pu&190~MeV&50&5n&0.4~pb \\
$^{274}$Ds& &&&&&&& \\
$^{275}$Ds& &&&&&&& \\
$^{276}$Ds& &&&&&&& \\
$^{277}$Ds& P.A. Ellison &\cite{2010Eli01}&\multicolumn{6}{l}{ Not directly populated } \\
$^{278}$Ds& &&&&&&& \\
$^{279}$Ds& Yu.T. Oganessian &\cite{2004Oga03}&\multicolumn{6}{l}{ Not directly populated } \\
$^{280}$Ds& &&&&&&& \\
$^{281}$Ds& Yu.T. Oganessian &\cite{2004Oga03}&\multicolumn{6}{l}{ Not directly populated } \\
&&&&&&&& \\
&&&&&&&& \\
$^{272}$Rg& S. Hofmann &\cite{1995Hof02}&$^{64}$Ni&$^{209}$Bi&320~MeV&12.5&1n&3.5$^{+4.6}_{-2.3}$~pb \\
$^{273}$Rg& &&&&&&& \\
$^{274}$Rg& K. Morita &\cite{2004Mor01}&\multicolumn{6}{l}{ Not directly populated } \\
$^{275}$Rg& &&&&&&& \\
$^{276}$Rg& &&&&&&& \\
$^{277}$Rg& &&&&&&& \\
$^{278}$Rg& Yu.T. Oganessian &\cite{2007Oga01}&\multicolumn{6}{l}{ Not directly populated } \\
$^{279}$Rg& Yu.T. Oganessian &\cite{2004Oga01}&\multicolumn{6}{l}{ Not directly populated } \\
$^{280}$Rg& Yu.T. Oganessian &\cite{2004Oga01}&\multicolumn{6}{l}{ Not directly populated } \\
$^{281}$Rg& Yu.T. Oganessian &\cite{2010Oga01}&\multicolumn{6}{l}{ Not directly populated } \\
$^{282}$Rg& Yu.T. Oganessian &\cite{2010Oga01}&\multicolumn{6}{l}{ Not directly populated } \\
&&&&&&&& \\
&&&&&&&& \\
$^{277}$Cn& S. Hofmann &\cite{1996Hof02}&$^{70}$Zn&$^{208}$Pb&344~MeV&10.1&1n&1.1$^{+1.2}_{-0.4}$~pb \\
$^{278}$Cn& &&&&&&& \\
$^{279}$Cn& &&&&&&& \\
$^{280}$Cn& &&&&&&& \\
$^{281}$Cn& P.A. Ellison &\cite{2010Eli01}&\multicolumn{6}{l}{ Not directly populated } \\
$^{282}$Cn& Yu.T. Oganessian &\cite{2004Oga03}&\multicolumn{6}{l}{ Not directly populated } \\
$^{283}$Cn& Yu.T. Oganessian &\cite{2004Oga03}&\multicolumn{6}{l}{ Not directly populated } \\
$^{284}$Cn& Yu.T. Oganessian &\cite{2004Oga03}&\multicolumn{6}{l}{ Not directly populated } \\
$^{285}$Cn& Yu.T. Oganessian &\cite{2004Oga03}&\multicolumn{6}{l}{ Not directly populated } \\
&&&&&&&& \\
&&&&&&&& \\
$^{278}$113& K. Morita &\cite{2004Mor01}&$^{70}$Zn&$^{209}$Bi&349.0~MeV&14.1$\pm$2.0&1n&55$^{+150}_{-45}$~fb \\
$^{279}$113& &&&&&&& \\
$^{280}$113& &&&&&&& \\
$^{281}$113& &&&&&&& \\
$^{282}$113& Yu.T. Oganessian &\cite{2007Oga01}&$^{48}$Ca&$^{237}$Np&244~MeV&39.1$\pm$2.2&3n&0.9$^{+1.6}_{-0.6}$~pb \\
$^{283}$113& Yu.T. Oganessian &\cite{2004Oga01}&\multicolumn{6}{l}{ Not directly populated } \\
$^{284}$113& Yu.T. Oganessian &\cite{2004Oga01}&\multicolumn{6}{l}{ Not directly populated } \\
$^{285}$113& Yu.T. Oganessian &\cite{2010Oga01}&\multicolumn{6}{l}{ Not directly populated } \\
$^{286}$113& Yu.T. Oganessian &\cite{2010Oga01}&\multicolumn{6}{l}{ Not directly populated } \\
&&&&&&&& \\
&&&&&&&& \\
$^{285}$114& P.A. Ellison &\cite{2010Eli01}&$^{48}$Ca&$^{242}$Pu&256~MeV&50&5n&0.6$^{+0.9}_{-0.5}$~pb \\
$^{286}$114& Yu.T. Oganessian &\cite{2004Oga03}&\multicolumn{6}{l}{ Not directly populated } \\
$^{287}$114& Yu.T. Oganessian &\cite{2004Oga03}&$^{48}$Ca&$^{244}$Pu&257~MeV&52.6&5n&1.1$^{+2.6}_{-0.9}$~pb \\
$^{288}$114& Yu.T. Oganessian &\cite{2004Oga03}&$^{48}$Ca&$^{244}$Pu&243~MeV&41&4n&5.3$^{+3.6}_{-2.1}$~pb \\
$^{289}$114& Yu.T. Oganessian &\cite{2004Oga03}&$^{48}$Ca&$^{244}$Pu&243~MeV&41&3n&1.7$^{+2.5}_{-1.1}$~pb \\
&&&&&&&& \\
&&&&&&&& \\
$^{287}$115& Yu.T. Oganessian &\cite{2004Oga01}&$^{48}$Ca&$^{243}$Am&253~MeV&42.4-46.5&4n&0.9$^{+3.2}_{-0.8}$~pb \\
$^{288}$115& Yu.T. Oganessian &\cite{2004Oga01}&$^{48}$Ca&$^{243}$Am&248~MeV&38.0-42.3&3n&2.7$^{+4.8}_{-1.6}$~pb \\
$^{289}$115& Yu.T. Oganessian &\cite{2010Oga01}&\multicolumn{6}{l}{ Not directly populated } \\
$^{290}$115& Yu.T. Oganessian &\cite{2010Oga01}&\multicolumn{6}{l}{ Not directly populated } \\
&&&&&&&& \\
&&&&&&&& \\
$^{290}$116& Yu.T. Oganessian &\cite{2004Oga03}&$^{48}$Ca&$^{245}$Cm&243~MeV&30.9-35.0&3n&1.3~pb \\
$^{291}$116& Yu.T. Oganessian &\cite{2004Oga03}&$^{48}$Ca&$^{245}$Cm&243~MeV&30.9-35.0&2n&0.9~pb \\
$^{292}$116& Yu.T. Oganessian &\cite{2004Oga02}&$^{48}$Ca&$^{248}$Cm&247~MeV&36.8-41.1&4n&3.3$^{+2.5}_{-1.4}$~pb \\
$^{293}$116& Yu.T. Oganessian\footnote{Reassigned from $^{292}$116 to $^{293}$116 in \cite{2004Oga03}} &\cite{2002Oga02}&$^{48}$Ca&$^{248}$Cm&240~MeV&28.9-37.2&3n&$\sim$1~pb \\
&&&&&&&& \\
&&&&&&&& \\
$^{293}$117& Yu.T. Oganessian &\cite{2010Oga01}&$^{48}$Ca&$^{249}$Bk&252~MeV&39&4n&1.3$^{+1.5}_{-0.6}$~pb \\
$^{294}$117& Yu.T. Oganessian &\cite{2010Oga01}&$^{48}$Ca&$^{249}$Bk&247~MeV&35&3n&0.5$^{+1.1}_{-0.4}$~pb \\
&&&&&&&& \\
&&&&&&&& \\
$^{294}$118& Yu.T. Oganessian &\cite{2006Oga01}&$^{48}$Ca&$^{249}$Cf&251~MeV&32.1-36.6&3n&0.5$^{+1.6}_{-0.3}$~pb \\
 \\
\end{longtable}

\end{document}